%% file: main.tex
\documentclass[12pt]{article}

\input{settings.tex}
\usepackage[utf8]{inputenc}
\usepackage{fullpage}
\usepackage{graphicx}
\usepackage{amsmath}
\usepackage{amssymb}
\usepackage{float}
\usepackage{mathtools}
\usepackage[font=small,labelfont=bf]{caption}
\usepackage{breqn}
\usepackage[nottoc,numbib]{tocbibind}

\graphicspath{ {images/} }
\usepackage{listings}
\usepackage{color} 
\definecolor{mygreen}{RGB}{28,172,0} 
\definecolor{mylilas}{RGB}{170,55,241}
\usepackage{pdfpages}
\usepackage{pdfpages}

\DeclarePairedDelimiter\abs{\lvert}{\rvert}%

\usepackage [english]{babel}
\usepackage [autostyle, english = american]{csquotes}
\MakeOuterQuote{"}

\title{Design and Control of a Photonic Neural Network Applied to High-Bandwidth Classification}
\author{Ethan Gordon '17 \\ \em{egordon@princeton.edu} \\ \\ Advisor: Paul R. Prucnal \\ \em{prucnal@princeton.edu} \\ \\ Submitted in partial fulfillment \\ of the requirements for the degree of \\ Bachelor of Science in Engineering \\ Department of Electrical Engineering \\ Princeton University \\ \\}
\date{May 8, 2017}

\DeclareFixedFont{\ttb}{T1}{txtt}{bx}{n}{12} 
\DeclareFixedFont{\ttm}{T1}{txtt}{m}{n}{12}  

\definecolor{deepblue}{rgb}{0,0,0.5}
\definecolor{deepred}{rgb}{0.6,0,0}
\definecolor{deepgreen}{rgb}{0,0.5,0}

\begin{document}

\maketitle

\newpage

\section*{Honor Statement}
I hereby declare that this Independent Work report represents my own work in accordance with University regulations.
\begin{flushright}
\includegraphics[width=0.2\textwidth]{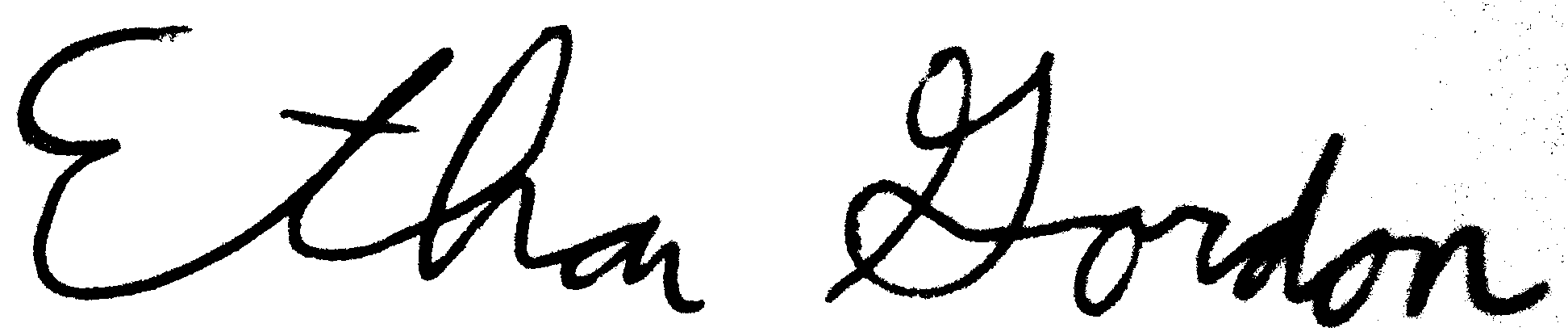} \\
Ethan K. Gordon '17
\end{flushright}

\newpage
\begin{centering}
{\LARGE Design and Control of a Photonic Neural Network Applied to High-Bandwidth Classification} \\
\vspace*{10px}
{\large Ethan Gordon '17} \\
{\em egordon@princeton.edu} \\
\end{centering}

\section*{Abstract}

Neural networks can very effectively perform multidimensional nonlinear classification. However, electronic networks suffer from significant bandwidth limitations due to carrier lifetimes and capacitive coupling. This project investigates photonic neural networks that can get around these limitations by performing both the activation function and weighted addition in the optical domain using microring resonators. These optical microring resonators provide both nonlinearity and superior fan-in without compromising bandwidth. The ability to thermally calibrate networks of cascaded axons and dendrites and train such a network to solve nonlinear classification problems are demonstrated using theory and simulations. The former is also demonstrated experimentally on a two-channel axon cascaded into a two-channel dendrite, showing good agreement between simulation and experiment. In addition, the use of transverse modes to increase the size of each photonic layer is examined. Simulations that determined the optimal waveguide geometry for using these modes were experimentally validated.

\newpage

\section*{Acknowledgements}
First and foremost, I would like to express my deepest gratitude to Alex Tait 'GS, my primary day-to-day mentor and advisor. Without him, absolutely none of this work would have been possible. I would also like to extend my thanks to the entire Lightwave Communications Laboratory for their advice, support, and infrastructure. This project was also made possible by generous funding from the School of Enginering and Applied Science and the Department of Electrical Engineering. Finally, I would like to acknowledge all of my friends and peers: the fighting ELE Class of 2017, off which I could always bounce ideas, and my friends and roommates, who were willing to listen to my ramblings and help edit this document. Thank you to everybody for helping make this a success.  \\

\noindent {\em This work is dedicated to the loving memory of my parents, Lauren and David Gordon.}
\newpage
\tableofcontents

\newpage

\section{Background}
\subsection{Motivation}
\label{sec:motivation}

Neural networks have captured the public imagination. Very large, deep networks like Google's AlphaGo have harnessed these nonlinear systems to perform impressive feats on human timescales: the order of seconds or minutes \cite{go}. However, software neural networks in general become impractical as the speed of the input signal increases, assuming the computation needs to be completed in real-time. There hardware networks can step in to fill the niche. While they cannot scale infinitely by definition, moderately-sized networks (on the order of hundreds of neurons or smaller) can provide useful, high-speed computation.

\begin{figure}[!ht]
\centering
\includegraphics[width=0.8\textwidth]{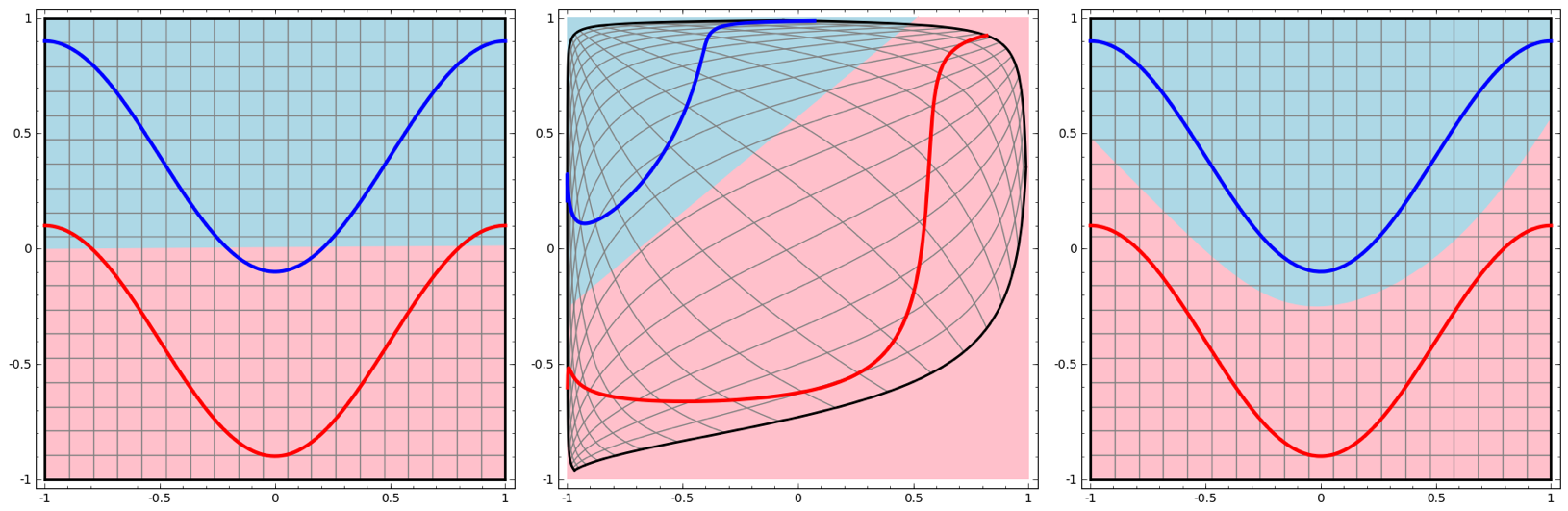}
\caption{{\bf (Left)} An optimal linear classifier for the provided data set. {\bf (Middle)} Example of using a 2-2-1 feed-forward neural network. The hidden (middle) layer warps the euclidean space, and the last layer performs linear classification in this distorted space. {\bf (Right)} The resulting effective nonlinear classification curve. Image Credit: \cite{colah}}
\label{fig:fnn}
\end{figure}

This thesis focuses on nonlinear classification: the problem of separating data into N different classes when the optimal curve of separation is more complicated than a straight line. As shown in Figure \ref{fig:fnn}, a neural network accomplishes this by using the nonlinear {\em activation function} in each layer to map the previous space onto a new, warped space. In this warped space, the different classes are linearly separable, and the last neurons draw this optimal line. Linear classification in curved space is equivalent to nonlinear classification in uncurved space. In terms of speed, a layered network's latency is the sum of the latency of each layer, but since each layer can operate in parallel, the total throughput of the network is limited by the throughput of the smallest layer.

High-speed nonlinear classification can have applications in scientific computing (such as CERN's L1 trigger) and radio (where even simple digital demodulation requires nonlinear classification). However, as the frequency of the input signal increases, even hardware networks begin to run into problems. Existing electronic networks \cite{hicann}\cite{neurogrid} are fundamentally limited by a trade-off between the bandwidth of the input signal and the size of each layer of the network. Adding more neurons in a layer allows for processing more information at once, but the increased fan-in also decreases the bandwidth of each neuron proportionally. The total throughput remains approximately constant. For example, the TrueNorth \cite{truenorth} network design has a layer size of 256 neurons, but each layer updates at a speed of 1kHz (which then limits the input signal bandwidth). The product is 256kHz, which is the total effective maximum throughput of the network. For example, this value can be achieved with audio by having the network operate on 256 audio samples at once. TrueNorth could increase the bandwidth by decreasing the layer size, but the total throughput would remain about the same.

\begin{figure}[!ht]
\centering
\includegraphics[width=0.6\textwidth]{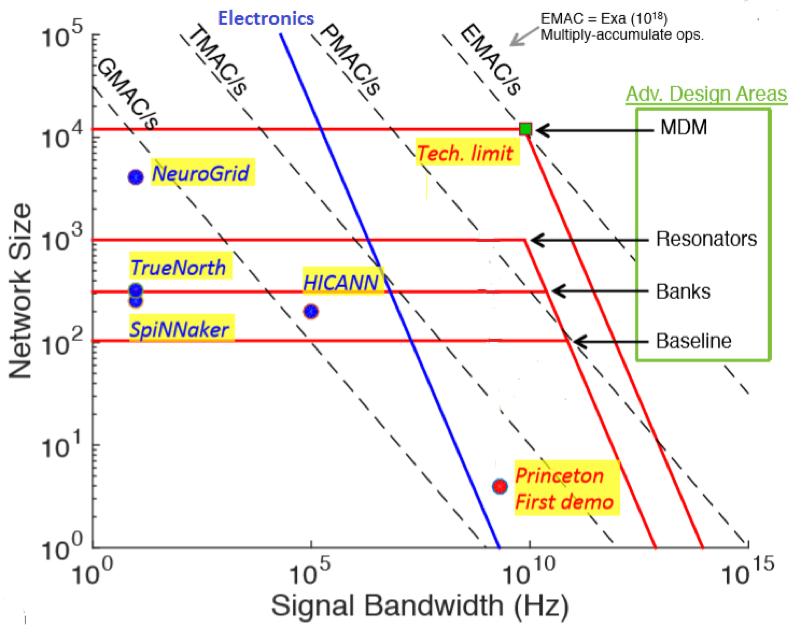}
\caption{Trade-off between the size of a network layer and the input signal bandwidth for different electronic and proposed photonic networks. Marked are dashed lines of constant throughput. A point is marked for the Lightwave Communication Lab's first neuron demonstration, covered in Section \ref{sec:prev}. Adding weighted addition to the optical domain is discussed in Section \ref{sec:design}, and expanding network size through mode-division-multiplexing (MDM) is discussed in Section \ref{sec:mdm}. }
\label{fig:bandSize}
\end{figure}

Figure \ref{fig:bandSize} demonstrates this trade-off in existing electronic networks. Such systems have trouble reaching terahertz throughput. Here is the motivation for switching to optical-electronic-optical (OEO) networks. By moving the interconnects between neurons into the optical domain, this trade-off can be beaten, allowing for large signal bandwidths for a given network size. Such networks would be even more useful in applications requiring high-speed nonlinear computation.

\subsection{Previous Work}
\label{sec:prev}
\begin{figure}[!ht]
\centering
\includegraphics[width=0.6\textwidth]{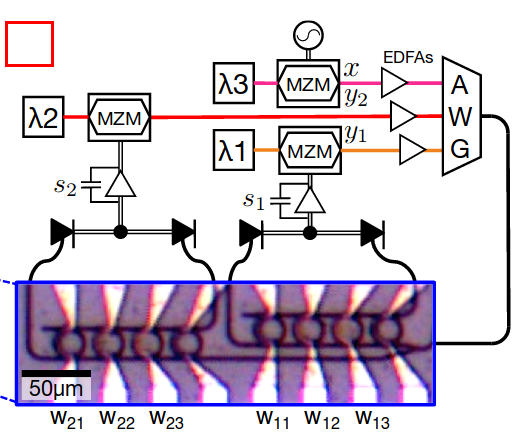}
\caption{Diagram of the components in Princeton's first neural network demonstration using wavelength division multiplexing (WDM) to distinguish between different channels. The network produced the results in Figure \ref{fig:prevHyst}. Note how the weight banks are the only integrated components, leaving the nonlinear modulation to happen off-chip.}
\label{fig:prevNet}
\end{figure}

The first proof of concept for an OEO neural network came from Princeton \cite{demo} and demonstrated neuromorphic dynamics. An outline of the network can be seen in Figure \ref{fig:prevNet}. In this demonstration, the weight banks making up the dendrites (described in more detail in Section \ref{sec:design}) were the only integrated component. These banks, combined with a pair of balanced photodiodes, carried out the weighted addition equivalent to a dot product with a weight matrix. This current signal was then amplified and fed into an off-chip Mach-Zahnder modulator, whose saturation provided the nonlinear activation function feeding back into the optical network.

\begin{figure}[!ht]
\centering
\includegraphics[width=0.6\textwidth]{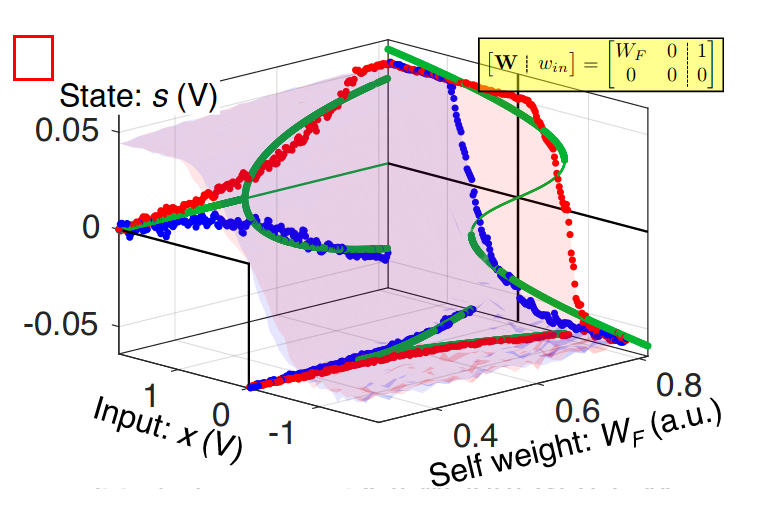}
\caption{Evidence of neuromorphic dynamics of a single-neuron in feedback with itself. As the self-weight increases, an intermediate input signal produces a bistable state, creating the hysteresis loop shown at the back of the graph (a self-weight of 0.8).}
\label{fig:prevHyst}
\end{figure}

Figure \ref{fig:prevHyst} provides evidence of neuromorphic dynamics. Modeling the activation function as being similar to hyperbolic tangent, the neuron output as $y(t)$, the self-weight as $W_f$, and the external input as $x$, we get the following network dynamics:
\begin{equation}
y_{t+1} = c\tanh(W_f*y_t + x)
\end{equation}

In the steady state, we set $y_{t+1}=y_{t}=y$ and solve for $y$. As a function of $W_f$, $y(W_f)$ bifurcates when $W_f*c>1.0$, leading to the image on the back-left of Figure \ref{fig:prevHyst}. Once the weight is high enough to reach bifurcation, the relation $y(x)$ ceases to be a function, forming instead a hysteresis loop where the value of $y$ is dependent upon the history of $x$. This leads to the S-shaped function seen on the back-right of Figure \ref{fig:prevHyst}.

While a good proof of concept, this first network demonstration suffered from quite a few drawbacks. The activation function was applied off-chip on external Mach-Zehnder interferometers, a scheme which is in general not scalable. The definition of positive and negative optical signals relative to some base positive optical power made for difficult network dynamics and required biases to be {\em weight-dependent}, making any possible experiments in future plasticity very difficult. Finally, while a control algorithm was demonstrated for this grouping of dendrites \cite{control}, that algorithm was incapable of working with different network topologies and was left untested on different examples of the same topology.

This thesis seeks to expand upon this previous work by looking at possible fixes to these limitations: investigating integraded axons in the form of microring resonators, looking at different network topologies, attempting to show nonlinear classification with strictly positive signals, and modifying the control algorithm to extend to all of these new additions.

\newpage

\section{Network Design}
\label{sec:design}
\subsection{Operating Principle}
\label{sec:principle}

\begin{figure}[!ht]
\centering
\includegraphics[width=0.9\textwidth]{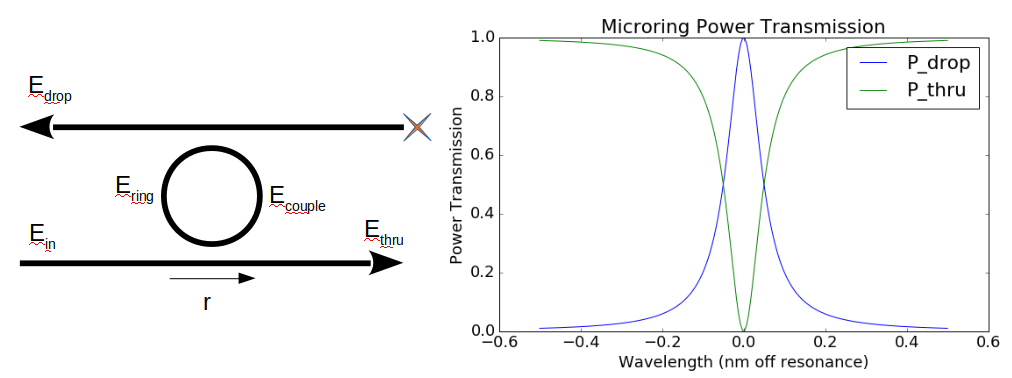}
\caption{{\bf(Left)} Layout of a microring resonator. {\bf (Right)} Power transmission spectra between the input port and the drop and thru ports. The drop port spectrum, to the second order, matches a Cauchy-Lorentz distribution. }
\label{fig:principle}
\end{figure}

To create a neural network, each neuron operates with an activation function on a weighted sum of fanned-in inputs and produces a single fanned-out output. For a neuron input $\vec{x}$, a weight vector $\vec{w}$, a bias $b$, an activation function $f$, and an output $y$, a single neuron can be modeled as:
\begin{equation}
y = f(\vec{w}\cdot\vec{x} + b)
\end{equation}
The basic unit of computation in the optical domain is the microring resonator, shown in Figure \ref{fig:principle}. Fortunately, this unit can accomplish both weighted addition and the nonlinear transfer funciton. The useful power transfer characteristics are derived here.

First consider a generic optical coupler made of two parallel waveguides, taking two optical inputs $\vec{a} = \begin{bmatrix} a_1 \\ a_2 \end{bmatrix}$ and generating two outputs $\vec{b} = \begin{bmatrix} b_1 \\ b_2 \end{bmatrix}$. In the continuous wave approximation, we assume that all operations occur independently of the frequency of the light. Therefore, each input $a_i$ and output $b_i$ can be represented as a simple complex number containing the amplitude and phase of the wave. We can model this optical coupler as a scattering matrix:
\begin{equation}
\vec{b} = S\vec{a}
\end{equation}
Assuming a lossless coupler, our matrix should also be unitary. Therefore, we can write $S$ generally as:
\begin{equation}
S = \begin{bmatrix}
r_c & t_c \\
-e^{i\theta}t_c^* & e^{i\theta}r_c^*
\end{bmatrix}
\end{equation}
where $|r_c|^2 + |t_c|^2 = 1$. Taking the limit as $t\rightarrow 0$, which is equivalent to an isolated waveguide, we know that the incoming light should experience no phase shift. Therefore, we take $r_c = r$ to be real. Additionally, due to the physical symmetry of the system, we expect this matrix to be both symmetric and persymmetric. Therefore, we have $r = e^{i\theta}r \implies e^{i\theta} = 1$ and $t_c = -t_c^*$. This last statement implies $t_c$ is purely imaginary, and can then be written as $it, t\in \mathbb{R}$. We now write our scattering matrix as:
\begin{equation}
S = \begin{bmatrix}
r & it \\
it & r
\end{bmatrix}
\end{equation}
The important takeaway is that coupled power in an optical coupler picks up a $\frac{\pi}{2}$ phase shift. For the rest of the derivation, $t$ will be replaced with $\sqrt{1-r^2}$. \\

\noindent Now consider a microring resonator with an input amplitude and phase $E_{in}$.  This input couples with the wave $E_{ring}$ already inside of the ring. We start by writing the equation for the wave coupling into the ring:
\begin{equation}
\label{eq:coupler}
E_{couple} = rE_{ring} + i\sqrt{1-r^2}E_{in}
\end{equation}
This wave is attenuated by $r$ due to the optical coupler on the other side of the ring and picks up a phase shift $\phi$ from the optical path length of the ring, allowing us to write a feedback equation for $\beta$:
\begin{equation}
E_{ring} = E_{couple} * r * e^{i\phi} = (rE_{ring} + i\sqrt{1 - r^2}E_{in})re^{i\phi} \implies E_{ring} = \frac{i\sqrt{1-r^2}E_{in} r e^{i\phi}}{1 - r^2e^{i\phi}}
\end{equation}
We can now look at the other side of the optical coupler from Equation \ref{eq:coupler} to determine the field transmission at the thru port:
\begin{equation}
T_{thru} = \frac{E_{thru}}{E_{in}} = \frac{1}{E_{in}}(r\alpha + i\sqrt{1-r^2}\beta) = r - \frac{(1-r^2)r e^{i\phi}}{1 - r^2e^{i\phi}} = \frac{r(1-e^{i\phi})}{1 - r^2e^{i\phi}}
\end{equation}
From here, we can derive the power transmission to each port of the microring.
\begin{equation}
P_{thru} = |T_{thru}|^2 = \frac{2r^2(1-cos(\phi))}{1 + r^4 - 2r^2cos(\phi)}
\end{equation}
\begin{equation}
P_{drop} = 1-P_{thru} = \frac{(1-r^2)^2}{1 + r^4 - 2r^2cos(\phi)}
\end{equation}
We can find $\phi$ with respect to the perimeter of the ring $L$, the index of refraction $n$ (creating the optical path length $nL$), and the wavelength of the light $\lambda$:
\begin{equation}
\phi = \frac{2\pi n L}{\lambda}
\end{equation}
The system is in resonance when $\phi = 2m\pi$, $m \in \mathbb{Z}$, giving us the resonant wavelengths:
\begin{equation}
\lambda_{0} = \frac{nL}{m},\ m\in\mathbb{Z}
\end{equation}
Finally, we replace $cos(\phi)$ with its second-order Taylor approximation around one of these resonant frequencies, yielding:
\begin{equation}
P_{drop} \approx \frac{\gamma^2}{\gamma^2 + (\lambda - \lambda_0)^2}
\end{equation}
\begin{equation}
P_{thru} \approx \frac{(\lambda - \lambda_0)^2}{\gamma^2 + (\lambda - \lambda_0)^2}
\end{equation}
Where:
\begin{equation}
\gamma = \frac{(1-r^2)nL}{2\pi r m^2}
\end{equation}
This shows that the power transmission spectrum of a microring around a single resonance peak can be approximated by a nonlinear Cauchy-Lorentz distribution. Especially important is the fact that this center frequency of the distribution $\lambda_0$ is proportional to the index of refraction $n$ in the medium. Silicon's index of refraction is temperature-dependent. Therefore, by using a current heater around the perimeter of the ring, the center frequency of the distribution can be shifted to any desired frequency.
\subsection{Components and Topology}

The first component of each neuron is the dendrite, charged with performing weighted addition on the optical inputs. In a WDM network, each input is an analog optical power on a given wavelength, and each weight is applied by a single microring resonator slightly detuned from the wavelength channel. Even though these microrings are cascaded together, if the tuning wavelenghts are sufficiently far apart, the continuous-wave approximation from the previous section holds, and each microring can be considered separately. After the weight bank, the power from the input is split between the combined drop port and the combined thru port. Each port then feeds into a photodiode. The current output from the photodiode is effectively independent of wavelength for the wavelengths used; it is proportional to the sum of all optical power entering the photodiode. By putting the photodiode on the drop port and the photodiode on the thru port in a balanced configuration (see Figure \ref{fig:circuit} and the explanation in Section \ref{sec:231}), the output electrical current can be described as being proportional to the difference between the thru port and the drop port optical powers.

This give us positive and negative weights. When the microring is on-resonance ($P_{drop} \approx 1$), all of the optical power on that channel goes into the drop photodiode, creating a positive elctrical current. When the microring is completely detuned ($P_{drop} \approx 0$), all the optical power on that channel goes to the thru photodiode, creating a negative electrical current. 50\% power transfer corresponds with a weight of 0.

After weighted addition, the next step is to feed the signal into some nonlinear element: the axon. Here, we can actually take advantage of the nonlinear power transfer characteristic of the microring resonator to use it as an axon as well. Cascading another microring tuned to the same channel as the microring in the dendrite would cause the continuous-wave approximation to break down. However, the axon can be configured to only connect to the rest of the network via the thru port, leaving dropped power to dissipate. This preserves the analysis done in the previous section and creates the inverted Cauchy-Lorentz transfer-function used in training, as described in Section \ref{sec:train}.

\begin{figure}[!ht]
\centering
\includegraphics[width=0.5\textwidth]{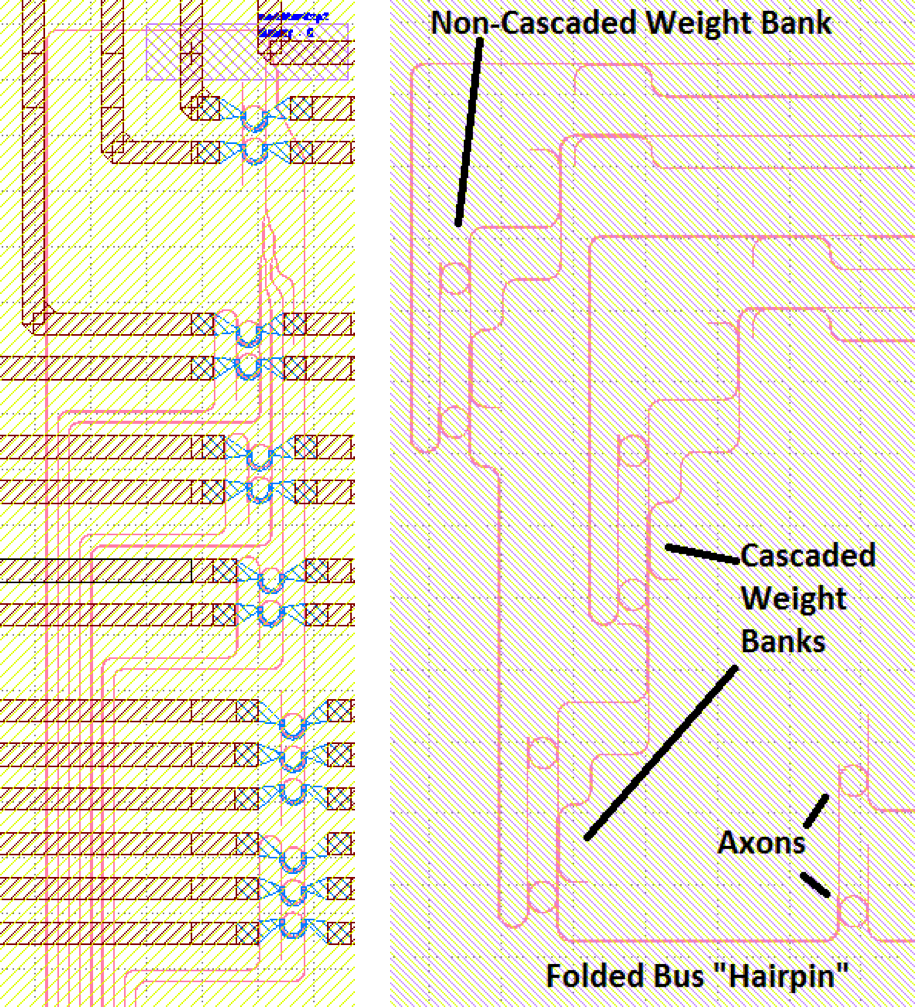}
\caption{(left) Two-branch network in star topology. Optical power for all dendrites in a branch, as well as all branches, originates at a single central point, hindering scalability. (right) One-branch network in hairpin topology. Intermediate weight banks pull power off the bus and shepherd it to each weight bank.}
\label{fig:topology}
\end{figure}

The next system-level problem is to determine the best configuration of dendrite-microrings and axon-microrings. Two possible technologies are shown in Figure \ref{fig:topology}. First, in the optical domain, let a {\em branch} be defined as optical transmission from an axon into a dendrite network. In the star topology, which was used the previous OEO neuron demonsration, each branch has optical power flowing from the axon into a splitter, dividing the power up equally between all dendrite weight banks. While easier to design and control, such a network runs into issues of scalability as the number of dendrites increases, since each dendrite needs to connect directly to a single point on the network.

A possibly better solution is a folded bus, or "hairpin," topology. Here, each branch starts out with the axon rings, which drop optical power onto a single bus. This bus folds into the dendrite network. Each neuron then consists of two cascaded weight banks: one pulls optical power off of the bus, and the second does the actual weighting. For an $n$-dendrite branch, the first dendrite can pull $1/n$ power off of the bus, the second can pull $1/(n-1)$, and so on, making sure each dendrite gets the same amount of optical power to weight. The benefits of a hairpin topology include scalability and configurability. On the first front, since each dendrite need only connect directly to the previous dendrite rather than a star point, the amount of waveguide required per dendrite is constant, rather than increasing linearly with the number of dendrites. Additionally, since the system uses microrings to drop power from the bus before weighting, the amount of dropped power can be configured. This, for example, can allow one dendrite to have magnified weights at the expense of other dendrites.

In addition, the hairpin topology is necessary for extension into Mode-Division Multiplexing (MDM) networks, discussed further in Section \ref{sec:mdm}.

\subsection{Experimental Design: A 2-3-1 Feed-Forward Network}
\label{sec:231}

\begin{figure}[!ht]
\centering
\includegraphics[width=0.8\textwidth]{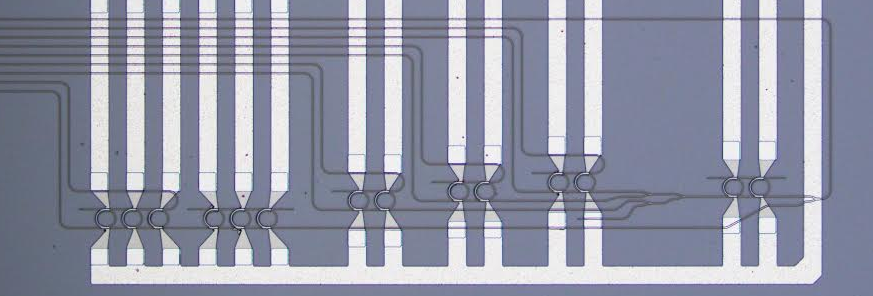}
\caption{Optical portion of 2-3-1 Feed Forward Network. From the right: optical pump power splits into two branches. An $n$x$m$ layer starts with $n$ axons to modulate the signal from the previous layer (or input) onto the bus. There are then $m$ weight banks to perform the weighted addition.}
\label{fig:231}
\end{figure}

While the hairpin topology makes the most sense long-term, the problems of control, calibration, and training are easier to initially solve on a star network. To demonstrate nonlinear classification, a 2-3-1 feed-forward network topology was chosen, and the optical components can be seen in Figure \ref{fig:231}. The input optical power, a combination of three wavelength channels, is split into two branches. On the first branch, two axons modulate the input signal onto two of the wavelength channels. These in turn are split and transmitted through three dendrite weight banks, creating three sets of complementary optical outputs to biased photodetectors. Note how, in this first branch, there is an unhandled optical wavelength channel, and this will show up as an effective extra optical bias in the training algorithm. The second optical branch operates similarly, with three axon microring-resonators modulating the three wavelength channels and feeding into a single dendrite weight bank to produce the final output.

\begin{figure}[!ht]
\centering
\includegraphics[width=0.8\textwidth]{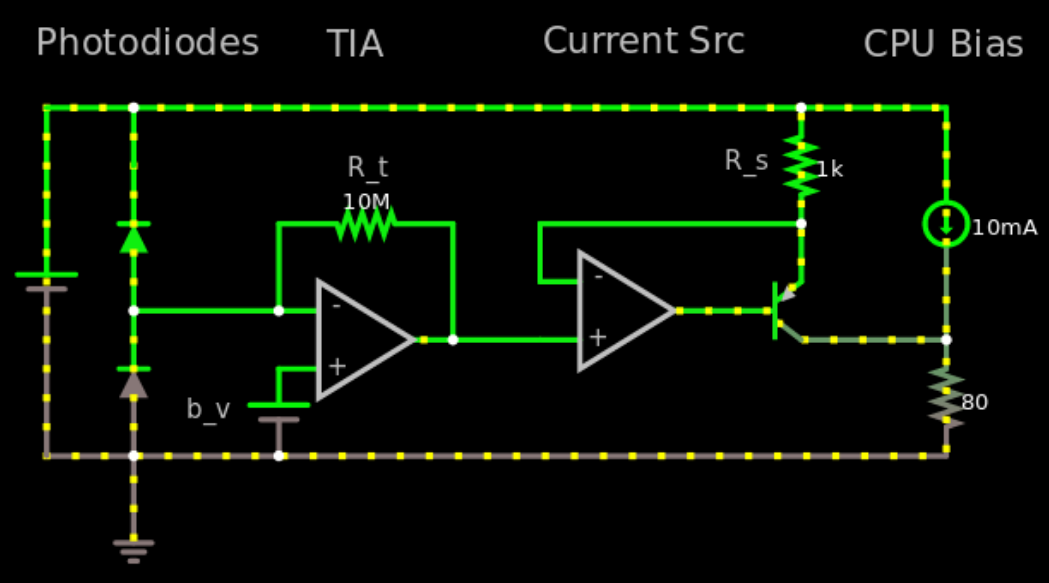}
\caption{Electrical component of 2-3-1 feed forward network. One circuit is required per dendrite weight bank.}
\label{fig:circuit}
\end{figure}

The relatively simple circuitry involved in the electrical portion of the network is shown in Figure \ref{fig:circuit}. For each weight bank, the two optical outputs are captured by two balanced photodiodes. Due to the flat frequency response of the diodes, the output current for each diode is essentially proportional to the sum of the input optical powers. By placing two diodes in series and measuring the output current from the intermediate node, the output becomes the difference between the sum of optical powers at the thru port and the sum of optical powers at the drop port. This is what allows for both positive and negative weights. From here, we can write the expression for the current output $c$ of a given weight bank as
\begin{equation}
\label{eq:current}
c = R(\vec{w}\cdot\vec{x} + b_p) = R((2\vec{t}-\vec{1})\cdot\vec{x} + b_p)
\end{equation}
where $x$ is the input optical power on each channel, $t\in[0,1]$ is the transmission through the tuned microring, $b_p$ encompasses any bias optical power from any unweighted channels, and $R$ is the responsitivity of the photodiodes (i.e. the proportionality between input optical power and output current).

This output current signal is then amplified and converted to a voltage through a transimpedence amplifier. The gain is set by the resistor $R_t$ and the offset is set by $b_v$. In practice, since the voltage $b_v$ is reflected backwards to the node between the photodiodes, it is best to fix it at $V_{dd}/2$ so that both photodiodes have the same reverse bias. This ensures that the photodiodes operate symmetrically with the same responsivity. The resulting voltage is fed through a voltage follower into a PNP transistor that to thre first order acts as an ideal current source. This current signal is attenuated by the resistor $R_s$ and is combined with the bias current controlled directly by an external CPU before feeding into the corresponding axon. Combining all of these circuit elements, the final circuit transfer function is:
\begin{equation}
\label{eq:circuit}
c_{out} = \frac{R_t*R(\vec{w}\cdot\vec{x} + b_p) + b_v}{R_s} + b_c
\end{equation}
Where $b_c$ is the current bias from the CPU. How this circuit's transfer function applies to a backpropagation algorithm is discussed with Network Training in Section \ref{sec:train}, along with a full simulation of this network design.

\newpage

\section{Calibration, Control, and Training}

Before building the physical network described in the previous section, it is necessary to construct a procedure to thermally calibrate and train it.

The goal of thermal calibration is to create an invertible map between the input current vector and the location of the resonance peaks (and by extension power transmission) for every microring in the system. In other words, it becomes possible to determine how much current to send to each microring to realize a given weight vector $\vec{w}$ that has be mapped to the power transmission vector $\vec{t}$ via Equation \ref{eq:current}. Meanwhile, the orthogonal problem of training involves using backpropagation to determine the optimal weights and biases for a given input dataset. 

The following three subsections discuss both procedures and provide the results of their application on simulated devices.\footnote{While the creation of simulation models for both the optical devices and laboratory instruments was an extensive part of this thesis, that code does little to enhance understanding of these procedures. Documented code can be requested directly from the Lightwave Comunications Laboratory.} To demonstrate good agreement between simulation and experiment, empirical results of thermal calibration are provided in the final subsection. Experimental validation of network training was not done here, but it is the immediate next step in future work.

\subsection{Basic Calibration Procedure}

\begin{figure}[!ht]
\centering
\includegraphics[width=\textwidth]{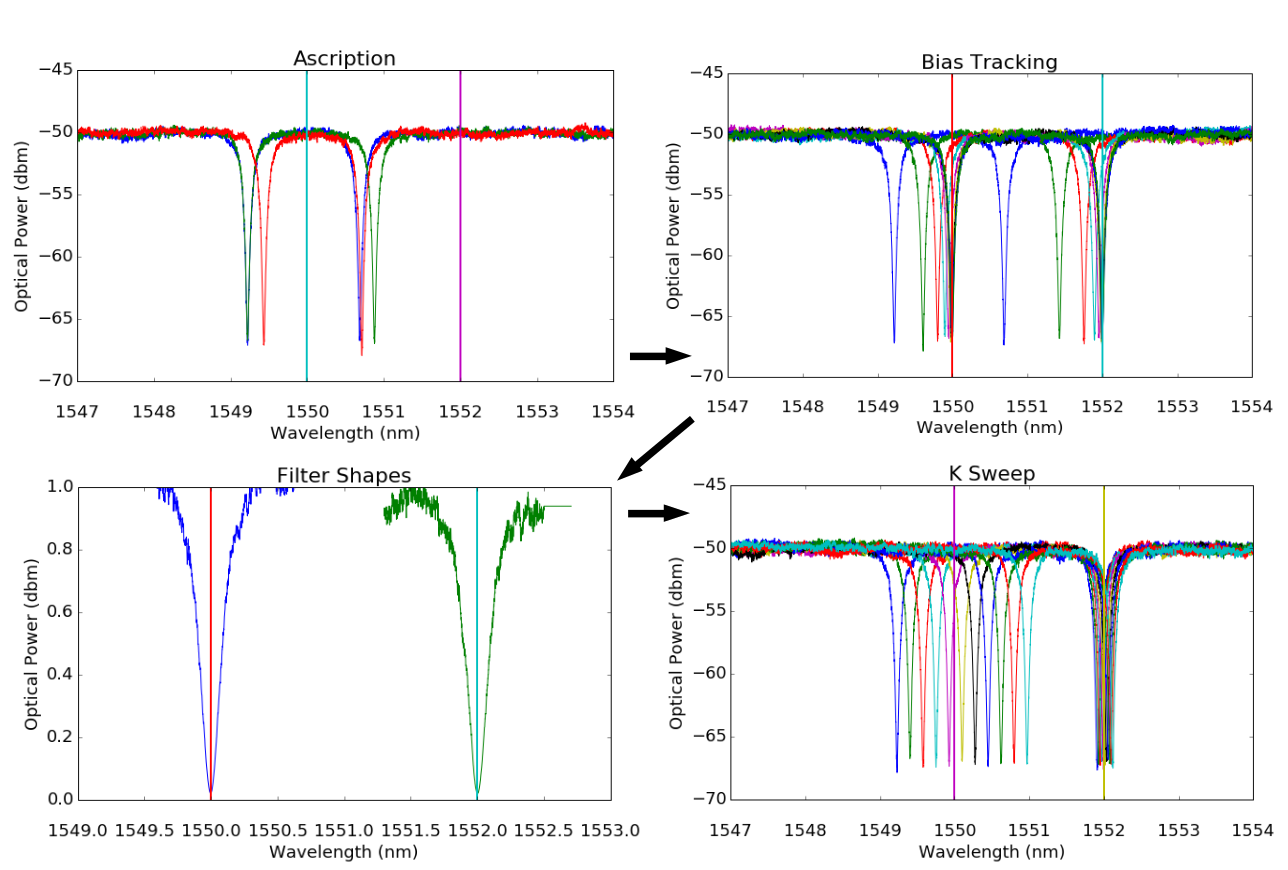}
\caption{Basic calibration procedure performed on a 2-microring dendrite without axons. From upper-left to lower-right: send a small current to determine primary mapping between filaments and microrings, use feedback control to direct the resonance peaks to the input channels (marked with vertical lines), pull the shape of each peak for use in backpropagation, and sweep around resonance to pull the proportionality constants between dissipated heat and wavelength shift. Note how, even though only one current channel is being swept, the other peak moves due to thermal cross-talk. These are the off-diagonal elements of the $K$ matrix described in the section.}
\label{fig:procedure}
\end{figure}

As stated before, the goal of thermal calibration is to provide a map between the state of electrical currents sent to the on-chip heaters and the power transmission (or, equivalently, the position of the resonance peak) of each microring. The annotated code to carry out this procedure in simulation can be found in Appendix \ref{ap:dendrite}, and an outline of the procedure in terms of its effect on a simulated spectrum through a single dendrite is shown in Figure \ref{fig:procedure}. 

In general, testing calibration code in simulation involves first creating a simulated model meant to act as the "real" system. The parameters of this module are set manually and remain hidden from the calibration code at all times. It is the goal of the calibration code to initialize an empty model of the same type and fill it with parameters that match those of the "hidden" model. In principle, running an inverse simulation of the calibration model will yield input currents from desired spectra. These currents can then be used to control the "hidden" model with predictable effects.

Recall from Section \ref{sec:principle} that the resonance wavelength of each microring is temperature-dependent. The heat supplied to the chip is proportional to the square of the current sent to the heater. Therefore, we can linearize the system around a bias current, creating the following system model:
\begin{equation}
(\lambda_j - \lambda_{0_{j}}) = K_{jk}[(i_k + i_{b_{k}})^2 - (i_{b_{k}})^2]
\end{equation}
Where $i_k$ are the currents sent to each microring, $i_{b_{k}}$ are the bias currents that set each microring's resonance to the bias wavelengths $\lambda_{0_{j}}$, and $\lambda_j$ are the resulting resonance wavelengths of each microring. Diagonal elements of the matrix $K$ are the proportionality constants relating each microring's resonance peak and its own heater, while the off-diagonal matrix represents the effect heaters have on adjacent microrings due to thermal cross-talk. In practice, the magnitude of the off-diagonal elements tend to be about 5\% the magnitude of the diagonal elements. Together, the $K$ matrix, bias wavelengths $\lambda_0$ and currents $i_0$, the approximate shape of each microring Cauchy-Lorentz distribution, and the global attenuation of the system (from effects including insertion loss and scattering) are all the model parameters that the calibration procedure must determine.

\noindent The description of the calibration procedure is as follows:
\begin{enumerate}
\item {\bf Ascription:} Initially, the calibrator does not know which current channel corresponds to which resonance. In this step, send a small amount of current to each current channel and see which resonance moves the most. Also use this step to estimate the diagonals of the $K$ matrix.
\item {\bf Background Removal:} While the noise in the simulation is zero-mean pink noise, the real system could have some non-random noise. Tune the resonances off to see the background spectrum behind them, and then subtract this background spectrum from each successive spectrum measurement. Also use this step to determine total system attenuation and store in the calibration model.
\item {\bf Track to Bias:} The wavelength biases are fixed to the wavelengths of the input channels. Use proportional feedback control on the supplied currents to adjust each resonance to the wavelength bias. The diagonals of the control matrix is proportional to the estimated diagonals of $K$, decreasing convergence time. This leads to the controller:
\begin{equation}
\Delta i_k^2 = k_p*K_{kk}\text{err}_k
\end{equation}
When done, store the wavelength and current biases in the calibration model.
\item {\bf Pull Filter Shapes:} Take a spectrum isolating each Cauchy-Lorentz distribution, and store that distribution in the calibration model to estimate the power transfer as a function of detuned wavelength.
\item {\bf Determine the $K$ Matrix:} Sweep each current channel around the bias current and determine the best linear fit for each resonance. Store the slope of this fit as the element of the $K$ matrix.
\end{enumerate}
\subsection{Cascaded Calibration and Experimental Results}
\begin{figure}[!ht]
\centering
\includegraphics[width=0.8\textwidth]{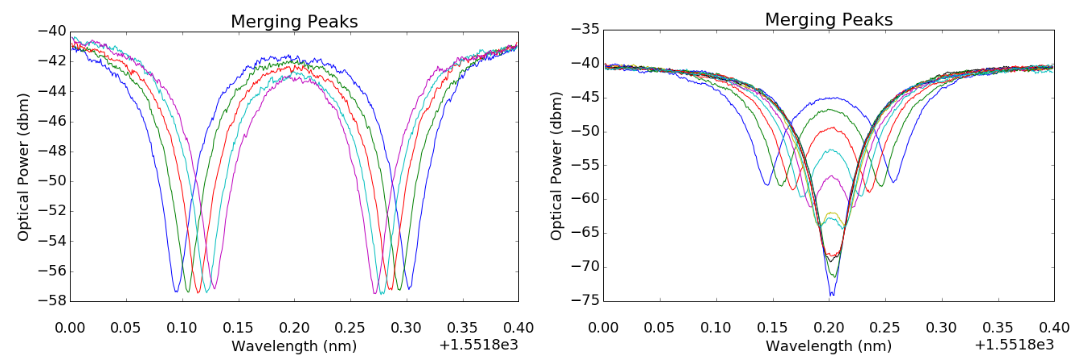}
\caption{Example of moving one axon peak and one dendrite peak onto the same resonance. {\bf (Left)} The two peaks are brought together until the peak-finding algorithm cannot distinguish them from a single trough. {\bf (Right)} A feedback controller attempts to minimize the full-width half-maximum (i.e. the width) of the combined trough, bringing both troughs onto the same resonance.}
\label{fig:merging}
\end{figure}

The difficulty with calibrating a real star-topology network is that it is only possible to measure the spectra of optical power passing through the axon {\em and} the dendrite weight banks, returning a spectrum product. This would normally not be an issue, except for the fact that each axon microring needs to be biased to the same wavelenegth resonance as a dendrite microring. Two superimposed troughs or peaks in a spectrum become difficult to distinguish from a single larger peak or trough. As such, an extra step needs to be added during the track to bias (the third step of the basic procedure), as illustrated in Figure \ref{fig:merging}. One dendrite and one axon are brought to wavelengths immediately below and immediately above the desired resonance through the same mechanism as the previous section. Then, the troughs are moved closer and closer until our peak-finding algorithm begins to mistake them for a single peak. From there, a new feedback control loop takes hold, trying to minimize the width of this double-peak while keeping it on center. This leads to the following controller:
\begin{equation}
\Delta i^2 = \pm k_p*K*(\text{err}_{fwhm}+\text{err}_{\lambda})
\end{equation}
Where the right most peak uses the minus sign, $\text{err}_{fwhm}$ is the width of the trough, and $\text{err}_{\lambda}$ is the deviation of the center of the double-trough. After this slight change to the bias calculation, the rest of the calibration procedure is carried out similarly, except for making sure the two peaks are slightly detuned during the sweep for the $K$ matrix so their trough locations can be distinguished. An annotated procedure can be found in Appendix \ref{ap:cascade}.

\begin{figure}[!ht]
\centering
\includegraphics[width=0.8\textwidth]{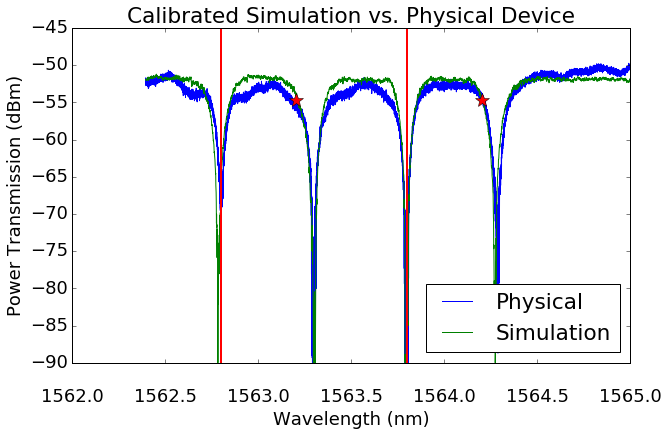}
\caption{Spectrum comparison between a simulation of the calibrated model and real cascaded microrings. The vertical lines show the requested axon wavelengths, and the red stars show the requested dendrite transmission (-3dBm relative to the maximum, or about 50\% transmission). Note the good agreement between calibrated simulation and experiment.}
\label{fig:sim_phy}
\end{figure}
\begin{figure}[!ht]
\centering
\includegraphics[width=0.8\textwidth]{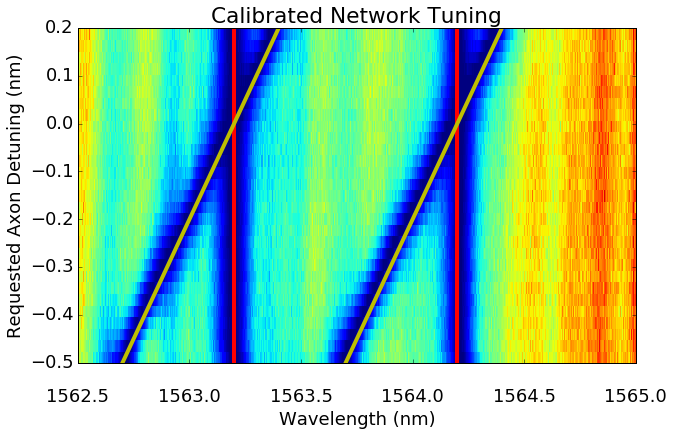}
\caption{Evidence of successful axon tuning. The yellow line corresponds to the requested axon tuning, and the red line corresponds with the requested dendrite tuning. The colormap represents the physical spectrum data (imagine Figure \ref{fig:sim_phy} flipped with the troughs pointing into the page). Note how the axon trough (dark blue) very closely matches the requested trough location.}
\label{fig:tune}
\end{figure}

\begin{figure}[!ht]
\centering
\includegraphics[width=0.6\textwidth]{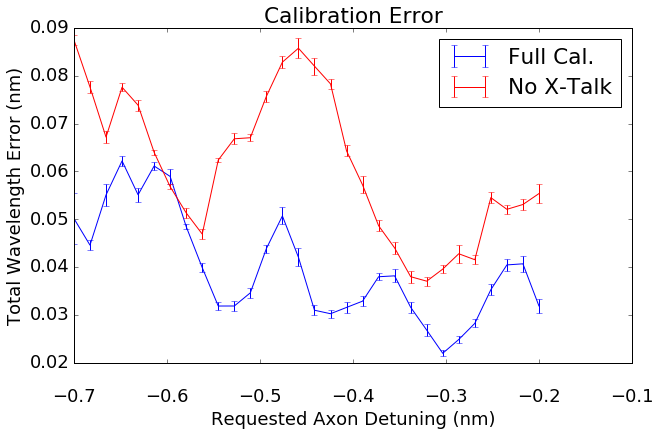}
\caption{Comparison of total wavelength error between using a fully calibrated model and a model that neglects thermal cross-talk (a more naive calibration algorithm).}
\label{fig:xtalk}
\end{figure}

Both simple and cascaded calibration procedures were carried out on simulated devices to great success, but to test the accuracy of those simulations, the more difficult cascaded procedure was also tested on a physical 2-axon, 2-dendrite network branch. After determining all the physical parameters and loading them into the calibration model, the model was used to control the physical device. Initial results are shown in Figures \ref{fig:sim_phy} and \ref{fig:tune}. In the former, the calibration model was asked for the electrical current vector that would allow the axons to be detuned by 0.5nm and the dendrites in order to allow 50\% optical power transmission (equivalent to -3dBm relative to maximum). These currents were then fed into both a simulation of the calibration device and the physical device, and the resonances ended up very close to their requested locations. Figure \ref{fig:tune} sweeps through axon location requests between -0.5nm and 0.2nm relative to bias. Note both how the dark blue axon trough follows the requested trajectory very closely and how the dendrite is held at a constant resonance despite the presence of thermal cross-talk. To quantify this, Figure \ref{fig:xtalk} compares the total wavelength errors between a fully calibrated model and one that neglected thermal cross-talk. Factoring in cross-talk decreases errors significantly, showing that the thermal calibration procedure reasonably estimates the off-diagonal elements of the $K$ matrix.

\subsection{Network Training and Simulated Results}
\label{sec:train}

\begin{figure}[!ht]
\centering
\includegraphics[width=0.8\textwidth]{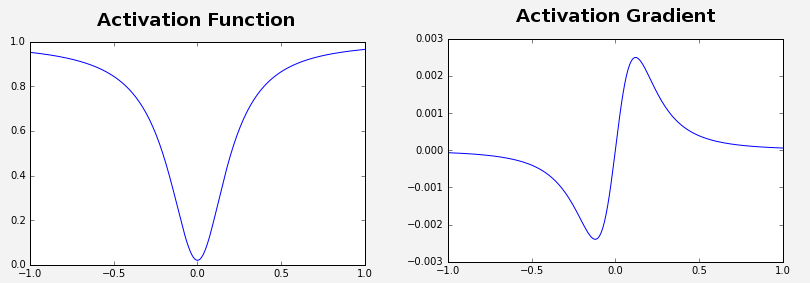}
\caption{{\bf (Left)} Activation function used in backpropagation and {\bf (Right)} its gradient. This is a composition of a quadratic to convert current to temperature and the upside-down Cauchy-Lorentz distribution of the axon.}
\label{fig:activation}
\end{figure}

With network calibration mapping the electrical state of the network to optical transmission through each weight bank, the next step is the determine the optimal transmissions for nonlinear classification of a given dataset. This optimization problem is generally solved using {\em backpropagation}, which is stochastic gradient descent applied to a large composition of functions. The first step is to determine the nonlinear activation function in each layer. For the photonic network, this is the axon, where the relationship between the shift in wavelength and optical power transmission is the nonlinear Cauchy-Lorentz distribution:
\begin{equation}
h(\Delta\lambda) = \frac{\Delta\lambda^2}{\gamma^2 + \Delta\lambda^2}
\end{equation}
However, $\Delta\lambda$ itself has a {\em quadratic} dependence on the current. This second nonlinear function is:
\begin{equation}
\Delta\lambda(i) = K[(i^2 + I_0)^2 - I_0^2]
\end{equation}
Where $I_0$ is the bias current where the microring is on resonance with the wavelength channel. Here, it is assumed that the current is small enough that thermal cross-talk is negligible, so $K$ is just the {\em diagonal} of the $K$ matrix. The final activation function $f(i)=h(\Delta\lambda(i))$, is therefore a composition of these two nonlinear functions. This function and its gradient, both of which are required for backpropagation, are shown in Figure \ref{fig:activation}.

The output of the activation function is power transmission in the domain $(0, 1)$, so there are many physical quantities in each layer that go into converting the previous layer's power transmission into the next layer's current. Using the transfer function of the amplification circuit (Equation \ref{eq:circuit}) and the attenuated optical pump power $p$, we can construct the entire composition of functions that maps the optical input $x_j^0 \in (0, 1)$ to the hidden layer $x_k^1$:
\begin{equation}
x_k^1 = f(\frac{R_t^0*R(\sum_{j}w^0_{kj}x_{j} + b_p^0) + b_v^0}{R_s^0} + b_k^0)
\end{equation}
where $j=1,2$ for the two inputs and $k=1,2,3$ for the three outputs. The second layer of the network is simply a single weight-bank dendrite, as the final signal is a scalar that does not need to go back into the optical domain. In addition, this layer does not have any extra optical bias $b_p$, since all three channels are used. The final voltage output $y$ can therefore be written as:
\begin{equation}
y = R_t^1*R(\sum_{k}w^1_{k}x_k^1) + b_v^1
\end{equation}
\begin{figure}[!ht]
\centering
\includegraphics[width=0.8\textwidth]{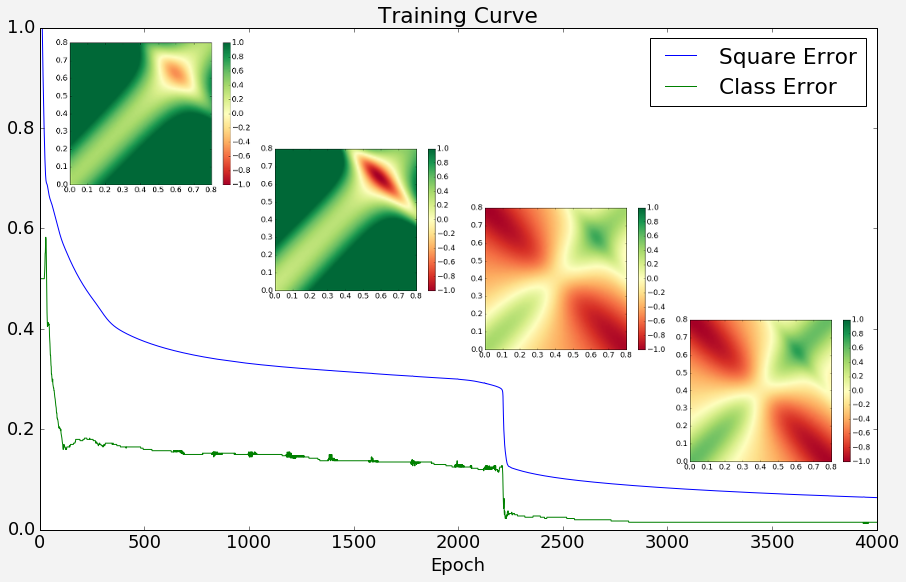}
\caption{Backpropagation training trajectory on XOR data. Square error is the difference between the continuous variable $y$ and perfect values of $-1$ and $1$, while the class error is the percentage of data that is completely mis-classified due to $y$ being the wrong sign. A straight, diagonal line is a metastable state approached in the first couple thousand epochs. Then, the system is knocked into the more stable, more accurate classification state.}
\label{fig:traincurve}
\end{figure}
The final binary classificaiton is simply the sign of $y$. While these equations provide a full description of the network, the extra physical constants make the backpropagation algorithm unnecessarily tedious and more sensitive to parameters such as learning rate and initial conditions. One solution is to distill the network dynamics to isolate just the activation function by introducing "virtual" weights and biases:
\begin{equation}
W_{kj}^0 = \frac{R_t^0*R*w^0_{kj}}{R_s^0}
\end{equation}
\begin{equation}
W_{k}^1 = R_t^1*R*w^1_{k}
\end{equation}
\begin{equation}
B_k^0 = f(\frac{R_t^0*R*b_p^0 + b_v^0}{R_s^0} + b_k^0)
\end{equation}
\begin{equation}
B^1 = b_v^1
\end{equation}

Then, the total composition of functions then distills to:
\begin{equation}
x_k^1 = f(\sum_jW_{kj}x_{j}^0 + B^0_k)
\end{equation}
\begin{equation}
y = \sum_kW_{k}x_{k}^1 + B^1_k
\end{equation}
It were these simplified functions that were fed into the backprop algorithm to find the optimal values for $W$ and $B$. This algorithm was tested on a generalized XOR dataset, where opposite quadrants are classified the same, and managed to find weights and biases that achieved greater than 99\% classification accuracy. The training curve can be seen in Figure \ref{fig:traincurve}, and the final training results are in Figure \ref{fig:backpropRes}. The backprop code, which used the TensorFlow framework, can be found in Appendix \ref{ap:backprop}, but a more complete description of the backprop algorithm can be found in Appendix \ref{apmat:backprop}.

\begin{figure}[!ht]
\centering
\includegraphics[width=0.8\textwidth]{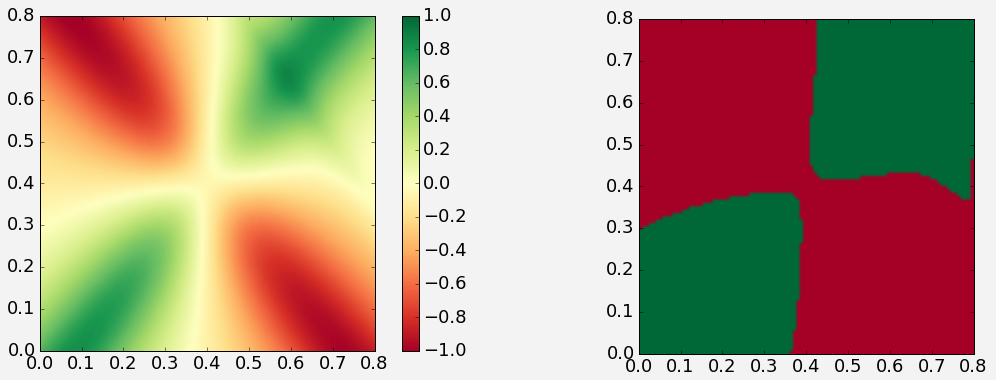}
\caption{Final results of XOR backpropagation on the abstract network model and activation function. {\bf (Left)} Continuous output variable. {\bf (Right)} Output variable thresholded for binary classification.}
\label{fig:backpropRes}
\end{figure}

\begin{figure}[!ht]
\centering
\includegraphics[width=0.8\textwidth]{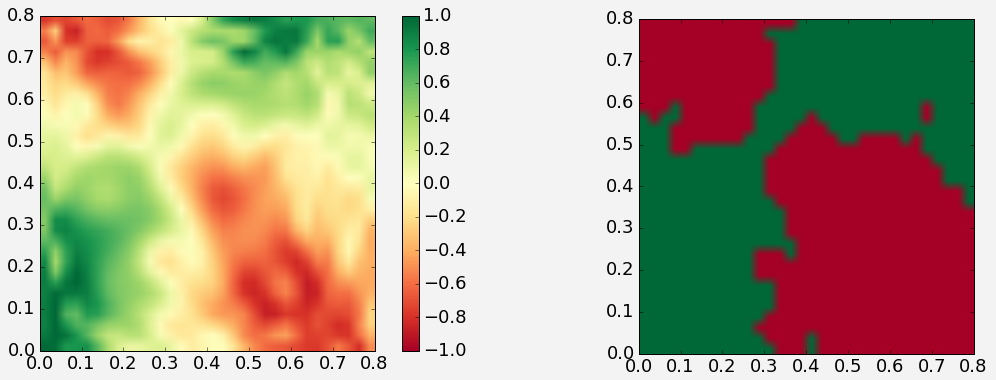}
\caption{Final results feeding the backpropagation weights into the simulated network after these weights were converted to non-abstract, physical weights. {\bf (Left)} Continuous output variable. {\bf (Right)} Output variable thresholded for binary classification.}
\label{fig:simRes}
\end{figure}

After backpropagation, the resulting virtual weights and biases were converted back into the physical weights and biases. Resistance values were fixed such that the physical weights were in the interval $[-1, 1]$ and the physical biases were at reasonable currents and voltages. These final parameters were loaded into a pre-calibrated network and the network was simulated. Annotated simulation code can be found in Appendix \ref{ap:sim}. The resulting classification space is shown in Figure \ref{fig:simRes}, showing decent agreement with the ideal backprop output. Performance was somewhat degraded due to noise, cross-talk, and calibration parameters not exactly matching the parameters used in backprop.

\newpage

\section{Next Steps: Mode Division Multiplexing}
\label{sec:mdm}

\subsection{Motivation and Operating Principle}

Looking back at Figure \ref{fig:bandSize}, the main draw of OEO networks is their ability to handle larger signal bandwidths for a given layer size. However, with the network scheme investigated in this thesis (labeled "Resonators" in Figure \ref{fig:bandSize}), there exists a hard limit on the size of each optical layer that is independent of input signal bandwidth. From Section \ref{sec:principle}, the Cauchy-Lorentz distribution is only a valid approximation of the microring spectrum around a single resonant wavelength. In reality, a microring has multiple resonant wavelengths, one for each integer multiple of the fundamental wavelength of the ring, and each resonance has its own Lorentz trough. The gap between these resonances is called the {\em free spectral range} of the microring. Since microring resonance troughs have a finite spectral width, only a finite amount of them can fit effectively in this free spectral range. This limits the total amount of wavelength channels that can be used in a single axon-dendrite branch.

\begin{figure}[!ht]
\centering
\includegraphics[width=0.6\textwidth]{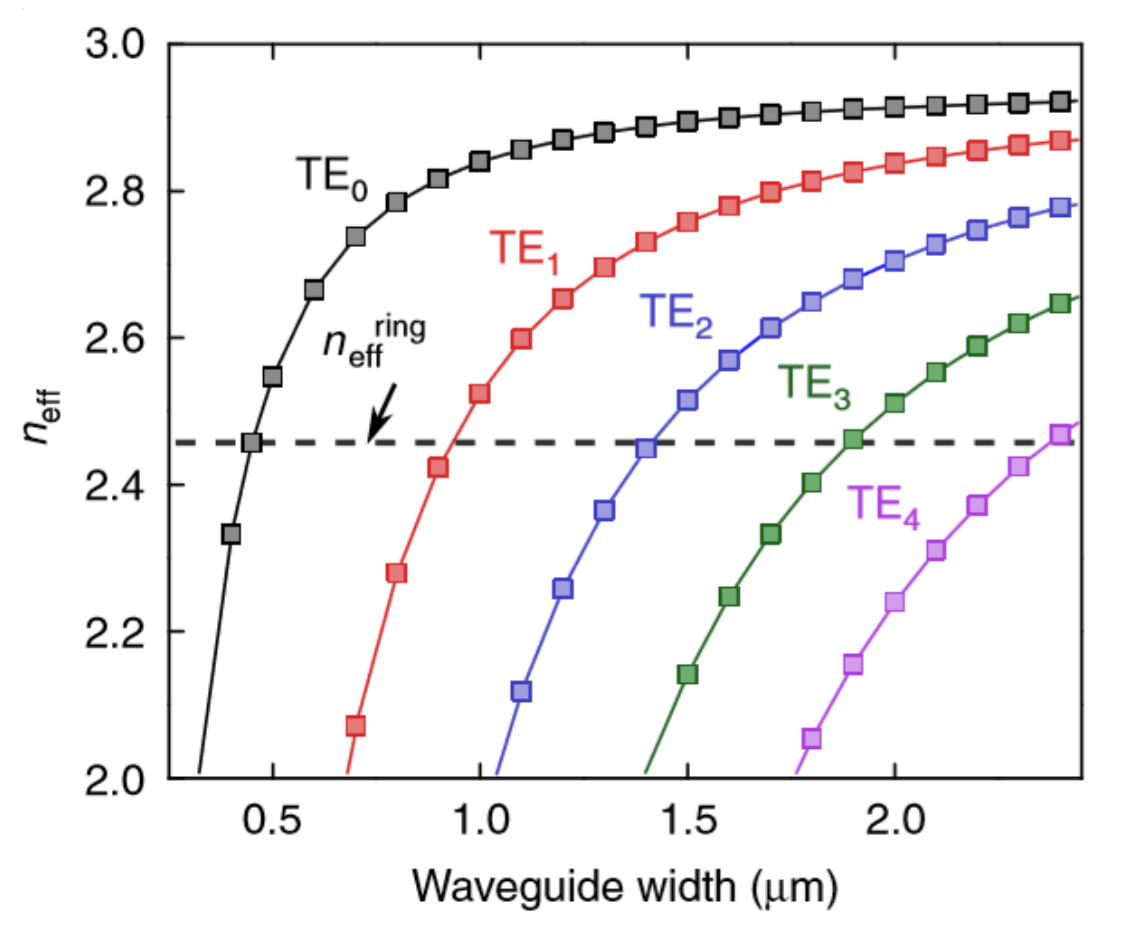}
\caption{Graph of the effective index for each transverse mode as a function of the waveguide width. The dashed horizontal line is the effective index of a standard-width single-mode waveguide used in all of the microrings in this project. Graph Credit: \cite{mdm}}
\label{fig:mdm_index}
\end{figure}

One way around this limitation is to expand the network to use both wavelengths and {\em transverse spatial modes} to differentiate between channels. Transverse modes look at how the magnitude of the electric fields changes across the waveguide in the transverse direction relative to the direction of propagation. Most integrated optics operate in single-mode waveguides, where the electric field is strongest in the center of the waveguide and tapers off towards the edges. This is called the fundamental mode. However, if the waveguide is sufficiently wide, other possible solutions exist with the magnitude of the electric field peaking more than once. The electric field peaks twice in the first excited mode, three times in the second excited mode, and so on. Since these modes are orthogonal, channels of a given wavelength in two separate transverse modes do not mix in a straight waveguide, allowing them to be distinguished. The total number of channels in a layer is now the product of the number of wavelength channels and the number of modes, theoretically increasing the layer size by an order of magnitude.

\begin{figure}[!ht]
\centering
\includegraphics[width=\textwidth]{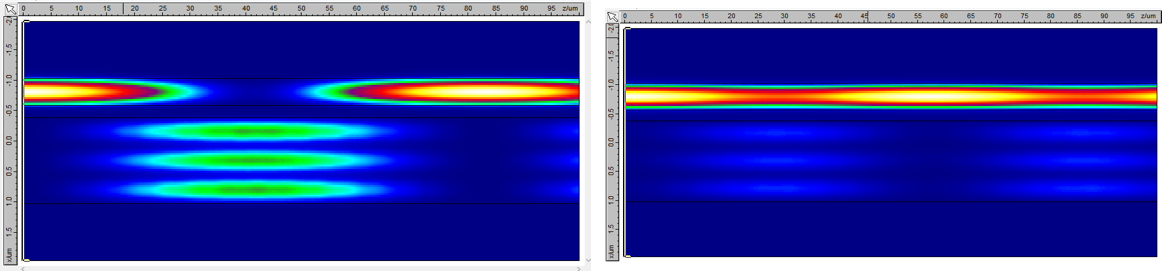}
\caption{Simulated power coupling between a single-mode waveguide and the second excited mode of a larger waveguide. {\bf (Left)} Full coupling when the effective indices of refraction match. {\bf (Right)} Significantly diminished coupling when the effective indices do not match.}
\label{fig:mdm_sim}
\end{figure}

The process of adding multiple single-mode channels onto a multi-mode waveguide is called {\em mode-division multiplexing} (MDM) and relies on the fact that optical coupling can only happen effectively between two media that have the same index of refraction. Each mode experiences a different index of refraction which is dependent on the width of the waveguide, as shown in Figure \ref{fig:mdm_index}. By adiabatically varying the width of the waveguide, the index of refraction for the desired mode can be adjusted to match the index of the fundamental mode in a single-mode waveguide, allowing for light to couple back and forth. Power oscillation occurs over the course of one {\em beat length,} so it is necessary to separate the two waveguides after half a beat length for maximum coupling. An example of coupling between the fundamental mode and the second excited mode for a matched width and an unmatched width is shown in Figure \ref{fig:mdm_sim}.

\subsection{Experimental Validation}
\begin{figure}[!ht]
\centering
\includegraphics[width=0.7\textwidth]{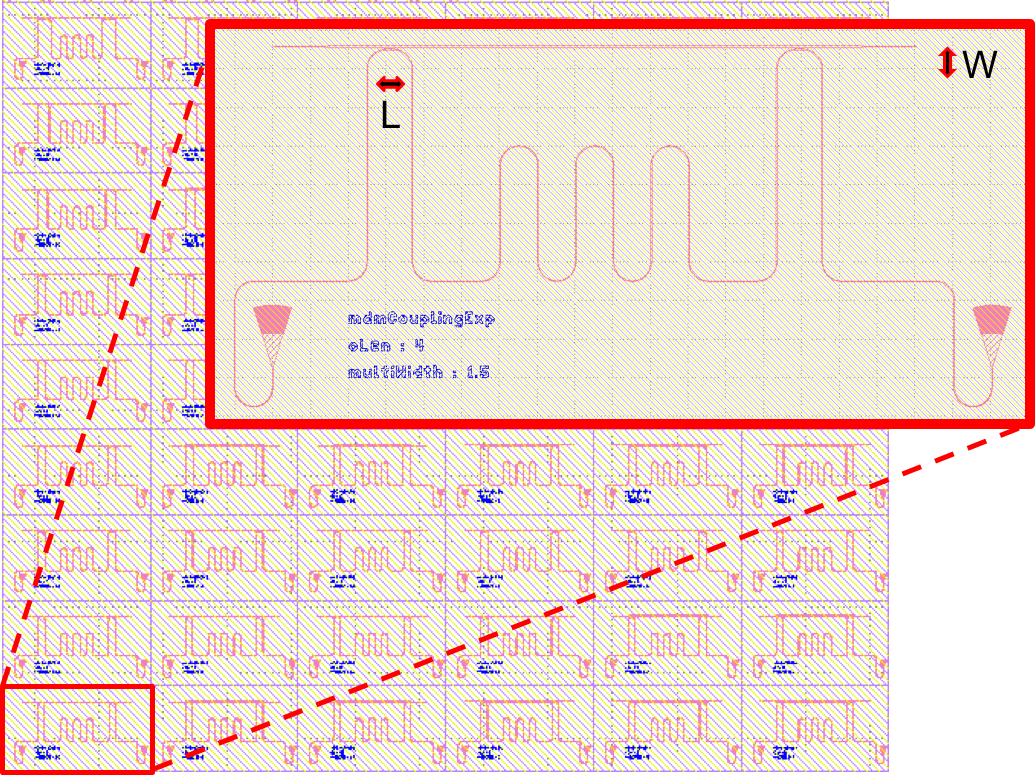}
\caption{Asymmetric Mach-Zehnder interferometer. This experiments sweeps over different coupling lengths and waveguide widths in order to determine the optimal geometry for coupling to each excited mode.}
\label{fig:interfermoeter}
\end{figure}

\begin{figure}[!ht]
\centering
\includegraphics[width=0.8\textwidth]{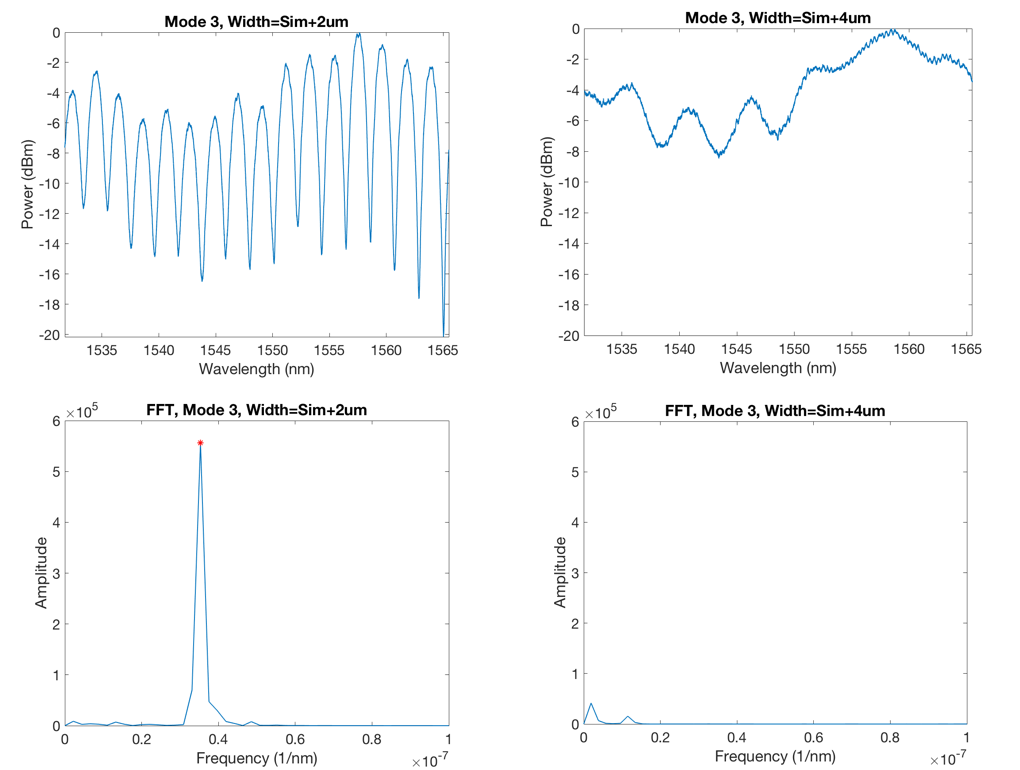}
\caption{Example of transmission spectra from a device with 50\% coupling {\bf (Left)} and a high extinction ration, and 0\% coupling {\bf (Right)} and a low extinction ratio due to mismatched widths.}
\label{fig:data}
\end{figure}

Because the coupling strength is so strongly dependent upon the waveguide width and the coupling length, it was necessary to construct a physical experiment to verify the accuracy of the simulations. This experiment, shown in Figure \ref{fig:interfermoeter}, is an asymmetric Mach-Zehnder interferometer, which provides a way to estimate the coupling coefficient for a given length and width. The exact mechanism is detailed in Appendix \ref{apmat:mdm}, but suffice it to say here that the magnitude of the oscillations in the spectrum of this device (called the extinction ratio) is large oscillations for 50\% coupling and small for 0\% and 100\% coupling. By setting the coupling length to approximately a quarter of the beat length (50\% coupling), the optimal width maximizes the extinction ratio. An example of this part of data from this part of the procedure can be seen in Figure \ref{fig:data}. Once at the optimal width, the optimal coupling length of a half beat length will minimize the extinction ratio. 

\begin{figure}[!ht]
\centering
\includegraphics[width=0.6\textwidth]{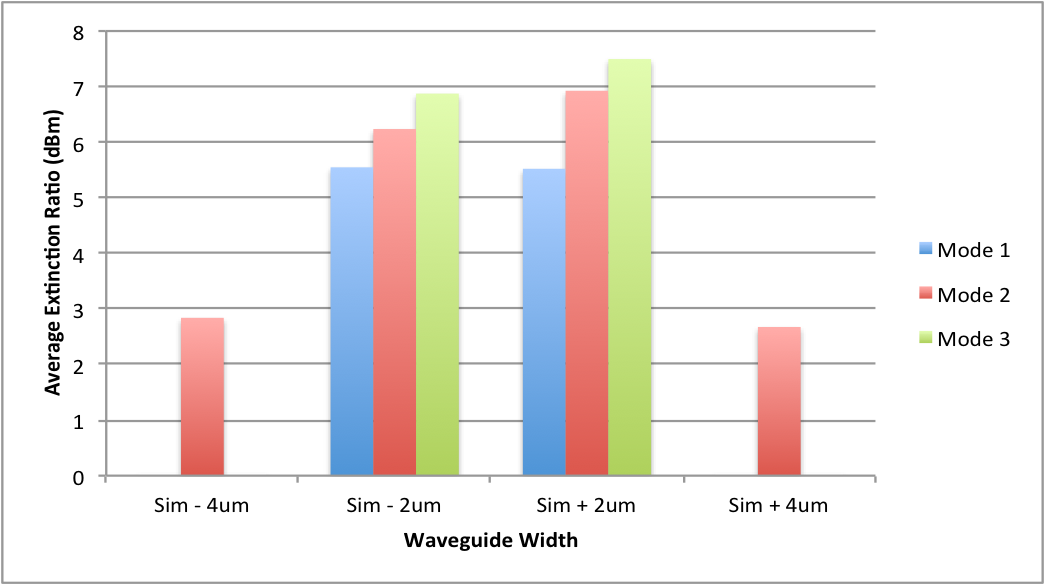}
\caption{Aggregate data across all coupling lengths, comparing the extinction ratio of different waveguide widths for each mode. Note that the waveguide widths closer to the simulated optimal width have a higher extinction ratio. This is evidence consistent with simulated results.}
\label{fig:aggregate}
\end{figure}

While the parameter space was too coarse and the devices too noisy to hone in on the exact optimal widths and lengths, aggregate data (as shown in Figure \ref{fig:aggregate}) was consistent with simulated results. Therefore, the next step is to look into repeating the results of previous network experiments using MDM channels.

\subsection{Challenges}
Even with optimal coupling geometry, MDM is not without further challenges. First and foremost, standard split waveguides do not split power evenly on each mode channel. It will be impossible to use a star topology network with MDM channels, necessitating a switch to the more-difficult-to-control hairpin topology. This is not really a limitation in itself, but it necessitates the challenge of switching topologies.

\begin{figure}[!ht]
\centering
\includegraphics[width=0.8\textwidth]{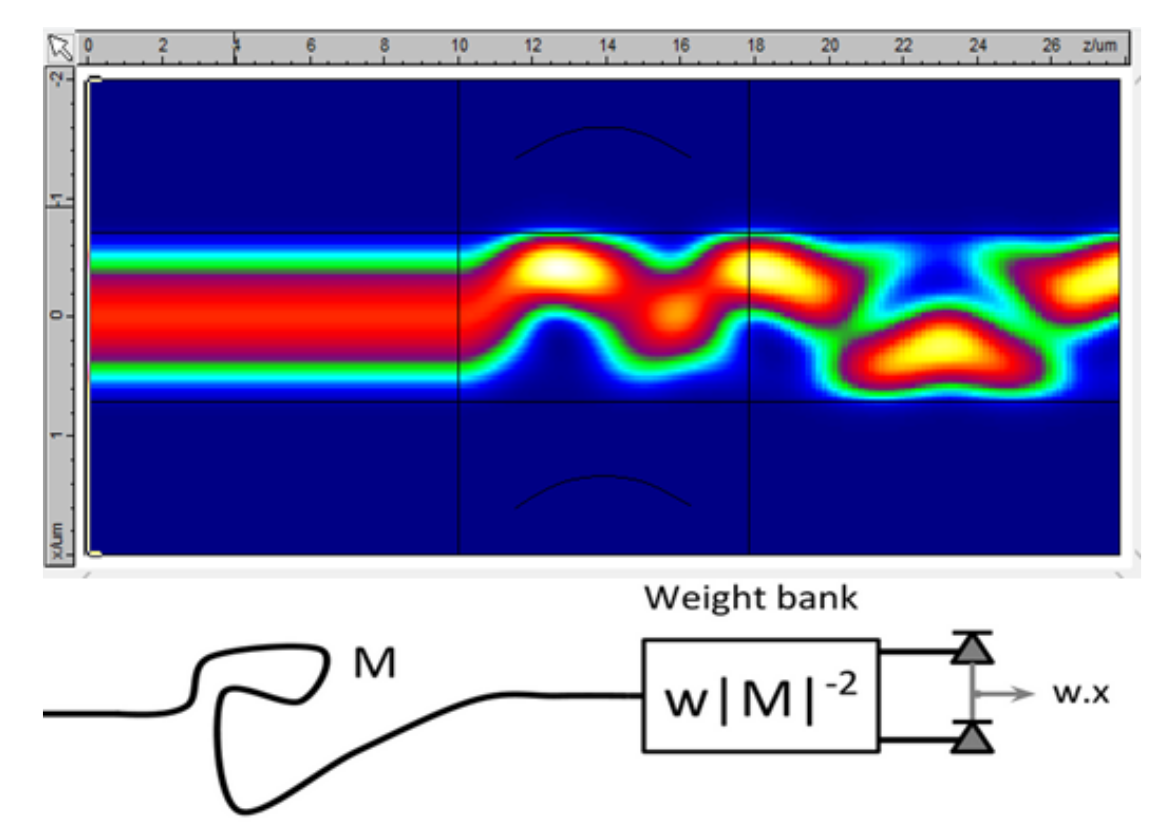}
\caption{{\bf (Top)} Simulation of the fundamental mode of a multi-mode waveguide mixing with the first excited mode. {\bf (Bottom)} Since this mixing matrix is unitary (assuming minimal attenuation), it can be inverted by the weight bank to yields the desired weights. }
\label{fig:intermodal}
\end{figure}

Secondly, spatial modes become less orthogonal as the spatial geometry changes, such as when the waveguide needs to bend. As shown in Figure \ref{fig:intermodal}, even if optical power starts out only in the fundamental mode, it will mix into other modes in anything but a straight waveguide. Assuming minimal attenuation, such {\em intermodal mixing} can be modeled as a unitary matrix $M$, and it is extremely sensitive to initial fabricaiton conditions, requiring direct measurement. Fortunately, once $M$ is determined, its inverse can just be added to the weight bank to yield the desired weights, but having to determine $M$ in the first place could make the calibration procedure exceedingly complicated. That said, increasing network throughput by an order of magnitude is worth pursuing MDM channels in future work.

\newpage

\section{Discussion and Future Work}

The work in this thesis demonstrated the ability to calibrate a feed-forward WDM network and train it to solve nonlinear classification problems. The immediate next step is to demonstrate a feed-forward network on a physical device. Fortunately, experimental verification of the calibration procedure also demonstrated a good agreement between simulation and experiment. Therefore, it is reasonable to deduce from the successful simulation of the 2-3-1 feed-forward network that the physical device will be equally successful.

However, these networks still face many practical limitations that could prevent them from reaching their full potential throughput. Even before moving to MDM networks, the current WDM scheme has plenty of room for improvement. Most pressingly, thermal tuning is simply a slow way of adjusting the index of refraction of silicon. This is fine for applying the large, stable biases and weights, but puts a significant limit on signal bandwidth when used in the path of actual computation. All of this is in addition to degraded performance due to the thermal cross-talk between the active network signal and all of the network weights. Other faster methods of modifying the index of refraction of silicon, such as by carrier depletion, need to be investigated.

Finally, future work should include an application study. Possible applications in radio frequency real-time computing and scientific computing were mentioned in Section \ref{sec:motivation}, but it is currently unknown which of those applications would most benefit from the extra throughput that OEO networks can provide. This is especially important considering that it will be difficult to keep OEO networks as power-efficient as existing electronic networks, a problem that has yet to be tackled.

Overall, photonic neural networks have the potential to benefit applications with high bandwidth demand, and this thesis took steps towards its realization. There are still many more problems to solve to finish crossing the gap from curiosity into utility, but that also leaves plenty of room for future research.

\newpage
\appendix
\section{Math Appendices}
\subsection{Backpropagation}
\label{apmat:backprop}

Given a neural network model, the goal of backpropagation is to determine the weights and biases that minimize some cost function using stochastic gradient descent. In the 2-3-1 neural network performing XOR classification, for a given input $x_j^0$, there is some output $y$ and a desired output $d=-1, 1$. The network model is as follows:
\begin{equation}
x_k^1 = f(\sum_jW_{kj}^0x_{j}^0 + B^0_k)
\end{equation}
\begin{equation}
y = \sum_kW_{k}^1x_{k}^1 + B^1
\end{equation}
One popular cost function is the square of the error between $y$ and $d$:
\begin{equation}
    E = \frac{1}{2}(d-y)^2
\end{equation}
In stochastic gradient ascent, the weights and biases are updated in the direction of the gradient of $E$ at some sample pair $x_j^0$ and $y$. Equivalently, each individual weight or bias is updated by an amount proportional to the partial derivative of $E$ with respect to that weight. Because the cost function needs to be minimized, the update happens opposite the gradient, or with a negative sign in front of the partial derivative.
\begin{equation}
\Delta W_{jk}^l = -\eta\frac{\partial E}{\partial W_{jk}^l}
\end{equation}
\begin{equation}
\Delta B_k^l = -\eta\frac{\partial E}{\partial B_{k}^l}
\end{equation}
Here $\eta$ is called the {\em learning rate}, and it is set manually prior to training. If the learning rate is too small, the network will take a long time to converge on the minimum. However, if the learning rate is too large, the network will never converge, jumping back and forth across the minimum.

Using the chain rule, it is possible to write out these partial derivatives in terms of known quantities:
\begin{dmath}
\Delta W_k^1 = -\eta\frac{\partial E}{\partial y}\frac{\partial y}{\partial W_k^1} = 
                \eta (d-y) x_k^1
\end{dmath}
\begin{dmath}
\Delta B^1 = -\eta\frac{\partial E}{\partial y}\frac{\partial y}{\partial B^1} = 
                \eta (d-y)
\end{dmath}
\begin{dmath}
\Delta W_{jk}^0 = -\eta\frac{\partial E}{\partial y}\frac{\partial y}{\partial x_k^1}\frac{\partial x_k^1}{\partial W_{jk}^0} =
                \eta (d-y) W_k^1 f'(\sum_{j}W_{jk}^0x_{j}^0 + B_{k}^0) x_{j}^0
\end{dmath}
\begin{dmath}
\Delta B_{k}^0 = -\eta\frac{\partial E}{\partial y}\frac{\partial y}{\partial x_k^1}\frac{\partial x_k^1}{\partial B_{k}^0} =
                \eta (d-y) W_k^1 f'(\sum_{j}W_{jk}^0x_{j}^0 + B_{k}^0)
\end{dmath}

One update is performed for each point in the dataset, and exhausting all points in the training dataset is called an {\em epoch}. A learning curve (Figure \ref{fig:traincurve}) plots the change in the average value of the error function (and the binary classification error) for each epoch.

\subsection{Multi-Mode Interferometer}
\label{apmat:mdm}

The following experiment was designed as part of the Junior Independent Work: "Mode division multiplexing (MDM) weight bank design for Use in photonic neural networks."

The asymmetric mach-zehnder interferometer has an input waveform $E_{in}$ and produces the output $E_{out}$ and an unused output $E_{taper}$. The device's optical transfer function of the device can be written as:
\begin{equation}
\begin{bmatrix}
    E_{out} \\
    E_{taper} \\
\end{bmatrix} = 
M \begin{bmatrix}
    E_{in} \\
    0 \\
\end{bmatrix}
\end{equation}
Where M is a composition of both single-mode to multi-mode couplers and a phase shift $\Delta\phi$ due to the difference in optical length of the two paths through the interferometer.
\begin{equation}
M = M_{coupler} * M_{\Delta\phi}*M_{coupler}
\end{equation}
$M_{coupler}$ takes on the same form as the coupler in Section \ref{sec:principle}, where the cross coupling term $\alpha$ is used instead of the self-coupling term $r$.
\begin{equation}
M_{coupler}=
\begin{bmatrix}
    \sqrt{1-\alpha} & i\sqrt{\alpha} \\
    i\sqrt{\alpha} & \sqrt{1-\alpha} \\
\end{bmatrix}
\end{equation}

\noindent Combined with the phase shift $\Delta\phi=k*\Delta L$, where $k=2\pi/\lambda$ is the wavenumber of the light, the rest of the matrix can be determined:
\begin{equation}
M = 
\begin{bmatrix}
    \sqrt{1-\alpha} & i\sqrt{\alpha} \\
    i\sqrt{\alpha} & \sqrt{1-\alpha} \\
\end{bmatrix}
\begin{bmatrix}
    e^{ik\Delta L} & 0 \\
    0 & 1 \\
\end{bmatrix}
\begin{bmatrix}
    \sqrt{1-\alpha} & i\sqrt{\alpha} \\
    i\sqrt{\alpha} & \sqrt{1-\alpha} \\
\end{bmatrix}
\end{equation}
\begin{equation}
\abs*{\frac{E_{out}}{E_{in}}}^{2} = \abs*{(1-\alpha)e^{ik\Delta L} - \alpha}^{2} = \alpha^{2} + (1-\alpha)^{2} - 2\alpha(1-\alpha)cos(k\Delta L)
\end{equation}
\noindent Therefore, a power spectrum of the device should exhibit sinusoidal oscillations ontop of a constant offset. The amplitude of these oscillations, called the {\em extinction ratio}, is only dependent on the coupling coefficient $\alpha$ between the single-mode and multi-mode waveguides. It reaches its maximum value when $\alpha=0.5$ and its minimum value when $\alpha=0,1$.

\newpage
\section{Code Appendices}

\subsection{Dendrite Calibration Procedure}
\label{ap:dendrite}

\input{FilterBank_Calibration_NEW.tex}

\newpage

\subsection{Cascaded Calibration Procedure}
\label{ap:cascade}

\input{Phys_Cascade_Calibration.tex}
\newpage

\subsection{2-3-1 Network Backpropagation}
\label{ap:backprop}

\input{2-3-1_FFNet_Backprop.tex}
\newpage

\subsection{2-3-1 Network Simulation}
\label{ap:sim}
\input{2-3-1_FFNet_Simulation.tex}

The thing about formatting code is that it always feels like perfectionless effort.

\end{document}

%% file: settings.tex
    \usepackage{graphicx}
    \makeatletter
    \def\maxwidth{\ifdim\Gin@nat@width>\linewidth\linewidth
    \else\Gin@nat@width\fi}
    \makeatother
    \usepackage{caption}

    \usepackage{adjustbox} 
    \usepackage{xcolor} 
    \usepackage{enumerate} 
    \usepackage{geometry} 
    \usepackage{amsmath} 
    \usepackage{amssymb} 
    \usepackage{textcomp} 
    \AtBeginDocument{%
        \def\PYZsq{\textquotesingle}
    }
    \usepackage{upquote} 
    \usepackage{eurosym} 
    \usepackage{fancyvrb} 
    \usepackage{grffile} 
    \usepackage{hyperref}
    \usepackage{longtable} 
    \usepackage{booktabs}  
    \usepackage[inline]{enumitem} 
    \usepackage[normalem]{ulem} 

    \definecolor{urlcolor}{rgb}{0,.145,.698}
    \definecolor{linkcolor}{rgb}{0.0,0.0,0.0}
    \definecolor{citecolor}{rgb}{.12,.54,.11}

    \definecolor{ansi-black}{HTML}{3E424D}
    \definecolor{ansi-black-intense}{HTML}{282C36}
    \definecolor{ansi-red}{HTML}{E75C58}
    \definecolor{ansi-red-intense}{HTML}{B22B31}
    \definecolor{ansi-green}{HTML}{00A250}
    \definecolor{ansi-green-intense}{HTML}{007427}
    \definecolor{ansi-yellow}{HTML}{DDB62B}
    \definecolor{ansi-yellow-intense}{HTML}{B27D12}
    \definecolor{ansi-blue}{HTML}{208FFB}
    \definecolor{ansi-blue-intense}{HTML}{0065CA}
    \definecolor{ansi-magenta}{HTML}{D160C4}
    \definecolor{ansi-magenta-intense}{HTML}{A03196}
    \definecolor{ansi-cyan}{HTML}{60C6C8}
    \definecolor{ansi-cyan-intense}{HTML}{258F8F}
    \definecolor{ansi-white}{HTML}{C5C1B4}
    \definecolor{ansi-white-intense}{HTML}{A1A6B2}

    \providecommand{\tightlist}{%
      \setlength{\itemsep}{0pt}\setlength{\parskip}{0pt}}
    \DefineVerbatimEnvironment{Highlighting}{Verbatim}{commandchars=\\\{\}}

\makeatletter
\def\PY@reset{\let\PY@it=\relax \let\PY@bf=\relax%
    \let\PY@ul=\relax \let\PY@tc=\relax%
    \let\PY@bc=\relax \let\PY@ff=\relax}
\def\PY@tok#1{\csname PY@tok@#1\endcsname}
\def\PY@toks#1+{\ifx\relax#1\empty\else%
    \PY@tok{#1}\expandafter\PY@toks\fi}
\def\PY@do#1{\PY@bc{\PY@tc{\PY@ul{%
    \PY@it{\PY@bf{\PY@ff{#1}}}}}}}
\def\PY#1#2{\PY@reset\PY@toks#1+\relax+\PY@do{#2}}

\expandafter\def\csname PY@tok@gs\endcsname{\let\PY@bf=\textbf}
\expandafter\def\csname PY@tok@gp\endcsname{\let\PY@bf=\textbf\def\PY@tc##1{\textcolor[rgb]{0.00,0.00,0.50}{##1}}}
\expandafter\def\csname PY@tok@gr\endcsname{\def\PY@tc##1{\textcolor[rgb]{1.00,0.00,0.00}{##1}}}
\expandafter\def\csname PY@tok@cp\endcsname{\def\PY@tc##1{\textcolor[rgb]{0.74,0.48,0.00}{##1}}}
\expandafter\def\csname PY@tok@ow\endcsname{\let\PY@bf=\textbf\def\PY@tc##1{\textcolor[rgb]{0.67,0.13,1.00}{##1}}}
\expandafter\def\csname PY@tok@mi\endcsname{\def\PY@tc##1{\textcolor[rgb]{0.40,0.40,0.40}{##1}}}
\expandafter\def\csname PY@tok@mf\endcsname{\def\PY@tc##1{\textcolor[rgb]{0.40,0.40,0.40}{##1}}}
\expandafter\def\csname PY@tok@nd\endcsname{\def\PY@tc##1{\textcolor[rgb]{0.67,0.13,1.00}{##1}}}
\expandafter\def\csname PY@tok@vc\endcsname{\def\PY@tc##1{\textcolor[rgb]{0.10,0.09,0.49}{##1}}}
\expandafter\def\csname PY@tok@s1\endcsname{\def\PY@tc##1{\textcolor[rgb]{0.73,0.13,0.13}{##1}}}
\expandafter\def\csname PY@tok@vm\endcsname{\def\PY@tc##1{\textcolor[rgb]{0.10,0.09,0.49}{##1}}}
\expandafter\def\csname PY@tok@vi\endcsname{\def\PY@tc##1{\textcolor[rgb]{0.10,0.09,0.49}{##1}}}
\expandafter\def\csname PY@tok@il\endcsname{\def\PY@tc##1{\textcolor[rgb]{0.40,0.40,0.40}{##1}}}
\expandafter\def\csname PY@tok@sc\endcsname{\def\PY@tc##1{\textcolor[rgb]{0.73,0.13,0.13}{##1}}}
\expandafter\def\csname PY@tok@na\endcsname{\def\PY@tc##1{\textcolor[rgb]{0.49,0.56,0.16}{##1}}}
\expandafter\def\csname PY@tok@kn\endcsname{\let\PY@bf=\textbf\def\PY@tc##1{\textcolor[rgb]{0.00,0.50,0.00}{##1}}}
\expandafter\def\csname PY@tok@mo\endcsname{\def\PY@tc##1{\textcolor[rgb]{0.40,0.40,0.40}{##1}}}
\expandafter\def\csname PY@tok@kt\endcsname{\def\PY@tc##1{\textcolor[rgb]{0.69,0.00,0.25}{##1}}}
\expandafter\def\csname PY@tok@gi\endcsname{\def\PY@tc##1{\textcolor[rgb]{0.00,0.63,0.00}{##1}}}
\expandafter\def\csname PY@tok@m\endcsname{\def\PY@tc##1{\textcolor[rgb]{0.40,0.40,0.40}{##1}}}
\expandafter\def\csname PY@tok@se\endcsname{\let\PY@bf=\textbf\def\PY@tc##1{\textcolor[rgb]{0.73,0.40,0.13}{##1}}}
\expandafter\def\csname PY@tok@cs\endcsname{\let\PY@it=\textit\def\PY@tc##1{\textcolor[rgb]{0.25,0.50,0.50}{##1}}}
\expandafter\def\csname PY@tok@ne\endcsname{\let\PY@bf=\textbf\def\PY@tc##1{\textcolor[rgb]{0.82,0.25,0.23}{##1}}}
\expandafter\def\csname PY@tok@kr\endcsname{\let\PY@bf=\textbf\def\PY@tc##1{\textcolor[rgb]{0.00,0.50,0.00}{##1}}}
\expandafter\def\csname PY@tok@gu\endcsname{\let\PY@bf=\textbf\def\PY@tc##1{\textcolor[rgb]{0.50,0.00,0.50}{##1}}}
\expandafter\def\csname PY@tok@ge\endcsname{\let\PY@it=\textit}
\expandafter\def\csname PY@tok@gt\endcsname{\def\PY@tc##1{\textcolor[rgb]{0.00,0.27,0.87}{##1}}}
\expandafter\def\csname PY@tok@mh\endcsname{\def\PY@tc##1{\textcolor[rgb]{0.40,0.40,0.40}{##1}}}
\expandafter\def\csname PY@tok@nt\endcsname{\let\PY@bf=\textbf\def\PY@tc##1{\textcolor[rgb]{0.00,0.50,0.00}{##1}}}
\expandafter\def\csname PY@tok@ss\endcsname{\def\PY@tc##1{\textcolor[rgb]{0.10,0.09,0.49}{##1}}}
\expandafter\def\csname PY@tok@gd\endcsname{\def\PY@tc##1{\textcolor[rgb]{0.63,0.00,0.00}{##1}}}
\expandafter\def\csname PY@tok@vg\endcsname{\def\PY@tc##1{\textcolor[rgb]{0.10,0.09,0.49}{##1}}}
\expandafter\def\csname PY@tok@nn\endcsname{\let\PY@bf=\textbf\def\PY@tc##1{\textcolor[rgb]{0.00,0.00,1.00}{##1}}}
\expandafter\def\csname PY@tok@sb\endcsname{\def\PY@tc##1{\textcolor[rgb]{0.73,0.13,0.13}{##1}}}
\expandafter\def\csname PY@tok@sh\endcsname{\def\PY@tc##1{\textcolor[rgb]{0.73,0.13,0.13}{##1}}}
\expandafter\def\csname PY@tok@nl\endcsname{\def\PY@tc##1{\textcolor[rgb]{0.63,0.63,0.00}{##1}}}
\expandafter\def\csname PY@tok@ch\endcsname{\let\PY@it=\textit\def\PY@tc##1{\textcolor[rgb]{0.25,0.50,0.50}{##1}}}
\expandafter\def\csname PY@tok@go\endcsname{\def\PY@tc##1{\textcolor[rgb]{0.53,0.53,0.53}{##1}}}
\expandafter\def\csname PY@tok@s2\endcsname{\def\PY@tc##1{\textcolor[rgb]{0.73,0.13,0.13}{##1}}}
\expandafter\def\csname PY@tok@kc\endcsname{\let\PY@bf=\textbf\def\PY@tc##1{\textcolor[rgb]{0.00,0.50,0.00}{##1}}}
\expandafter\def\csname PY@tok@w\endcsname{\def\PY@tc##1{\textcolor[rgb]{0.73,0.73,0.73}{##1}}}
\expandafter\def\csname PY@tok@sa\endcsname{\def\PY@tc##1{\textcolor[rgb]{0.73,0.13,0.13}{##1}}}
\expandafter\def\csname PY@tok@k\endcsname{\let\PY@bf=\textbf\def\PY@tc##1{\textcolor[rgb]{0.00,0.50,0.00}{##1}}}
\expandafter\def\csname PY@tok@nf\endcsname{\def\PY@tc##1{\textcolor[rgb]{0.00,0.00,1.00}{##1}}}
\expandafter\def\csname PY@tok@dl\endcsname{\def\PY@tc##1{\textcolor[rgb]{0.73,0.13,0.13}{##1}}}
\expandafter\def\csname PY@tok@nb\endcsname{\def\PY@tc##1{\textcolor[rgb]{0.00,0.50,0.00}{##1}}}
\expandafter\def\csname PY@tok@fm\endcsname{\def\PY@tc##1{\textcolor[rgb]{0.00,0.00,1.00}{##1}}}
\expandafter\def\csname PY@tok@o\endcsname{\def\PY@tc##1{\textcolor[rgb]{0.40,0.40,0.40}{##1}}}
\expandafter\def\csname PY@tok@cm\endcsname{\let\PY@it=\textit\def\PY@tc##1{\textcolor[rgb]{0.25,0.50,0.50}{##1}}}
\expandafter\def\csname PY@tok@cpf\endcsname{\let\PY@it=\textit\def\PY@tc##1{\textcolor[rgb]{0.25,0.50,0.50}{##1}}}
\expandafter\def\csname PY@tok@c\endcsname{\let\PY@it=\textit\def\PY@tc##1{\textcolor[rgb]{0.25,0.50,0.50}{##1}}}
\expandafter\def\csname PY@tok@sd\endcsname{\let\PY@it=\textit\def\PY@tc##1{\textcolor[rgb]{0.73,0.13,0.13}{##1}}}
\expandafter\def\csname PY@tok@no\endcsname{\def\PY@tc##1{\textcolor[rgb]{0.53,0.00,0.00}{##1}}}
\expandafter\def\csname PY@tok@sr\endcsname{\def\PY@tc##1{\textcolor[rgb]{0.73,0.40,0.53}{##1}}}
\expandafter\def\csname PY@tok@mb\endcsname{\def\PY@tc##1{\textcolor[rgb]{0.40,0.40,0.40}{##1}}}
\expandafter\def\csname PY@tok@gh\endcsname{\let\PY@bf=\textbf\def\PY@tc##1{\textcolor[rgb]{0.00,0.00,0.50}{##1}}}
\expandafter\def\csname PY@tok@kp\endcsname{\def\PY@tc##1{\textcolor[rgb]{0.00,0.50,0.00}{##1}}}
\expandafter\def\csname PY@tok@c1\endcsname{\let\PY@it=\textit\def\PY@tc##1{\textcolor[rgb]{0.25,0.50,0.50}{##1}}}
\expandafter\def\csname PY@tok@si\endcsname{\let\PY@bf=\textbf\def\PY@tc##1{\textcolor[rgb]{0.73,0.40,0.53}{##1}}}
\expandafter\def\csname PY@tok@bp\endcsname{\def\PY@tc##1{\textcolor[rgb]{0.00,0.50,0.00}{##1}}}
\expandafter\def\csname PY@tok@nc\endcsname{\let\PY@bf=\textbf\def\PY@tc##1{\textcolor[rgb]{0.00,0.00,1.00}{##1}}}
\expandafter\def\csname PY@tok@s\endcsname{\def\PY@tc##1{\textcolor[rgb]{0.73,0.13,0.13}{##1}}}
\expandafter\def\csname PY@tok@kd\endcsname{\let\PY@bf=\textbf\def\PY@tc##1{\textcolor[rgb]{0.00,0.50,0.00}{##1}}}
\expandafter\def\csname PY@tok@sx\endcsname{\def\PY@tc##1{\textcolor[rgb]{0.00,0.50,0.00}{##1}}}
\expandafter\def\csname PY@tok@ni\endcsname{\let\PY@bf=\textbf\def\PY@tc##1{\textcolor[rgb]{0.60,0.60,0.60}{##1}}}
\expandafter\def\csname PY@tok@nv\endcsname{\def\PY@tc##1{\textcolor[rgb]{0.10,0.09,0.49}{##1}}}
\expandafter\def\csname PY@tok@err\endcsname{\def\PY@bc##1{\setlength{\fboxsep}{0pt}\fcolorbox[rgb]{1.00,0.00,0.00}{1,1,1}{\strut ##1}}}

\def\PYZbs{\char`\\}
\def\PYZus{\char`\_}
\def\PYZob{\char`\{}
\def\PYZcb{\char`\}}

\def\PYZam{\char`\&}
\def\PYZlt{\char`\<}
\def\PYZgt{\char`\>}
\def\PYZsh{\char`\#}
\def\PYZpc{\char`\%}

\def\PYZhy{\char`\-}
\def\PYZsq{\char`\'}
\def\PYZdq{\char`\"}

\makeatother

\definecolor{incolor}{rgb}{0.0, 0.0, 0.5}
\definecolor{outcolor}{rgb}{0.545, 0.0, 0.0}

\hypersetup{
	breaklinks=true,  
    colorlinks=true,
    urlcolor=urlcolor,
    linkcolor=linkcolor,
    citecolor=citecolor,
}
    

%% file: FilterBank_Calibration_NEW.tex
    \subsubsection{Step 1: Initialization}\

Import all classes and set up global parameters

    \begin{Verbatim}[commandchars=\\\{\}]
{\color{incolor}In [{\color{incolor}1}]:} \PY{c+c1}{\PYZsh{} Imports and reservations}
        \PY{k+kn}{import} \PY{n+nn}{numpy} \PY{k}{as} \PY{n+nn}{np}
        \PY{k+kn}{import} \PY{n+nn}{matplotlib}\PY{n+nn}{.}\PY{n+nn}{pyplot} \PY{k}{as} \PY{n+nn}{plt}
        \PY{k+kn}{import} \PY{n+nn}{lightlab}\PY{n+nn}{.}\PY{n+nn}{instruments} \PY{k}{as} \PY{n+nn}{inst}
        \PY{k+kn}{import} \PY{n+nn}{lightlab}\PY{n+nn}{.}\PY{n+nn}{model} \PY{k}{as} \PY{n+nn}{m}
        \PY{k+kn}{from} \PY{n+nn}{lightlab}\PY{n+nn}{.}\PY{n+nn}{util}\PY{n+nn}{.}\PY{n+nn}{modeling} \PY{k}{import} \PY{n}{kOSAPwr}\PY{p}{,} \PY{n}{dbm2lin}\PY{p}{,}
            \PY{n}{kOSASpacing}\PY{p}{,} \PY{n}{CurrentUnit}\PY{p}{,} \PY{n}{MrrOut}
        \PY{k+kn}{from} \PY{n+nn}{lightlab}\PY{n+nn}{.}\PY{n+nn}{util}\PY{n+nn}{.}\PY{n+nn}{calibrating} \PY{k}{import} \PY{n}{SpectrumMeasurementAssistant}
        \PY{k+kn}{from} \PY{n+nn}{lightlab}\PY{n+nn}{.}\PY{n+nn}{util} \PY{k}{import} \PY{n}{io}
\end{Verbatim}

    \begin{Verbatim}[commandchars=\\\{\}]
{\color{incolor}In [{\color{incolor}2}]:} \PY{c+c1}{\PYZsh{} Global Parameters, Available to Calibration Model}
        \PY{n}{wlChannels} \PY{o}{=} \PY{p}{[}\PY{l+m+mi}{1550}\PY{p}{,} \PY{l+m+mi}{1552}\PY{p}{,} \PY{l+m+mi}{1554}\PY{p}{,} \PY{l+m+mi}{1556}\PY{p}{]}
        \PY{n}{wlRange} \PY{o}{=} \PY{p}{[}\PY{l+m+mi}{1530}\PY{p}{,} \PY{l+m+mi}{1559}\PY{p}{]}
        \PY{n}{currentChan} \PY{o}{=} \PY{p}{[}\PY{l+m+mi}{5}\PY{p}{,} \PY{l+m+mi}{3}\PY{p}{,} \PY{l+m+mi}{6}\PY{p}{,} \PY{l+m+mi}{4}\PY{p}{]}
        \PY{n}{numRings} \PY{o}{=} \PY{l+m+mi}{4}
        \PY{n}{minPeakDist} \PY{o}{=} \PY{l+m+mf}{0.5} \PY{o}{/} \PY{n}{kOSASpacing} \PY{c+c1}{\PYZsh{} 0.5nm spacing for peak\PYZhy{}finder}
\end{Verbatim}

    \subsubsection{Step 2: Initialize Hidden
Model}

Create a virtual model that will mimic the instruments. This will remain
hidden during calibration.

In this case, we are making a 4-ring Lorentzian Filter Bank. Then,
register it to the Instruments Module.

    \begin{Verbatim}[commandchars=\\\{\}]
{\color{incolor}In [{\color{incolor}3}]:} \PY{c+c1}{\PYZsh{} Thermal Group, Uses 4 current channels}
        \PY{n}{therm} \PY{o}{=} \PY{n}{m}\PY{o}{.}\PY{n}{ThermalGroup}\PY{p}{(}\PY{n}{currentChan}\PY{p}{)}
        
        \PY{c+c1}{\PYZsh{} Model Parameters, NOT available to calibration model}
        \PY{n}{attenuation} \PY{o}{=} \PY{l+m+mf}{0.0001} \PY{c+c1}{\PYZsh{} 40dB baseline attenuation}
        \PY{n}{lfwhm} \PY{o}{=} \PY{l+m+mf}{0.2}          \PY{c+c1}{\PYZsh{} 0.1nm fwhm of lorentzian}
        \PY{n}{latten} \PY{o}{=} \PY{l+m+mf}{0.98}        \PY{c+c1}{\PYZsh{} Attenuation at resonance}
        \PY{n}{heatBias} \PY{o}{=} \PY{p}{\PYZob{}}\PY{n}{currentChan}\PY{p}{[}\PY{l+m+mi}{0}\PY{p}{]}\PY{p}{:} \PY{l+m+mf}{2.5}\PY{p}{,} \PY{n}{currentChan}\PY{p}{[}\PY{l+m+mi}{1}\PY{p}{]}\PY{p}{:} \PY{l+m+mf}{2.0}\PY{p}{,} 
            \PY{n}{currentChan}\PY{p}{[}\PY{l+m+mi}{2}\PY{p}{]}\PY{p}{:} \PY{l+m+mf}{1.5}\PY{p}{,} \PY{n}{currentChan}\PY{p}{[}\PY{l+m+mi}{3}\PY{p}{]}\PY{p}{:} \PY{l+m+mf}{1.0}\PY{p}{\PYZcb{}}
        
        \PY{c+c1}{\PYZsh{} Random K with extra on diagonal to denote primary filament}
        \PY{n}{K} \PY{o}{=} \PY{n}{np}\PY{o}{.}\PY{n}{multiply}\PY{p}{(}\PY{n}{np}\PY{o}{.}\PY{n}{absolute}\PY{p}{(}\PY{p}{(}\PY{l+m+mf}{0.1} \PY{o}{*} \PY{n}{np}\PY{o}{.}\PY{n}{random}\PY{o}{.}\PY{n}{randn}\PY{p}{(}\PY{n}{numRings}\PY{p}{,} 
            \PY{n+nb}{len}\PY{p}{(}\PY{n}{currentChan}\PY{p}{)}\PY{p}{)} \PY{o}{+} \PY{l+m+mf}{0.1}\PY{p}{)} \PY{o}{+} \PY{l+m+mf}{0.5}\PY{o}{*}\PY{n}{np}\PY{o}{.}\PY{n}{eye}\PY{p}{(}\PY{n}{numRings}\PY{p}{)}\PY{p}{)}\PY{p}{,} \PY{l+m+mi}{200}\PY{p}{)}
        
        \PY{c+c1}{\PYZsh{} FilterBank Module}
        \PY{n}{fb} \PY{o}{=} \PY{n}{m}\PY{o}{.}\PY{n}{FilterBank}\PY{p}{(}\PY{n}{therm}\PY{p}{,} \PY{n}{numRings}\PY{p}{)}
        \PY{n}{fb}\PY{o}{.}\PY{n}{setAttenuation}\PY{p}{(}\PY{n}{attenuation}\PY{p}{)}
        
        \PY{n}{fb}\PY{o}{.}\PY{n}{setBiasParams}\PY{p}{(}\PY{n}{wlChannels}\PY{p}{,} \PY{n}{latten} \PY{o}{*} \PY{n}{np}\PY{o}{.}\PY{n}{ones}\PY{p}{(}\PY{n}{numRings}\PY{p}{)}\PY{p}{,}
            \PY{n}{lfwhm} \PY{o}{*} \PY{n}{np}\PY{o}{.}\PY{n}{ones}\PY{p}{(}\PY{n}{numRings}\PY{p}{)}\PY{p}{)}
        
        \PY{c+c1}{\PYZsh{} Set K and bias for Thermal Group}
        \PY{n}{therm}\PY{o}{.}\PY{n}{setK}\PY{p}{(}\PY{n}{K}\PY{p}{)}
        \PY{n}{therm}\PY{o}{.}\PY{n}{setHeatBias}\PY{p}{(}\PY{n}{heatBias}\PY{p}{)}
\end{Verbatim}

    \begin{Verbatim}[commandchars=\\\{\}]
{\color{incolor}In [{\color{incolor}4}]:} \PY{c+c1}{\PYZsh{} Reserve Current Channels, Add Current Channels}
        \PY{n}{inst}\PY{o}{.}\PY{n}{togglePhony}\PY{p}{(}\PY{k+kc}{True}\PY{p}{,} \PY{n}{fb}\PY{p}{)}
        \PY{n}{token} \PY{o}{=} \PY{n}{inst}\PY{o}{.}\PY{n}{reserveCurrentChan}\PY{p}{(}\PY{n}{currentChan}\PY{p}{)}
\end{Verbatim}

    \begin{Verbatim}[commandchars=\\\{\}]
{\color{incolor}In [{\color{incolor}5}]:} \PY{n}{fb}\PY{o}{.}\PY{n}{setOsaOut}\PY{p}{(}\PY{n}{MrrOut}\PY{o}{.}\PY{n}{kThru}\PY{o}{.}\PY{n}{value}\PY{p}{)}
        \PY{n}{nm}\PY{p}{,} \PY{n}{dbm} \PY{o}{=} \PY{n}{inst}\PY{o}{.}\PY{n}{spectrum}\PY{p}{(}\PY{n}{wlRange}\PY{p}{)}
        \PY{n}{plt}\PY{o}{.}\PY{n}{plot}\PY{p}{(}\PY{n}{nm}\PY{p}{,} \PY{n}{dbm}\PY{p}{)}
        \PY{n}{plt}\PY{o}{.}\PY{n}{show}\PY{p}{(}\PY{p}{)}
\end{Verbatim}

    { \hspace*{\fill} \\}
    
    \subsubsection{Step 3: Ascription}

Map primary filament to each peak.

    \begin{Verbatim}[commandchars=\\\{\}]
{\color{incolor}In [{\color{incolor}6}]:} \PY{c+c1}{\PYZsh{} Initialization}
        \PY{n}{calCurrentChan} \PY{o}{=} \PY{n}{np}\PY{o}{.}\PY{n}{array}\PY{p}{(}\PY{p}{[}\PY{l+m+mi}{3}\PY{p}{,} \PY{l+m+mi}{4}\PY{p}{,} \PY{l+m+mi}{5}\PY{p}{,} \PY{l+m+mi}{6}\PY{p}{]}\PY{p}{)}
        \PY{n}{calCurrentState} \PY{o}{=} \PY{n+nb}{dict}\PY{p}{(}\PY{p}{)} \PY{c+c1}{\PYZsh{} in mW}
        \PY{k}{for} \PY{n}{ch} \PY{o+ow}{in} \PY{n}{calCurrentChan}\PY{p}{:}
            \PY{n}{calCurrentState}\PY{p}{[}\PY{n}{ch}\PY{p}{]} \PY{o}{=} \PY{l+m+mi}{0}
            
        \PY{n}{spctAssist} \PY{o}{=} \PY{n}{SpectrumMeasurementAssistant}\PY{p}{(}\PY{n}{nChan}\PY{o}{=}\PY{n}{numRings}\PY{p}{,} 
            \PY{n}{arePeaks}\PY{o}{=}\PY{k+kc}{False}\PY{p}{,} \PY{n}{visualize}\PY{o}{=}\PY{k+kc}{False}\PY{p}{)}
\end{Verbatim}

    \begin{Verbatim}[commandchars=\\\{\}]
{\color{incolor}In [{\color{incolor}7}]:} \PY{c+c1}{\PYZsh{} Set Base Numbers}
        \PY{n}{baseTune} \PY{o}{=} \PY{n}{calCurrentState}
        \PY{n}{baseLams} \PY{o}{=} \PY{n}{np}\PY{o}{.}\PY{n}{array}\PY{p}{(}\PY{p}{[}\PY{n}{r}\PY{o}{.}\PY{n}{lam} \PY{k}{for} \PY{n}{r} \PY{o+ow}{in} \PY{n}{spctAssist}\PY{o}{.}\PY{n}{resonances}\PY{p}{(}\PY{p}{)}\PY{p}{]}\PY{p}{)}
        \PY{n}{tuneBy} \PY{o}{=} \PY{l+m+mf}{0.01} \PY{c+c1}{\PYZsh{} in mW/Ohm\PYZdq{}}
        \PY{n}{ascrBuilder} \PY{o}{=} \PY{n}{np}\PY{o}{.}\PY{n}{arange}\PY{p}{(}\PY{n+nb}{len}\PY{p}{(}\PY{n}{calCurrentChan}\PY{p}{)}\PY{p}{)}
        \PY{n}{kEstBuilder} \PY{o}{=} \PY{n}{np}\PY{o}{.}\PY{n}{arange}\PY{p}{(}\PY{n+nb}{len}\PY{p}{(}\PY{n}{calCurrentChan}\PY{p}{)}\PY{p}{)}
\end{Verbatim}

    \begin{Verbatim}[commandchars=\\\{\}]
{\color{incolor}In [{\color{incolor}8}]:} \PY{k+kn}{from} \PY{n+nn}{time} \PY{k}{import} \PY{n}{sleep}
        \PY{c+c1}{\PYZsh{} Run Ascription}
        \PY{k}{for} \PY{n}{iChan}\PY{p}{,} \PY{n}{ch} \PY{o+ow}{in} \PY{n+nb}{enumerate}\PY{p}{(}\PY{n}{calCurrentChan}\PY{p}{)}\PY{p}{:}
            \PY{n}{inst}\PY{o}{.}\PY{n}{setCurrentChanTuning}\PY{p}{(}\PY{p}{\PYZob{}}\PY{n}{ch}\PY{p}{:} 
                \PY{n}{baseTune}\PY{p}{[}\PY{n}{ch}\PY{p}{]} \PY{o}{+} \PY{n}{tuneBy}\PY{p}{\PYZcb{}}\PY{p}{,} \PY{n}{token}\PY{p}{,} \PY{n}{CurrentUnit}\PY{o}{.}\PY{n}{mW}\PY{p}{)}
            \PY{n}{spect} \PY{o}{=} \PY{n}{spctAssist}\PY{o}{.}\PY{n}{fgSpect}\PY{p}{(}\PY{p}{)}
            \PY{n}{presLams} \PY{o}{=} \PY{n}{np}\PY{o}{.}\PY{n}{array}\PY{p}{(}\PY{p}{[}\PY{n}{r}\PY{o}{.}\PY{n}{lam} 
                \PY{k}{for} \PY{n}{r} \PY{o+ow}{in} \PY{n}{spctAssist}\PY{o}{.}\PY{n}{resonances}\PY{p}{(}\PY{n}{spect}\PY{p}{)}\PY{p}{]}\PY{p}{)}
            \PY{n}{inst}\PY{o}{.}\PY{n}{setCurrentChanTuning}\PY{p}{(}\PY{p}{\PYZob{}}\PY{n}{ch}\PY{p}{:} \PY{n}{baseTune}\PY{p}{[}\PY{n}{ch}\PY{p}{]}\PY{p}{\PYZcb{}}\PY{p}{,} \PY{n}{token}\PY{p}{)}
            \PY{n}{shifts} \PY{o}{=} \PY{n}{presLams} \PY{o}{\PYZhy{}} \PY{n}{baseLams}
            \PY{n}{ascrBuilder}\PY{p}{[}\PY{n}{iChan}\PY{p}{]} \PY{o}{=} \PY{n}{np}\PY{o}{.}\PY{n}{argmax}\PY{p}{(}\PY{n}{shifts}\PY{p}{)}
            \PY{n}{kEstBuilder}\PY{p}{[}\PY{n}{iChan}\PY{p}{]} \PY{o}{=} \PY{n+nb}{max}\PY{p}{(}\PY{n}{shifts}\PY{p}{)} \PY{o}{/} \PY{n}{tuneBy}
            
        \PY{c+c1}{\PYZsh{} Re\PYZhy{}Order Current Channels}
        \PY{n}{newCalCurrentChan} \PY{o}{=} \PY{p}{[}\PY{n}{calCurrentChan}\PY{p}{[}\PY{n}{j}\PY{p}{]} 
            \PY{k}{for} \PY{n}{j} \PY{o+ow}{in} \PY{n}{np}\PY{o}{.}\PY{n}{argsort}\PY{p}{(}\PY{n}{ascrBuilder}\PY{p}{)}\PY{p}{]}
        \PY{n}{kEstTemp} \PY{o}{=} \PY{n}{kEstBuilder}\PY{p}{[}\PY{n}{np}\PY{o}{.}\PY{n}{argsort}\PY{p}{(}\PY{n}{ascrBuilder}\PY{p}{)}\PY{p}{]}
\end{Verbatim}

    \begin{Verbatim}[commandchars=\\\{\}]
{\color{incolor}In [{\color{incolor}9}]:} \PY{c+c1}{\PYZsh{} Validation, make sure re\PYZhy{}ordered list is }
        \PY{k}{for} \PY{n}{j} \PY{o+ow}{in} \PY{n+nb}{range}\PY{p}{(}\PY{n+nb}{len}\PY{p}{(}\PY{n}{newCalCurrentChan}\PY{p}{)}\PY{p}{)}\PY{p}{:}
            \PY{k}{assert} \PY{n}{newCalCurrentChan}\PY{p}{[}\PY{n}{j}\PY{p}{]} \PY{o}{==} \PY{n}{currentChan}\PY{p}{[}\PY{n}{j}\PY{p}{]}
\end{Verbatim}

    \subsubsection{Step 4: Create Calibration
Model}

Create an empty calibration model to be filled, using the ascribed
channels for validation purposes.

    \begin{Verbatim}[commandchars=\\\{\}]
{\color{incolor}In [{\color{incolor}10}]:} \PY{n}{calTherm} \PY{o}{=} \PY{n}{m}\PY{o}{.}\PY{n}{ThermalGroup}\PY{p}{(}\PY{n}{newCalCurrentChan}\PY{p}{)}
         \PY{n}{calFB} \PY{o}{=} \PY{n}{m}\PY{o}{.}\PY{n}{FilterBank}\PY{p}{(}\PY{n}{calTherm}\PY{p}{,} \PY{n}{numRings}\PY{p}{)}
\end{Verbatim}

    \subsubsection{Step 5: Background Removal and
Tracking}

PI-Controller to move channels onto wavelengths.

At the end, we should be able to set the \textbf{biases.}

    \begin{Verbatim}[commandchars=\\\{\}]
{\color{incolor}In [{\color{incolor}11}]:} \PY{c+c1}{\PYZsh{}\PYZsh{}\PYZsh{} Background Removal}
         
         \PY{n}{avgOnSpect}\PY{o}{=}\PY{l+m+mi}{3}
         \PY{n}{detuneByFwhms} \PY{o}{=} \PY{l+m+mi}{3}
         
         \PY{c+c1}{\PYZsh{} Get resonance FWHMs.}
         \PY{n}{resFwhms} \PY{o}{=} \PY{n}{np}\PY{o}{.}\PY{n}{array}\PY{p}{(}\PY{p}{[}\PY{n}{r}\PY{o}{.}\PY{n}{fwhm} \PY{k}{for} \PY{n}{r} \PY{o+ow}{in} \PY{n}{spctAssist}\PY{o}{.}\PY{n}{resonances}\PY{p}{(}\PY{p}{)}\PY{p}{]}\PY{p}{)}
         \PY{n}{displacedWls} \PY{o}{=} \PY{n}{detuneByFwhms} \PY{o}{*} \PY{n}{resFwhms}
         
         \PY{c+c1}{\PYZsh{} Get Raw Spectrum}
         \PY{n}{baseRawSpct} \PY{o}{=} \PY{n}{spctAssist}\PY{o}{.}\PY{n}{fgSpect}\PY{p}{(}\PY{n}{avgCnt}\PY{o}{=}\PY{n}{avgOnSpect}\PY{p}{,}
            \PY{n}{bgType}\PY{o}{=}\PY{l+s+s1}{\PYZsq{}}\PY{l+s+s1}{smoothed}\PY{l+s+s1}{\PYZsq{}}\PY{p}{)}
         \PY{c+c1}{\PYZsh{} Tune to displace resonances, look at new spectrum, tune back}
         \PY{n}{displTuning} \PY{o}{=} \PY{n+nb}{dict}\PY{p}{(}\PY{p}{)}
         \PY{k}{for} \PY{n}{j} \PY{o+ow}{in} \PY{n+nb}{range}\PY{p}{(}\PY{n+nb}{len}\PY{p}{(}\PY{n}{displacedWls}\PY{p}{)}\PY{p}{)}\PY{p}{:}
             \PY{n}{displTuning}\PY{p}{[}\PY{n}{newCalCurrentChan}\PY{p}{[}\PY{n}{j}\PY{p}{]}\PY{p}{]} \PY{o}{=} 
                \PY{n}{displacedWls}\PY{p}{[}\PY{n}{j}\PY{p}{]} \PY{o}{/} \PY{n}{kEstTemp}\PY{p}{[}\PY{n}{j}\PY{p}{]}
         \PY{n}{inst}\PY{o}{.}\PY{n}{setCurrentChanTuning}\PY{p}{(}\PY{n}{displTuning}\PY{p}{,} \PY{n}{token}\PY{p}{,} \PY{n}{CurrentUnit}\PY{o}{.}\PY{n}{mW}\PY{p}{)}
         \PY{n}{displacedRawSpct} \PY{o}{=} \PY{n}{spctAssist}\PY{o}{.}\PY{n}{fgSpect}\PY{p}{(}\PY{n}{avgCnt}\PY{o}{=}\PY{n}{avgOnSpect}\PY{p}{,}
            \PY{n}{bgType}\PY{o}{=}\PY{l+s+s1}{\PYZsq{}}\PY{l+s+s1}{smoothed}\PY{l+s+s1}{\PYZsq{}}\PY{p}{)}
         
         \PY{c+c1}{\PYZsh{} Return to base}
         \PY{n}{inst}\PY{o}{.}\PY{n}{setCurrentChanTuning}\PY{p}{(}\PY{n}{baseTune}\PY{p}{,} \PY{n}{token}\PY{p}{,} \PY{n}{CurrentUnit}\PY{o}{.}\PY{n}{mW}\PY{p}{)}
         
         \PY{c+c1}{\PYZsh{} Update Background}
         \PY{n}{spctAssist}\PY{o}{.}\PY{n}{setBgTuned}\PY{p}{(}\PY{n}{baseRawSpct}\PY{p}{,} \PY{n}{displacedRawSpct}\PY{p}{)}
\end{Verbatim}

    \begin{Verbatim}[commandchars=\\\{\}]
{\color{incolor}In [{\color{incolor}12}]:} \PY{c+c1}{\PYZsh{} See Peaks w/ Background Removed}
         \PY{n}{spctAssist}\PY{o}{.}\PY{n}{fgResPlot}\PY{p}{(}\PY{p}{)}
         \PY{n}{plt}\PY{o}{.}\PY{n}{show}\PY{p}{(}\PY{p}{)}
\end{Verbatim}

    
    \begin{Verbatim}[commandchars=\\\{\}]
{\color{incolor}In [{\color{incolor}13}]:} \PY{c+c1}{\PYZsh{}\PYZsh{}\PYZsh{} Tracking}
         \PY{n}{precision} \PY{o}{=} \PY{l+m+mf}{0.005} \PY{c+c1}{\PYZsh{} Threshold}
         \PY{n}{propCoef} \PY{o}{=} \PY{l+m+mf}{0.5} \PY{c+c1}{\PYZsh{} kP}
         \PY{n}{avgCnt} \PY{o}{=} \PY{l+m+mi}{4}
         \PY{n}{targets} \PY{o}{=} \PY{n}{np}\PY{o}{.}\PY{n}{array}\PY{p}{(}\PY{n}{wlChannels}\PY{p}{)}
         \PY{n}{nowTune} \PY{o}{=} \PY{n}{baseTune}
         \PY{n}{appxThrmCoefs} \PY{o}{=} \PY{n}{kEstTemp}
         \PY{n}{inst}\PY{o}{.}\PY{n}{setCurrentChanTuning}\PY{p}{(}\PY{n}{baseTune}\PY{p}{,} \PY{n}{token}\PY{p}{,} \PY{n}{CurrentUnit}\PY{o}{.}\PY{n}{mW}\PY{p}{)}
         
         \PY{k}{for} \PY{n}{i} \PY{o+ow}{in} \PY{n+nb}{range}\PY{p}{(}\PY{l+m+mi}{100}\PY{p}{)}\PY{p}{:}
             \PY{n}{spect} \PY{o}{=} \PY{n}{spctAssist}\PY{o}{.}\PY{n}{fgSpect}\PY{p}{(}\PY{n}{bgType}\PY{o}{=}\PY{l+s+s1}{\PYZsq{}}\PY{l+s+s1}{tuned}\PY{l+s+s1}{\PYZsq{}}\PY{p}{,} \PY{n}{avgCnt} \PY{o}{=} \PY{n}{avgCnt}\PY{p}{)}
             \PY{n}{spect}\PY{o}{.}\PY{n}{simplePlot}\PY{p}{(}\PY{p}{)}
             \PY{n}{actualPeaks} \PY{o}{=} \PY{n}{np}\PY{o}{.}\PY{n}{array}\PY{p}{(}\PY{p}{[}\PY{n}{r}\PY{o}{.}\PY{n}{lam} 
                \PY{k}{for} \PY{n}{r} \PY{o+ow}{in} \PY{n}{spctAssist}\PY{o}{.}\PY{n}{resonances}\PY{p}{(}\PY{n}{spect}\PY{p}{)}\PY{p}{]}\PY{p}{)}
             \PY{n}{errs} \PY{o}{=} \PY{n}{targets} \PY{o}{\PYZhy{}} \PY{n}{actualPeaks}
             \PY{n}{io}\PY{o}{.}\PY{n}{printProgress}\PY{p}{(}\PY{l+s+s1}{\PYZsq{}}\PY{l+s+s1}{i=}\PY{l+s+s1}{\PYZsq{}}\PY{p}{,} \PY{n}{i}\PY{p}{,} \PY{l+s+s1}{\PYZsq{}}\PY{l+s+s1}{, error=}\PY{l+s+s1}{\PYZsq{}}\PY{p}{,} \PY{n+nb}{max}\PY{p}{(}\PY{n+nb}{abs}\PY{p}{(}\PY{n}{errs}\PY{p}{)}\PY{p}{)}\PY{p}{)}
             \PY{k}{if} \PY{n+nb}{max}\PY{p}{(}\PY{n+nb}{abs}\PY{p}{(}\PY{n}{errs}\PY{p}{)}\PY{p}{)} \PY{o}{\PYZlt{}} \PY{n}{precision}\PY{p}{:}
                 \PY{n+nb}{print}\PY{p}{(}\PY{l+s+s1}{\PYZsq{}}\PY{l+s+se}{\PYZbs{}n}\PY{l+s+s1}{Tracking complete}\PY{l+s+s1}{\PYZsq{}}\PY{p}{)}
                 \PY{k}{break}
             \PY{c+c1}{\PYZsh{} recenter wlRange to avoid other FSR resonances}
             \PY{n}{wlRangeTight} \PY{o}{=} \PY{n}{np}\PY{o}{.}\PY{n}{array}\PY{p}{(}\PY{p}{[}\PY{n+nb}{min}\PY{p}{(}\PY{n+nb}{min}\PY{p}{(}\PY{n}{actualPeaks}\PY{p}{)}\PY{p}{,} \PY{n+nb}{min}\PY{p}{(}\PY{n}{targets}\PY{p}{)}\PY{p}{)}\PY{p}{,}
                \PY{n+nb}{max}\PY{p}{(}\PY{n+nb}{max}\PY{p}{(}\PY{n}{actualPeaks}\PY{p}{)}\PY{p}{,} \PY{n+nb}{max}\PY{p}{(}\PY{n}{targets}\PY{p}{)}\PY{p}{)}\PY{p}{]}\PY{p}{)}
             \PY{n}{newwlRange} \PY{o}{=} \PY{n}{np}\PY{o}{.}\PY{n}{mean}\PY{p}{(}\PY{n}{wlRangeTight}\PY{p}{)} \PY{o}{+}
                \PY{n}{np}\PY{o}{.}\PY{n}{diff}\PY{p}{(}\PY{n}{wlRangeTight}\PY{p}{)} \PY{o}{*} \PY{n}{np}\PY{o}{.}\PY{n}{array}\PY{p}{(}\PY{p}{[}\PY{o}{\PYZhy{}}\PY{l+m+mi}{1}\PY{p}{,} \PY{l+m+mi}{1}\PY{p}{]}\PY{p}{)} \PY{o}{/} \PY{l+m+mi}{2} \PY{o}{*} \PY{l+m+mi}{2}
             \PY{n}{spctAssist}\PY{o}{.}\PY{n}{wlRange} \PY{o}{=} \PY{n}{newwlRange}
             \PY{k}{try}\PY{p}{:}
                 \PY{k}{for} \PY{n}{j} \PY{o+ow}{in} \PY{n+nb}{range}\PY{p}{(}\PY{n+nb}{len}\PY{p}{(}\PY{n}{newCalCurrentChan}\PY{p}{)}\PY{p}{)}\PY{p}{:}
                     \PY{n}{ch} \PY{o}{=} \PY{n}{newCalCurrentChan}\PY{p}{[}\PY{n}{j}\PY{p}{]}
                     \PY{n}{nowTune}\PY{p}{[}\PY{n}{ch}\PY{p}{]} \PY{o}{=} \PY{n}{nowTune}\PY{p}{[}\PY{n}{ch}\PY{p}{]} \PY{o}{+}
                        \PY{n}{propCoef}\PY{o}{*}\PY{n}{errs}\PY{p}{[}\PY{n}{j}\PY{p}{]} \PY{o}{/} \PY{n}{appxThrmCoefs}\PY{p}{[}\PY{n}{j}\PY{p}{]}
                 \PY{n}{inst}\PY{o}{.}\PY{n}{setCurrentChanTuning}\PY{p}{(}\PY{n}{nowTune}\PY{p}{,} \PY{n}{token}\PY{p}{,} \PY{n}{CurrentUnit}\PY{o}{.}\PY{n}{mW}\PY{p}{)}
             \PY{k}{except} \PY{n}{io}\PY{o}{.}\PY{n}{RangeError} \PY{k}{as} \PY{n}{err}\PY{p}{:}
                 \PY{n+nb}{print}\PY{p}{(}\PY{l+s+s1}{\PYZsq{}}\PY{l+s+s1}{Out of range during tracking. See plot}\PY{l+s+s1}{\PYZsq{}}\PY{p}{)}
                 \PY{n}{spctAssist}\PY{o}{.}\PY{n}{fgResPlot}\PY{p}{(}\PY{p}{)}
                 \PY{k}{raise} \PY{n}{err}
                 
         \PY{n}{plt}\PY{o}{.}\PY{n}{show}\PY{p}{(}\PY{p}{)}
\end{Verbatim}

    \begin{Verbatim}[commandchars=\\\{\}]

Tracking complete

    \end{Verbatim}

    
    \begin{Verbatim}[commandchars=\\\{\}]
{\color{incolor}In [{\color{incolor}14}]:} \PY{n}{spctAssist}\PY{o}{.}\PY{n}{fgResPlot}\PY{p}{(}\PY{p}{)}
         \PY{n}{plt}\PY{o}{.}\PY{n}{show}\PY{p}{(}\PY{p}{)}
\end{Verbatim}

    
    \begin{Verbatim}[commandchars=\\\{\}]
{\color{incolor}In [{\color{incolor}15}]:} \PY{n}{calTherm}\PY{o}{.}\PY{n}{setHeatBias}\PY{p}{(}\PY{n}{nowTune}\PY{p}{,} \PY{n}{CurrentUnit}\PY{o}{.}\PY{n}{mW}\PY{p}{)}
\end{Verbatim}

    \begin{Verbatim}[commandchars=\\\{\}]
{\color{incolor}In [{\color{incolor}16}]:} \PY{c+c1}{\PYZsh{} Validation}
         \PY{n}{threshold} \PY{o}{=} \PY{l+m+mf}{0.01}
         
         \PY{n}{calBias} \PY{o}{=} \PY{n}{np}\PY{o}{.}\PY{n}{array}\PY{p}{(}\PY{n+nb}{sorted}\PY{p}{(}\PY{p}{[}\PY{n}{CurrentUnit}\PY{o}{.}\PY{n}{toVolt}\PY{p}{(}\PY{n}{b}\PY{p}{,} \PY{n}{CurrentUnit}\PY{o}{.}\PY{n}{mW}\PY{p}{)}
            \PY{k}{for} \PY{n}{b} \PY{o+ow}{in} \PY{n}{calTherm}\PY{o}{.}\PY{n}{heatBias}\PY{p}{]}\PY{p}{)}\PY{p}{)}
         \PY{n}{virtBias} \PY{o}{=} \PY{n}{np}\PY{o}{.}\PY{n}{array}\PY{p}{(}\PY{n+nb}{sorted}\PY{p}{(}\PY{n+nb}{list}\PY{p}{(}\PY{n}{heatBias}\PY{o}{.}\PY{n}{values}\PY{p}{(}\PY{p}{)}\PY{p}{)}\PY{p}{)}\PY{p}{)}
         \PY{n}{diff} \PY{o}{=} \PY{n}{np}\PY{o}{.}\PY{n}{absolute}\PY{p}{(}\PY{n}{calBias} \PY{o}{\PYZhy{}} \PY{n}{virtBias}\PY{p}{)}
         \PY{n+nb}{print}\PY{p}{(}\PY{n}{diff}\PY{p}{)}
         \PY{k}{for} \PY{n}{d} \PY{o+ow}{in} \PY{n}{diff}\PY{p}{:}
             \PY{k}{assert} \PY{n}{d} \PY{o}{\PYZlt{}} \PY{n}{threshold}
\end{Verbatim}

    \begin{Verbatim}[commandchars=\\\{\}]
[ 0.00272099  0.00028576  0.00065551  0.00018863]

    \end{Verbatim}

    \begin{Verbatim}[commandchars=\\\{\}]
{\color{incolor}In [{\color{incolor}17}]:} \PY{n}{calFB}\PY{o}{.}\PY{n}{setBiasParams}\PY{p}{(}\PY{p}{[}\PY{n}{r}\PY{o}{.}\PY{n}{lam} \PY{k}{for} \PY{n}{r} \PY{o+ow}{in} \PY{n}{spctAssist}\PY{o}{.}\PY{n}{resonances}\PY{p}{(}\PY{p}{)}\PY{p}{]}\PY{p}{)}
\end{Verbatim}

    \begin{Verbatim}[commandchars=\\\{\}]
{\color{incolor}In [{\color{incolor}18}]:} \PY{c+c1}{\PYZsh{}\PYZsh{} Validation}
         \PY{n}{threshold} \PY{o}{=} \PY{l+m+mf}{0.01}
         
         \PY{n}{calBias} \PY{o}{=} \PY{p}{[}\PY{n}{r}\PY{o}{.}\PY{n}{lam} \PY{k}{for} \PY{n}{r} \PY{o+ow}{in} \PY{n}{spctAssist}\PY{o}{.}\PY{n}{resonances}\PY{p}{(}\PY{p}{)}\PY{p}{]}
         \PY{n}{virtBias} \PY{o}{=} \PY{n}{np}\PY{o}{.}\PY{n}{array}\PY{p}{(}\PY{n}{wlChannels}\PY{p}{)}
         \PY{n}{diff} \PY{o}{=} \PY{n}{np}\PY{o}{.}\PY{n}{absolute}\PY{p}{(}\PY{n}{calBias} \PY{o}{\PYZhy{}} \PY{n}{virtBias}\PY{p}{)}
         \PY{n+nb}{print}\PY{p}{(}\PY{n}{diff}\PY{p}{)}
         \PY{k}{for} \PY{n}{d} \PY{o+ow}{in} \PY{n}{diff}\PY{p}{:}
             \PY{k}{assert} \PY{n}{d} \PY{o}{\PYZlt{}} \PY{n}{threshold}
\end{Verbatim}

    \begin{Verbatim}[commandchars=\\\{\}]
[ 0.00202127  0.00264774  0.00487436  0.00730028]

    \end{Verbatim}

    \subsubsection{Step 6: Take Filter Shapes}
    
    \begin{Verbatim}[commandchars=\\\{\}]
{\color{incolor}In [{\color{incolor}19}]:} \PY{n}{spect} \PY{o}{=}\PY{n}{spctAssist}\PY{o}{.}\PY{n}{fgSpect}\PY{p}{(}\PY{n}{avgCnt}\PY{o}{=}\PY{l+m+mi}{5}\PY{p}{,} \PY{n}{bgType}\PY{o}{=}\PY{l+s+s1}{\PYZsq{}}\PY{l+s+s1}{tuned}\PY{l+s+s1}{\PYZsq{}}\PY{p}{)}
         \PY{n}{filtShapes} \PY{o}{=} \PY{n}{np}\PY{o}{.}\PY{n}{empty}\PY{p}{(}\PY{n}{numRings}\PY{p}{,} \PY{n}{dtype}\PY{o}{=}\PY{n+nb}{object}\PY{p}{)}
         \PY{n}{filtCurves} \PY{o}{=} \PY{p}{[}\PY{k+kc}{None}\PY{p}{]} \PY{o}{*} \PY{n}{numRings}
         \PY{k}{for} \PY{n}{i}\PY{p}{,} \PY{n}{r} \PY{o+ow}{in} \PY{n+nb}{enumerate}\PY{p}{(}\PY{n}{spctAssist}\PY{o}{.}\PY{n}{resonances}\PY{p}{(}\PY{n}{spect}\PY{p}{)}\PY{p}{)}\PY{p}{:}
             \PY{n}{relWindow} \PY{o}{=} \PY{l+m+mi}{7} \PY{o}{*} \PY{n}{r}\PY{o}{.}\PY{n}{fwhm} \PY{o}{*} \PY{n}{np}\PY{o}{.}\PY{n}{array}\PY{p}{(}\PY{p}{[}\PY{o}{\PYZhy{}}\PY{l+m+mi}{1}\PY{p}{,}\PY{l+m+mi}{1}\PY{p}{]}\PY{p}{)}\PY{o}{/}\PY{l+m+mi}{2}
             \PY{n}{proximitySpect} \PY{o}{=} \PY{n}{spect}\PY{o}{.}\PY{n}{shift}\PY{p}{(}\PY{o}{\PYZhy{}}\PY{n}{r}\PY{o}{.}\PY{n}{lam}\PY{p}{)}\PY{o}{.}\PY{n}{crop}\PY{p}{(}\PY{n}{relWindow}\PY{p}{)}
             \PY{n}{filtShapes}\PY{p}{[}\PY{n}{i}\PY{p}{]} \PY{o}{=} \PY{n}{proximitySpect}
             \PY{n}{nm}\PY{p}{,} \PY{n}{dbm} \PY{o}{=} \PY{n}{proximitySpect}\PY{o}{.}\PY{n}{shift}\PY{p}{(}\PY{n}{r}\PY{o}{.}\PY{n}{lam}\PY{p}{)}\PY{o}{.}\PY{n}{getData}\PY{p}{(}\PY{p}{)}
             \PY{n}{lin} \PY{o}{=} \PY{n}{np}\PY{o}{.}\PY{n}{clip}\PY{p}{(}\PY{n}{dbm2lin}\PY{p}{(}\PY{n}{dbm}\PY{p}{)}\PY{p}{,} \PY{l+m+mf}{0.0}\PY{p}{,} \PY{l+m+mf}{1.0}\PY{p}{)}
             \PY{n}{filtCurves}\PY{p}{[}\PY{n}{i}\PY{p}{]} \PY{o}{=} \PY{p}{(}\PY{n}{nm}\PY{p}{,}\PY{n}{lin}\PY{p}{)}
             \PY{n}{plt}\PY{o}{.}\PY{n}{plot}\PY{p}{(}\PY{n}{nm}\PY{p}{,} \PY{n}{lin}\PY{p}{)}
         
         \PY{c+c1}{\PYZsh{} Store in assistant}
         \PY{n}{spctAssist}\PY{o}{.}\PY{n}{filtShapesForConvolution} \PY{o}{=} \PY{n}{filtShapes}
         \PY{c+c1}{\PYZsh{} Store in model}
         \PY{n}{calFB}\PY{o}{.}\PY{n}{setCurve}\PY{p}{(}\PY{n}{filtCurves}\PY{p}{,} \PY{n}{MrrOut}\PY{o}{.}\PY{n}{kThru}\PY{p}{)}
         \PY{n}{plt}\PY{o}{.}\PY{n}{show}\PY{p}{(}\PY{p}{)}
\end{Verbatim}

    { \hspace*{\fill} \\}
    
    \begin{Verbatim}[commandchars=\\\{\}]
{\color{incolor}In [{\color{incolor}20}]:} \PY{c+c1}{\PYZsh{} Hot\PYZhy{}Swap in Calibration Module and make sure things look good}
         \PY{n}{inst}\PY{o}{.}\PY{n}{lockPhony}\PY{p}{(}\PY{n}{calFB}\PY{p}{)}
         \PY{n}{calFB}\PY{o}{.}\PY{n}{setOsaOut}\PY{p}{(}\PY{n}{MrrOut}\PY{o}{.}\PY{n}{kThru}\PY{o}{.}\PY{n}{value}\PY{p}{)}
         
         \PY{c+c1}{\PYZsh{} Assume we can pick this up easily}
         \PY{n}{calFB}\PY{o}{.}\PY{n}{setAttenuation}\PY{p}{(}\PY{n}{attenuation}\PY{p}{)}
         
         \PY{n}{spctAssist}\PY{o}{.}\PY{n}{fgResPlot}\PY{p}{(}\PY{p}{)}
         \PY{n}{plt}\PY{o}{.}\PY{n}{show}\PY{p}{(}\PY{p}{)}
         \PY{n}{inst}\PY{o}{.}\PY{n}{releasePhony}\PY{p}{(}\PY{p}{)}
         \PY{n}{biasTune} \PY{o}{=} \PY{n}{nowTune}\PY{o}{.}\PY{n}{copy}\PY{p}{(}\PY{p}{)}
\end{Verbatim}

    
    \subsubsection{Step 7: Fill K Matrix}

K is the coefficients between mW and deltaWL.

    \begin{Verbatim}[commandchars=\\\{\}]
{\color{incolor}In [{\color{incolor}21}]:} \PY{n}{nPts} \PY{o}{=} \PY{l+m+mi}{11}
         \PY{n}{avgCnt} \PY{o}{=} \PY{l+m+mi}{5}
         
         \PY{n}{nowTune} \PY{o}{=} \PY{n}{biasTune}\PY{o}{.}\PY{n}{copy}\PY{p}{(}\PY{p}{)}
         \PY{n}{biasWL} \PY{o}{=} \PY{n}{np}\PY{o}{.}\PY{n}{array}\PY{p}{(}\PY{p}{[}\PY{n}{r}\PY{o}{.}\PY{n}{lam} \PY{k}{for} \PY{n}{r} \PY{o+ow}{in} \PY{n}{spctAssist}\PY{o}{.}\PY{n}{resonances}\PY{p}{(}\PY{p}{)}\PY{p}{]}\PY{p}{)}
         \PY{n}{nowWL} \PY{o}{=} \PY{n}{np}\PY{o}{.}\PY{n}{copy}\PY{p}{(}\PY{n}{biasWL}\PY{p}{)}
         
         \PY{n}{newK} \PY{o}{=} \PY{n}{np}\PY{o}{.}\PY{n}{zeros}\PY{p}{(}\PY{p}{(}\PY{n}{numRings}\PY{p}{,} \PY{n+nb}{len}\PY{p}{(}\PY{n}{newCalCurrentChan}\PY{p}{)}\PY{p}{)}\PY{p}{)}
         
         \PY{c+c1}{\PYZsh{} For Each Current Channel}
         \PY{k}{for} \PY{n}{ich} \PY{o+ow}{in} \PY{n+nb}{range}\PY{p}{(}\PY{n+nb}{len}\PY{p}{(}\PY{n}{newCalCurrentChan}\PY{p}{)}\PY{p}{)}\PY{p}{:}
             \PY{n}{ch} \PY{o}{=} \PY{n}{newCalCurrentChan}\PY{p}{[}\PY{n}{ich}\PY{p}{]}
             
             \PY{c+c1}{\PYZsh{} Shift in a 1nm range around bias WL}
             \PY{n}{dB} \PY{o}{=} \PY{l+m+mf}{1.0} \PY{o}{/} \PY{n}{kEstTemp}\PY{p}{[}\PY{n}{ich}\PY{p}{]}
             \PY{n}{x} \PY{o}{=} \PY{n}{np}\PY{o}{.}\PY{n}{linspace}\PY{p}{(}\PY{n+nb}{max}\PY{p}{(}\PY{l+m+mf}{0.0}\PY{p}{,} \PY{p}{(}\PY{n}{biasTune}\PY{p}{[}\PY{n}{ch}\PY{p}{]}\PY{o}{\PYZhy{}}\PY{n}{dB}\PY{p}{)}\PY{p}{)}\PY{p}{,}
                \PY{n}{biasTune}\PY{p}{[}\PY{n}{ch}\PY{p}{]}\PY{o}{+}\PY{n}{dB}\PY{p}{,} \PY{n}{nPts}\PY{p}{)}
             \PY{n}{y} \PY{o}{=} \PY{n}{np}\PY{o}{.}\PY{n}{zeros}\PY{p}{(}\PY{p}{(}\PY{n}{numRings}\PY{p}{,} \PY{n}{nPts}\PY{p}{)}\PY{p}{)}
             
             \PY{k}{for} \PY{n}{ipt}\PY{p}{,} \PY{n}{pt} \PY{o+ow}{in} \PY{n+nb}{enumerate}\PY{p}{(}\PY{n}{x}\PY{p}{)}\PY{p}{:}
                 \PY{n}{nowTune}\PY{p}{[}\PY{n}{ch}\PY{p}{]} \PY{o}{=} \PY{n}{pt}
                 \PY{n}{inst}\PY{o}{.}\PY{n}{setCurrentChanTuning}\PY{p}{(}\PY{n}{nowTune}\PY{p}{,} \PY{n}{token}\PY{p}{,} \PY{n}{CurrentUnit}\PY{o}{.}\PY{n}{mW}\PY{p}{)}
                 \PY{n}{nowWL} \PY{o}{=} \PY{n}{np}\PY{o}{.}\PY{n}{array}\PY{p}{(}\PY{p}{[}\PY{n}{r}\PY{o}{.}\PY{n}{lam} \PY{k}{for} \PY{n}{r} \PY{o+ow}{in} \PY{n}{spctAssist}\PY{o}{.}\PY{n}{resonances}\PY{p}{(}\PY{p}{)}\PY{p}{]}\PY{p}{)}
                 \PY{n}{diff} \PY{o}{=} \PY{n}{nowWL} \PY{o}{\PYZhy{}} \PY{n}{biasWL}
                 \PY{n}{y}\PY{p}{[}\PY{p}{:}\PY{p}{,} \PY{n}{ipt}\PY{p}{]} \PY{o}{=} \PY{n}{diff}
             
             \PY{c+c1}{\PYZsh{} Make Linear Fit}
             
             \PY{k}{for} \PY{n}{r} \PY{o+ow}{in} \PY{n+nb}{range}\PY{p}{(}\PY{n}{numRings}\PY{p}{)}\PY{p}{:}
                 \PY{n}{yr} \PY{o}{=} \PY{n}{y}\PY{p}{[}\PY{n}{r}\PY{p}{,} \PY{p}{:}\PY{p}{]}
                 \PY{n}{p} \PY{o}{=} \PY{n}{np}\PY{o}{.}\PY{n}{polyfit}\PY{p}{(}\PY{n}{x}\PY{p}{,} \PY{n}{yr}\PY{p}{,} \PY{l+m+mi}{1}\PY{p}{)}
                 \PY{n}{newK}\PY{p}{[}\PY{n}{r}\PY{p}{,} \PY{n}{ich}\PY{p}{]} \PY{o}{=} \PY{n+nb}{max}\PY{p}{(}\PY{n}{p}\PY{p}{[}\PY{l+m+mi}{0}\PY{p}{]}\PY{p}{,} \PY{l+m+mf}{0.0}\PY{p}{)}

             \PY{n}{inst}\PY{o}{.}\PY{n}{setCurrentChanTuning}\PY{p}{(}\PY{n}{biasTune}\PY{p}{,} \PY{n}{token}\PY{p}{,} \PY{n}{CurrentUnit}\PY{o}{.}\PY{n}{mW}\PY{p}{)}
             \PY{n}{nowTune} \PY{o}{=} \PY{n}{biasTune}\PY{o}{.}\PY{n}{copy}\PY{p}{(}\PY{p}{)}
             
         \PY{n}{calTherm}\PY{o}{.}\PY{n}{setK}\PY{p}{(}\PY{n}{newK}\PY{p}{)}
\end{Verbatim}

    \begin{Verbatim}[commandchars=\\\{\}]
{\color{incolor}In [{\color{incolor}22}]:} \PY{c+c1}{\PYZsh{} Percent Error}
         \PY{n}{err} \PY{o}{=} \PY{n}{np}\PY{o}{.}\PY{n}{round}\PY{p}{(}\PY{n}{np}\PY{o}{.}\PY{n}{absolute}\PY{p}{(}\PY{l+m+mi}{100} \PY{o}{*}
            \PY{n}{np}\PY{o}{.}\PY{n}{divide}\PY{p}{(}\PY{n}{np}\PY{o}{.}\PY{n}{subtract}\PY{p}{(}\PY{n}{newK}\PY{p}{,} \PY{n}{K}\PY{p}{)}\PY{p}{,} \PY{n}{K}\PY{p}{)}\PY{p}{)}\PY{p}{)}
         
         \PY{k}{for} \PY{n}{i}\PY{p}{,} \PY{n}{e} \PY{o+ow}{in} \PY{n+nb}{enumerate}\PY{p}{(}\PY{n}{err}\PY{o}{.}\PY{n}{flatten}\PY{p}{(}\PY{p}{)}\PY{p}{)}\PY{p}{:}
             \PY{k}{if} \PY{n}{e} \PY{o}{\PYZgt{}} \PY{l+m+mi}{10}\PY{p}{:}
                 \PY{k}{if} \PY{n}{K}\PY{o}{.}\PY{n}{flatten}\PY{p}{(}\PY{p}{)}\PY{p}{[}\PY{n}{i}\PY{p}{]} \PY{o}{\PYZlt{}} \PY{l+m+mi}{10}\PY{p}{:}
                     \PY{n+nb}{print}\PY{p}{(}\PY{l+s+s2}{\PYZdq{}}\PY{l+s+s2}{Small K: }\PY{l+s+s2}{\PYZdq{}} \PY{o}{+} \PY{n+nb}{str}\PY{p}{(}\PY{n}{K}\PY{o}{.}\PY{n}{flatten}\PY{p}{(}\PY{p}{)}\PY{p}{[}\PY{n}{i}\PY{p}{]}\PY{p}{)}\PY{p}{)}
                     \PY{k}{continue}
             \PY{k}{assert} \PY{n}{e} \PY{o}{\PYZlt{}}\PY{o}{=} \PY{l+m+mi}{10}
             
\end{Verbatim}

    \begin{Verbatim}[commandchars=\\\{\}]
Percent Error:
[[ 0.  1.  2.  0.]
 [ 0.  0.  3.  0.]
 [ 0.  1.  0.  1.]
 [ 0.  7.  0.  0.]]

Avg K: 45.3534028127

If you see an error >10, that means we made a normal mistake on a small K.

    \end{Verbatim}

    \textbf{And calFB is a calibrated device!}

Validation was done at the following points: * Ascription is guaranteed
to be correct. * HeatBias is within 0.01mW * Wavelength bias is within
0.01nm * Significant K cross-terms have less than 10\% error

%% file: Phys_Cascade_Calibration.tex
    \subsubsection{Step 1: Initialization}\label{step-1-initialization}

Import all classes and set up global parameters

    \begin{Verbatim}[commandchars=\\\{\}]
{\color{incolor}In [{\color{incolor}1}]:} \PY{c+c1}{\PYZsh{} Imports and reservations}
        \PY{k+kn}{import} \PY{n+nn}{numpy} \PY{k}{as} \PY{n+nn}{np}
        \PY{k+kn}{import} \PY{n+nn}{matplotlib}\PY{n+nn}{.}\PY{n+nn}{pyplot} \PY{k}{as} \PY{n+nn}{plt}
        \PY{k+kn}{import} \PY{n+nn}{lightlab}\PY{n+nn}{.}\PY{n+nn}{instruments} \PY{k}{as} \PY{n+nn}{inst}
        \PY{k+kn}{import} \PY{n+nn}{lightlab}\PY{n+nn}{.}\PY{n+nn}{model} \PY{k}{as} \PY{n+nn}{m}
        \PY{k+kn}{from} \PY{n+nn}{lightlab}\PY{n+nn}{.}\PY{n+nn}{util}\PY{n+nn}{.}\PY{n+nn}{modeling} \PY{k}{import} \PY{n}{kOSAPwr}\PY{p}{,} 
            \PY{n}{dbm2lin}\PY{p}{,} \PY{n}{kOSASpacing}\PY{p}{,} \PY{n}{CurrentUnit}\PY{p}{,} \PY{n}{MrrOut}\PY{p}{,} \PY{n}{lin2dbm}
        \PY{k+kn}{from} \PY{n+nn}{lightlab}\PY{n+nn}{.}\PY{n+nn}{util}\PY{n+nn}{.}\PY{n+nn}{calibrating} \PY{k}{import} \PY{n}{SpectrumMeasurementAssistant}
        \PY{k+kn}{from} \PY{n+nn}{lightlab}\PY{n+nn}{.}\PY{n+nn}{util} \PY{k}{import} \PY{n}{io}
        \PY{k+kn}{from} \PY{n+nn}{bidict} \PY{k}{import} \PY{n}{bidict}
        \PY{k+kn}{from} \PY{n+nn}{time} \PY{k}{import} \PY{n}{sleep}
\end{Verbatim}

    \begin{Verbatim}[commandchars=\\\{\}]
{\color{incolor}In [{\color{incolor}2}]:} \PY{c+c1}{\PYZsh{} Global Parameters, Available to Calibration Model}
        \PY{n}{wlChannels} \PY{o}{=} \PY{n}{np}\PY{o}{.}\PY{n}{array}\PY{p}{(}\PY{p}{[}\PY{l+m+mi}{1550}\PY{p}{,} \PY{l+m+mi}{1552}\PY{p}{,} \PY{l+m+mi}{1554}\PY{p}{]}\PY{p}{)}
        \PY{n}{wlRange} \PY{o}{=} \PY{n}{np}\PY{o}{.}\PY{n}{array}\PY{p}{(}\PY{p}{[}\PY{l+m+mi}{1545}\PY{p}{,} \PY{l+m+mi}{1560}\PY{p}{]}\PY{p}{)}
        \PY{n}{axonChan} \PY{o}{=} \PY{n}{np}\PY{o}{.}\PY{n}{array}\PY{p}{(}\PY{p}{[}\PY{l+m+mi}{0}\PY{p}{,} \PY{l+m+mi}{1}\PY{p}{,} \PY{l+m+mi}{2}\PY{p}{]}\PY{p}{)}
        \PY{n}{dendriteChan} \PY{o}{=} \PY{n}{np}\PY{o}{.}\PY{n}{array}\PY{p}{(}\PY{p}{[}\PY{l+m+mi}{3}\PY{p}{,} \PY{l+m+mi}{4}\PY{p}{,} \PY{l+m+mi}{5}\PY{p}{]}\PY{p}{)}
        
        \PY{n}{numRings} \PY{o}{=} \PY{l+m+mi}{3}
\end{Verbatim}

    \subsubsection{Step 2: Initialize Hidden
Model}\label{step-2-initialize-hidden-model}

Create a virtual model that will mimic the instruments. This will remain
hidden during calibration.

In this case, we are cascading 2 2-ring Filter Banks.

    \begin{Verbatim}[commandchars=\\\{\}]
{\color{incolor}In [{\color{incolor} }]:} \PY{n}{axonMixed} \PY{o}{=} \PY{n+nb}{list}\PY{p}{(}\PY{p}{[}\PY{n}{axonChan}\PY{p}{[}\PY{l+m+mi}{1}\PY{p}{]}\PY{p}{,} \PY{n}{axonChan}\PY{p}{[}\PY{l+m+mi}{2}\PY{p}{]}\PY{p}{,} \PY{n}{axonChan}\PY{p}{[}\PY{l+m+mi}{0}\PY{p}{]}\PY{p}{]}\PY{p}{)}
        \PY{n}{dendriteMixed} \PY{o}{=}
            \PY{n+nb}{list}\PY{p}{(}\PY{p}{[}\PY{n}{dendriteChan}\PY{p}{[}\PY{l+m+mi}{1}\PY{p}{]}\PY{p}{,} \PY{n}{dendriteChan}\PY{p}{[}\PY{l+m+mi}{2}\PY{p}{]}\PY{p}{,} \PY{n}{dendriteChan}\PY{p}{[}\PY{l+m+mi}{0}\PY{p}{]}\PY{p}{]}\PY{p}{)}
        \PY{n}{currentChan} \PY{o}{=} \PY{n}{axonMixed} \PY{o}{+} \PY{n}{dendriteMixed}
        
        \PY{c+c1}{\PYZsh{} Thermal Group, Uses 4 current channels}
        \PY{n}{therm} \PY{o}{=} \PY{n}{m}\PY{o}{.}\PY{n}{ThermalGroup}\PY{p}{(}\PY{n}{currentChan}\PY{p}{)}
        
        \PY{c+c1}{\PYZsh{} Model Parameters, NOT available to calibration model}
        \PY{n}{attenuation} \PY{o}{=} \PY{l+m+mf}{0.0001} \PY{c+c1}{\PYZsh{} 40dB baseline attenuation}
        \PY{n}{lfwhm} \PY{o}{=} \PY{l+m+mf}{0.1}          \PY{c+c1}{\PYZsh{} 0.1nm fwhm of lorentzian}
        \PY{n}{latten} \PY{o}{=} \PY{l+m+mf}{0.98}        \PY{c+c1}{\PYZsh{} Attenuation at resonance}
        \PY{n}{heatBias} \PY{o}{=} \PY{n+nb}{dict}\PY{p}{(}\PY{p}{)}
        \PY{k}{for} \PY{n}{i} \PY{o+ow}{in} \PY{n+nb}{range}\PY{p}{(}\PY{n+nb}{len}\PY{p}{(}\PY{n}{axonMixed}\PY{p}{)}\PY{p}{)}\PY{p}{:}
            \PY{n}{heatBias}\PY{p}{[}\PY{n}{axonMixed}\PY{p}{[}\PY{n}{i}\PY{p}{]}\PY{p}{]} \PY{o}{=} \PY{l+m+mf}{2.0}
            \PY{n}{heatBias}\PY{p}{[}\PY{n}{dendriteMixed}\PY{p}{[}\PY{n}{i}\PY{p}{]}\PY{p}{]} \PY{o}{=} \PY{l+m+mf}{1.5}

        \PY{c+c1}{\PYZsh{} Random K with extra on diagonal to denote primary filament}
        \PY{n}{K} \PY{o}{=} \PY{n}{np}\PY{o}{.}\PY{n}{absolute}\PY{p}{(}\PY{n}{np}\PY{o}{.}\PY{n}{random}\PY{o}{.}\PY{n}{randn}\PY{p}{(}\PY{l+m+mi}{2}\PY{o}{*}\PY{n}{numRings}\PY{p}{,} \PY{l+m+mi}{2}\PY{o}{*}\PY{n}{numRings}\PY{p}{)} \PY{o}{+}
            \PY{l+m+mf}{20.0}\PY{o}{*}\PY{n}{np}\PY{o}{.}\PY{n}{eye}\PY{p}{(}\PY{l+m+mi}{2}\PY{o}{*}\PY{n}{numRings}\PY{p}{)}\PY{p}{)}
        \PY{n+nb}{print}\PY{p}{(}\PY{n}{K}\PY{p}{)}
        
        \PY{c+c1}{\PYZsh{} FilterBank Module}
        \PY{n}{axon} \PY{o}{=} \PY{n}{m}\PY{o}{.}\PY{n}{FilterBank}\PY{p}{(}\PY{n}{therm}\PY{p}{,} \PY{n}{numRings}\PY{p}{)}
        \PY{n}{axon}\PY{o}{.}\PY{n}{setAxon}\PY{p}{(}\PY{p}{)}
        \PY{n}{fb} \PY{o}{=} \PY{n}{m}\PY{o}{.}\PY{n}{FilterBank}\PY{p}{(}\PY{n}{therm}\PY{p}{,} \PY{n}{numRings}\PY{p}{)}
        
        \PY{n}{fb}\PY{o}{.}\PY{n}{setBiasParams}\PY{p}{(}\PY{n}{wlChannels}\PY{p}{,} \PY{n}{latten} \PY{o}{*} \PY{n}{np}\PY{o}{.}\PY{n}{ones}\PY{p}{(}\PY{n}{numRings}\PY{p}{)}\PY{p}{,} \PY{n}{lfwhm} \PY{o}{*}
            \PY{n}{np}\PY{o}{.}\PY{n}{ones}\PY{p}{(}\PY{n}{numRings}\PY{p}{)}\PY{p}{)}
        \PY{n}{axon}\PY{o}{.}\PY{n}{setBiasParams}\PY{p}{(}\PY{n}{wlChannels}\PY{p}{,} \PY{n}{latten} \PY{o}{*} \PY{n}{np}\PY{o}{.}\PY{n}{ones}\PY{p}{(}\PY{n}{numRings}\PY{p}{)}\PY{p}{,} \PY{n}{lfwhm} \PY{o}{*}
            \PY{n}{np}\PY{o}{.}\PY{n}{ones}\PY{p}{(}\PY{n}{numRings}\PY{p}{)}\PY{p}{)}
        
        \PY{n}{csc} \PY{o}{=} \PY{n}{m}\PY{o}{.}\PY{n}{Cascade}\PY{p}{(}\PY{n}{axon}\PY{p}{,} \PY{n}{fb}\PY{p}{)}
        
        \PY{c+c1}{\PYZsh{} Set K and bias for Thermal Group}
        \PY{n}{therm}\PY{o}{.}\PY{n}{setK}\PY{p}{(}\PY{n}{K}\PY{p}{)}
        \PY{n}{therm}\PY{o}{.}\PY{n}{setHeatBias}\PY{p}{(}\PY{n}{heatBias}\PY{p}{)}
\end{Verbatim}

    \begin{Verbatim}[commandchars=\\\{\}]
{\color{incolor}In [{\color{incolor}4}]:} \PY{c+c1}{\PYZsh{} Reserve Current Channels, Add Current Channels}
        \PY{n}{inst}\PY{o}{.}\PY{n}{togglePhony}\PY{p}{(}\PY{k+kc}{True}\PY{p}{,} \PY{n}{csc}\PY{p}{)}
        \PY{n}{token} \PY{o}{=} \PY{n}{inst}\PY{o}{.}\PY{n}{reserveCurrentChan}\PY{p}{(}\PY{n}{currentChan}\PY{p}{)}
\end{Verbatim}

    \begin{Verbatim}[commandchars=\\\{\}]
{\color{incolor}In [{\color{incolor}5}]:} \PY{n}{csc}\PY{o}{.}\PY{n}{setOsaOut}\PY{p}{(}\PY{n}{MrrOut}\PY{o}{.}\PY{n}{kThru}\PY{o}{.}\PY{n}{value}\PY{p}{)}
        
        \PY{c+c1}{\PYZsh{} Manually Set to Separate Peaks}
        \PY{n}{baseTune} \PY{o}{=} \PY{n+nb}{dict}\PY{p}{(}\PY{p}{)}
        \PY{n}{zeroTune} \PY{o}{=} \PY{n+nb}{dict}\PY{p}{(}\PY{p}{)}
        \PY{k}{for} \PY{n}{j} \PY{o+ow}{in} \PY{n+nb}{range}\PY{p}{(}\PY{n+nb}{len}\PY{p}{(}\PY{n}{currentChan}\PY{p}{)}\PY{p}{)}\PY{p}{:}
            \PY{n}{zeroTune}\PY{p}{[}\PY{n}{j}\PY{p}{]} \PY{o}{=} \PY{l+m+mf}{0.0}
        \PY{n}{baseTune}\PY{p}{[}\PY{l+m+mi}{1}\PY{p}{]} \PY{o}{=} \PY{l+m+mf}{0.0}
        \PY{n}{baseTune}\PY{p}{[}\PY{l+m+mi}{4}\PY{p}{]} \PY{o}{=} \PY{l+m+mf}{0.0}
        \PY{n}{baseTune}\PY{p}{[}\PY{l+m+mi}{2}\PY{p}{]} \PY{o}{=} \PY{l+m+mf}{0.0}
        \PY{n}{baseTune}\PY{p}{[}\PY{l+m+mi}{5}\PY{p}{]} \PY{o}{=} \PY{l+m+mf}{0.0}
        \PY{n}{baseTune}\PY{p}{[}\PY{l+m+mi}{0}\PY{p}{]} \PY{o}{=} \PY{l+m+mf}{0.0}
        \PY{n}{baseTune}\PY{p}{[}\PY{l+m+mi}{3}\PY{p}{]} \PY{o}{=} \PY{l+m+mf}{0.0}
        
        \PY{n}{inst}\PY{o}{.}\PY{n}{setCurrentChanTuning}\PY{p}{(}\PY{n}{zeroTune}\PY{p}{,} \PY{n}{token}\PY{p}{,} \PY{n}{CurrentUnit}\PY{o}{.}\PY{n}{mW}\PY{p}{)}
        \PY{n}{nm}\PY{p}{,} \PY{n}{dbm} \PY{o}{=} \PY{n}{inst}\PY{o}{.}\PY{n}{spectrum}\PY{p}{(}\PY{n}{wlRange}\PY{p}{)}
        \PY{n}{plt}\PY{o}{.}\PY{n}{plot}\PY{p}{(}\PY{n}{nm}\PY{p}{,} \PY{n}{dbm}\PY{p}{)}\PY{p}{,} \PY{l+s+s1}{\PYZsq{}}\PY{l+s+s1}{b}\PY{l+s+s1}{\PYZsq{}}
        \PY{n}{inst}\PY{o}{.}\PY{n}{setCurrentChanTuning}\PY{p}{(}\PY{n}{baseTune}\PY{p}{,} \PY{n}{token}\PY{p}{,} \PY{n}{CurrentUnit}\PY{o}{.}\PY{n}{mW}\PY{p}{)}
        \PY{n}{nm}\PY{p}{,} \PY{n}{dbm} \PY{o}{=} \PY{n}{inst}\PY{o}{.}\PY{n}{spectrum}\PY{p}{(}\PY{n}{wlRange}\PY{p}{)}
        \PY{n}{plt}\PY{o}{.}\PY{n}{plot}\PY{p}{(}\PY{n}{nm}\PY{p}{,} \PY{n}{dbm}\PY{p}{,} \PY{l+s+s1}{\PYZsq{}}\PY{l+s+s1}{g}\PY{l+s+s1}{\PYZsq{}}\PY{p}{)}
        \PY{n}{plt}\PY{o}{.}\PY{n}{show}\PY{p}{(}\PY{p}{)}
\end{Verbatim}

    
    \subsubsection{Step 3: Ascription}\label{step-3-ascription}

Map primary axon and dendrite filaments to peaks.

    \begin{Verbatim}[commandchars=\\\{\}]
{\color{incolor}In [{\color{incolor}26}]:} \PY{n}{spctAssist} \PY{o}{=} \PY{n}{SpectrumMeasurementAssistant}\PY{p}{(}\PY{n}{nChan}\PY{o}{=}\PY{l+m+mi}{2}\PY{o}{*}\PY{n}{numRings}\PY{p}{,} 
    \PY{n}{arePeaks}\PY{o}{=}\PY{k+kc}{False}\PY{p}{,} \PY{n}{visualize}\PY{o}{=}\PY{k+kc}{False}\PY{p}{)}
         \PY{k}{def} \PY{n+nf}{ascribe}\PY{p}{(}\PY{n}{aChans}\PY{p}{,} \PY{n}{dChans}\PY{p}{,} \PY{n}{bTune}\PY{p}{,} \PY{n}{tuneBy}\PY{o}{=}\PY{l+m+mf}{0.01}\PY{p}{)}\PY{p}{:}
             
             \PY{n}{baseLams} \PY{o}{=} \PY{n}{np}\PY{o}{.}\PY{n}{array}\PY{p}{(}\PY{p}{[}\PY{n}{r}\PY{o}{.}\PY{n}{lam} 
                \PY{k}{for} \PY{n}{r} \PY{o+ow}{in} \PY{n}{spctAssist}\PY{o}{.}\PY{n}{resonances}\PY{p}{(}\PY{p}{)}\PY{p}{]}\PY{p}{)}
             \PY{n}{pMap} \PY{o}{=} \PY{n}{bidict}\PY{p}{(}\PY{p}{)}
             \PY{n}{tEstMap} \PY{o}{=} \PY{n}{bidict}\PY{p}{(}\PY{p}{)}
             
             \PY{n}{nowTune} \PY{o}{=} \PY{n}{bTune}\PY{o}{.}\PY{n}{copy}\PY{p}{(}\PY{p}{)}
             
             \PY{c+c1}{\PYZsh{} Run Ascription}
             \PY{k}{for} \PY{n}{iChan}\PY{p}{,} \PY{n}{ch} \PY{o+ow}{in} \PY{n+nb}{enumerate}\PY{p}{(}\PY{n}{aChans}\PY{p}{)}\PY{p}{:}
                 \PY{n}{nowTune}\PY{p}{[}\PY{n}{ch}\PY{p}{]} \PY{o}{=} \PY{n}{bTune}\PY{p}{[}\PY{n}{ch}\PY{p}{]} \PY{o}{+} \PY{n}{tuneBy}
                 \PY{n}{inst}\PY{o}{.}\PY{n}{setCurrentChanTuning}\PY{p}{(}\PY{n}{nowTune}\PY{p}{,} \PY{n}{token}\PY{p}{,} \PY{n}{CurrentUnit}\PY{o}{.}\PY{n}{mW}\PY{p}{)}
                 \PY{n}{spect} \PY{o}{=} \PY{n}{spctAssist}\PY{o}{.}\PY{n}{fgSpect}\PY{p}{(}\PY{p}{)}
                 \PY{n}{presLams} \PY{o}{=} \PY{n}{np}\PY{o}{.}\PY{n}{array}\PY{p}{(}\PY{p}{[}\PY{n}{r}\PY{o}{.}\PY{n}{lam} \PY{k}{for} \PY{n}{r} \PY{o+ow}{in}
                    \PY{n}{spctAssist}\PY{o}{.}\PY{n}{resonances}\PY{p}{(}\PY{n}{spect}\PY{p}{)}\PY{p}{]}\PY{p}{)}
                 \PY{n}{nowTune}\PY{p}{[}\PY{n}{ch}\PY{p}{]} \PY{o}{=} \PY{n}{bTune}\PY{p}{[}\PY{n}{ch}\PY{p}{]}
                 \PY{n}{inst}\PY{o}{.}\PY{n}{setCurrentChanTuning}\PY{p}{(}\PY{n}{nowTune}\PY{p}{,} \PY{n}{token}\PY{p}{)}
                 \PY{n}{shifts} \PY{o}{=} \PY{n}{presLams} \PY{o}{\PYZhy{}} \PY{n}{baseLams}
                 \PY{n}{pNum} \PY{o}{=} \PY{n}{np}\PY{o}{.}\PY{n}{argmax}\PY{p}{(}\PY{n}{shifts}\PY{p}{)}
                 \PY{n}{pMap}\PY{p}{[}\PY{n}{pNum}\PY{p}{]} \PY{o}{=} \PY{n}{ch}
                 \PY{n}{tEstMap}\PY{p}{[}\PY{n}{ch}\PY{p}{]} \PY{o}{=} \PY{n}{shifts}\PY{p}{[}\PY{n}{pNum}\PY{p}{]} \PY{o}{/} \PY{n}{tuneBy}
                 
             \PY{k}{for} \PY{n}{iChan}\PY{p}{,} \PY{n}{ch} \PY{o+ow}{in} \PY{n+nb}{enumerate}\PY{p}{(}\PY{n}{dChans}\PY{p}{)}\PY{p}{:}
                 \PY{n}{nowTune}\PY{p}{[}\PY{n}{ch}\PY{p}{]} \PY{o}{=} \PY{n}{bTune}\PY{p}{[}\PY{n}{ch}\PY{p}{]} \PY{o}{+} \PY{n}{tuneBy}
                 \PY{n}{inst}\PY{o}{.}\PY{n}{setCurrentChanTuning}\PY{p}{(}\PY{n}{nowTune}\PY{p}{,} \PY{n}{token}\PY{p}{,} \PY{n}{CurrentUnit}\PY{o}{.}\PY{n}{mW}\PY{p}{)}
                 \PY{n}{spect} \PY{o}{=} \PY{n}{spctAssist}\PY{o}{.}\PY{n}{fgSpect}\PY{p}{(}\PY{p}{)}
                 \PY{n}{presLams} \PY{o}{=} \PY{n}{np}\PY{o}{.}\PY{n}{array}\PY{p}{(}\PY{p}{[}\PY{n}{r}\PY{o}{.}\PY{n}{lam} \PY{k}{for} \PY{n}{r} \PY{o+ow}{in}
                    \PY{n}{spctAssist}\PY{o}{.}\PY{n}{resonances}\PY{p}{(}\PY{n}{spect}\PY{p}{)}\PY{p}{]}\PY{p}{)}
                 \PY{n}{nowTune}\PY{p}{[}\PY{n}{ch}\PY{p}{]} \PY{o}{=} \PY{n}{bTune}\PY{p}{[}\PY{n}{ch}\PY{p}{]}
                 \PY{n}{inst}\PY{o}{.}\PY{n}{setCurrentChanTuning}\PY{p}{(}\PY{n}{nowTune}\PY{p}{,} \PY{n}{token}\PY{p}{)}
                 \PY{n}{shifts} \PY{o}{=} \PY{n}{presLams} \PY{o}{\PYZhy{}} \PY{n}{baseLams}
                 \PY{n}{pNum} \PY{o}{=} \PY{n}{np}\PY{o}{.}\PY{n}{argmax}\PY{p}{(}\PY{n}{shifts}\PY{p}{)}
                 \PY{n}{pMap}\PY{p}{[}\PY{n}{pNum}\PY{p}{]} \PY{o}{=} \PY{n}{ch}
                 \PY{n}{tEstMap}\PY{p}{[}\PY{n}{ch}\PY{p}{]} \PY{o}{=} \PY{n}{shifts}\PY{p}{[}\PY{n}{pNum}\PY{p}{]} \PY{o}{/} \PY{n}{tuneBy}
             
             \PY{k}{try}\PY{p}{:}
                 \PY{k}{assert} \PY{n+nb}{len}\PY{p}{(}\PY{n+nb}{list}\PY{p}{(}\PY{n}{pMap}\PY{o}{.}\PY{n}{keys}\PY{p}{(}\PY{p}{)}\PY{p}{)}\PY{p}{)} \PY{o}{==} \PY{n+nb}{len}\PY{p}{(}\PY{n}{aChans}\PY{p}{)} \PY{o}{+} \PY{n+nb}{len}\PY{p}{(}\PY{n}{dChans}\PY{p}{)}
             \PY{k}{except}\PY{p}{:}
                 \PY{n+nb}{print}\PY{p}{(}\PY{n}{pMap}\PY{p}{)}
                 \PY{k}{assert} \PY{k+kc}{False}
                 
             \PY{k}{return} \PY{n}{pMap}\PY{p}{,} \PY{n}{tEstMap}
\end{Verbatim}

    \begin{Verbatim}[commandchars=\\\{\}]
{\color{incolor}In [{\color{incolor}27}]:} \PY{c+c1}{\PYZsh{} Run Ascription}
         \PY{n}{pMap}\PY{p}{,} \PY{n}{tEstMap} \PY{o}{=} \PY{n}{ascribe}\PY{p}{(}\PY{n}{axonChan}\PY{p}{,} \PY{n}{dendriteChan}\PY{p}{,} \PY{n}{baseTune}\PY{p}{)}
         
         \PY{n}{calAxonChan} \PY{o}{=} \PY{n+nb}{list}\PY{p}{(}\PY{p}{)}
         \PY{n}{calDendriteChan} \PY{o}{=} \PY{n+nb}{list}\PY{p}{(}\PY{p}{)}
         
         \PY{k}{for} \PY{n}{p} \PY{o+ow}{in} \PY{n+nb}{sorted}\PY{p}{(}\PY{n+nb}{list}\PY{p}{(}\PY{n}{pMap}\PY{o}{.}\PY{n}{keys}\PY{p}{(}\PY{p}{)}\PY{p}{)}\PY{p}{)}\PY{p}{:}
             \PY{n}{ch} \PY{o}{=} \PY{n}{pMap}\PY{p}{[}\PY{n}{p}\PY{p}{]}
             \PY{k}{if} \PY{n}{ch} \PY{o+ow}{in} \PY{n}{axonChan}\PY{p}{:}
                 \PY{n}{calAxonChan}\PY{o}{.}\PY{n}{append}\PY{p}{(}\PY{n}{ch}\PY{p}{)}
             \PY{k}{else}\PY{p}{:}
                 \PY{n}{calDendriteChan}\PY{o}{.}\PY{n}{append}\PY{p}{(}\PY{n}{ch}\PY{p}{)}
                 
         \PY{n}{calCurrentChan} \PY{o}{=} \PY{n}{calAxonChan} \PY{o}{+} \PY{n}{calDendriteChan}
\end{Verbatim}

    \begin{Verbatim}[commandchars=\\\{\}]
{\color{incolor}In [{\color{incolor}28}]:} \PY{c+c1}{\PYZsh{} Validation}
         \PY{k}{for} \PY{n}{j} \PY{o+ow}{in} \PY{n+nb}{range}\PY{p}{(}\PY{n+nb}{len}\PY{p}{(}\PY{n}{currentChan}\PY{p}{)}\PY{p}{)}\PY{p}{:}
             \PY{k}{try}\PY{p}{:}
                 \PY{k}{assert} \PY{n}{calCurrentChan}\PY{p}{[}\PY{n}{j}\PY{p}{]} \PY{o}{==} \PY{n}{currentChan}\PY{p}{[}\PY{n}{j}\PY{p}{]}
             \PY{k}{except}\PY{p}{:}
                 \PY{n+nb}{print}\PY{p}{(}\PY{l+s+s2}{\PYZdq{}}\PY{l+s+s2}{ERROR: Should be the same...}\PY{l+s+s2}{\PYZdq{}}\PY{p}{)}
                 \PY{n+nb}{print}\PY{p}{(}\PY{n}{currentChan}\PY{p}{)}
                 \PY{n+nb}{print}\PY{p}{(}\PY{n}{calCurrentChan}\PY{p}{)}
                 \PY{k}{break}
\end{Verbatim}

    \subsubsection{Step 4: Create Calibration
Model}\label{step-4-create-calibration-model}

Create an empty calibration model to be filled, using the ascribed
channels for validation purposes.

    \begin{Verbatim}[commandchars=\\\{\}]
{\color{incolor}In [{\color{incolor}29}]:} \PY{n}{calTherm} \PY{o}{=} \PY{n}{m}\PY{o}{.}\PY{n}{ThermalGroup}\PY{p}{(}\PY{n}{calCurrentChan}\PY{p}{)}
         
         \PY{n}{calAxon} \PY{o}{=} \PY{n}{m}\PY{o}{.}\PY{n}{FilterBank}\PY{p}{(}\PY{n}{calTherm}\PY{p}{,} \PY{n}{numRings}\PY{p}{)}
         \PY{n}{calAxon}\PY{o}{.}\PY{n}{setAxon}\PY{p}{(}\PY{p}{)}
         \PY{n}{calFb} \PY{o}{=} \PY{n}{m}\PY{o}{.}\PY{n}{FilterBank}\PY{p}{(}\PY{n}{calTherm}\PY{p}{,} \PY{n}{numRings}\PY{p}{)}
         
         \PY{n}{calCsc} \PY{o}{=} \PY{n}{m}\PY{o}{.}\PY{n}{Cascade}\PY{p}{(}\PY{n}{calAxon}\PY{p}{,} \PY{n}{calFb}\PY{p}{)}
         \PY{n}{calCsc}\PY{o}{.}\PY{n}{setOsaOut}\PY{p}{(}\PY{n}{MrrOut}\PY{o}{.}\PY{n}{kThru}\PY{o}{.}\PY{n}{value}\PY{p}{)}
\end{Verbatim}

    \subsubsection{Step 5: Background Removal}\label{step-5-background-removal}

Also sets the OSA attenuation in the calibration model.

    \begin{Verbatim}[commandchars=\\\{\}]
{\color{incolor}In [{\color{incolor}30}]:} \PY{n}{avgOnSpect}\PY{o}{=}\PY{l+m+mi}{3}
         \PY{n}{detuneByFwhms} \PY{o}{=} \PY{l+m+mi}{3}
         
         \PY{c+c1}{\PYZsh{} Get resonance FWHMs. The extra sweep is technically}
         \PY{c+c1}{  unnecessary, but code below is cleaner}
         \PY{n}{resFwhms} \PY{o}{=} \PY{n}{np}\PY{o}{.}\PY{n}{array}\PY{p}{(}\PY{p}{[}\PY{n}{r}\PY{o}{.}\PY{n}{fwhm} \PY{k}{for} \PY{n}{r} \PY{o+ow}{in} \PY{n}{spctAssist}\PY{o}{.}\PY{n}{resonances}\PY{p}{(}\PY{p}{)}\PY{p}{]}\PY{p}{)}
         \PY{n}{displacedWls} \PY{o}{=} \PY{n}{detuneByFwhms} \PY{o}{*} \PY{n}{resFwhms}
         
         \PY{c+c1}{\PYZsh{} Get Raw Spectrum}
         \PY{n}{baseRawSpct} \PY{o}{=} \PY{n}{spctAssist}\PY{o}{.}\PY{n}{fgSpect}\PY{p}{(}\PY{n}{avgCnt}\PY{o}{=}\PY{n}{avgOnSpect}\PY{p}{,}
            \PY{n}{bgType}\PY{o}{=}\PY{l+s+s1}{\PYZsq{}}\PY{l+s+s1}{smoothed}\PY{l+s+s1}{\PYZsq{}}\PY{p}{)}
         \PY{c+c1}{\PYZsh{} Tune to displace resonances, look at new spectrum, tune back}
         \PY{n}{displTuning} \PY{o}{=} \PY{n+nb}{dict}\PY{p}{(}\PY{p}{)}
         \PY{k}{for} \PY{n}{j} \PY{o+ow}{in} \PY{n+nb}{range}\PY{p}{(}\PY{n+nb}{len}\PY{p}{(}\PY{n}{displacedWls}\PY{p}{)}\PY{p}{)}\PY{p}{:}
             \PY{n}{ch} \PY{o}{=} \PY{n}{pMap}\PY{p}{[}\PY{n}{j}\PY{p}{]}
             \PY{n}{tEst} \PY{o}{=} \PY{n}{tEstMap}\PY{p}{[}\PY{n}{ch}\PY{p}{]}
             \PY{n}{displTuning}\PY{p}{[}\PY{n}{pMap}\PY{p}{[}\PY{n}{j}\PY{p}{]}\PY{p}{]} \PY{o}{=} \PY{n}{baseTune}\PY{p}{[}\PY{n}{ch}\PY{p}{]} \PY{o}{+} \PY{n}{displacedWls}\PY{p}{[}\PY{n}{j}\PY{p}{]} \PY{o}{/} \PY{n}{tEst}
         \PY{n}{inst}\PY{o}{.}\PY{n}{setCurrentChanTuning}\PY{p}{(}\PY{n}{displTuning}\PY{p}{,} \PY{n}{token}\PY{p}{,} \PY{n}{CurrentUnit}\PY{o}{.}\PY{n}{mW}\PY{p}{)}
         \PY{n}{displacedRawSpct} \PY{o}{=} \PY{n}{spctAssist}\PY{o}{.}\PY{n}{fgSpect}\PY{p}{(}\PY{n}{avgCnt}\PY{o}{=}\PY{n}{avgOnSpect}\PY{p}{,}
            \PY{n}{bgType}\PY{o}{=}\PY{l+s+s1}{\PYZsq{}}\PY{l+s+s1}{smoothed}\PY{l+s+s1}{\PYZsq{}}\PY{p}{)}
         
         \PY{c+c1}{\PYZsh{} Return to base}
         \PY{n}{inst}\PY{o}{.}\PY{n}{setCurrentChanTuning}\PY{p}{(}\PY{n}{baseTune}\PY{p}{,} \PY{n}{token}\PY{p}{,} \PY{n}{CurrentUnit}\PY{o}{.}\PY{n}{mW}\PY{p}{)}
         
         \PY{c+c1}{\PYZsh{} Update Background}
         \PY{n}{spctAssist}\PY{o}{.}\PY{n}{setBgTuned}\PY{p}{(}\PY{n}{baseRawSpct}\PY{p}{,} \PY{n}{displacedRawSpct}\PY{p}{)}
\end{Verbatim}

    \begin{Verbatim}[commandchars=\\\{\}]
{\color{incolor}In [{\color{incolor}31}]:} \PY{c+c1}{\PYZsh{} See Peaks w/ Background Removed}
         \PY{n}{spctAssist}\PY{o}{.}\PY{n}{fgResPlot}\PY{p}{(}\PY{p}{)}
         \PY{n}{plt}\PY{o}{.}\PY{n}{show}\PY{p}{(}\PY{p}{)}
\end{Verbatim}

    
    \begin{Verbatim}[commandchars=\\\{\}]
{\color{incolor}In [{\color{incolor}37}]:} \PY{c+c1}{\PYZsh{} Set Attenuation and Validate}
         \PY{n}{calAtten} \PY{o}{=} \PY{n}{spctAssist}\PY{o}{.}\PY{n}{getAtten}\PY{p}{(}\PY{p}{)}
         \PY{n}{err} \PY{o}{=} \PY{l+m+mi}{100}\PY{o}{*}\PY{n}{np}\PY{o}{.}\PY{n}{absolute}\PY{p}{(}\PY{n}{np}\PY{o}{.}\PY{n}{divide}\PY{p}{(}\PY{n}{calAtten}\PY{o}{\PYZhy{}}\PY{n}{attenuation}\PY{p}{,} 
            \PY{n}{attenuation}\PY{p}{)}\PY{p}{)}
         \PY{k}{try}\PY{p}{:}
             \PY{k}{assert} \PY{n}{err} \PY{o}{\PYZlt{}} \PY{l+m+mi}{4}
         \PY{k}{except}\PY{p}{:}
             \PY{n}{spctAssist}\PY{o}{.}\PY{n}{rawSpect}\PY{p}{(}\PY{p}{)}\PY{o}{.}\PY{n}{simplePlot}\PY{p}{(}\PY{p}{)}
             \PY{n}{plt}\PY{o}{.}\PY{n}{show}\PY{p}{(}\PY{p}{)}
             \PY{n+nb}{print}\PY{p}{(}\PY{l+m+mi}{10}\PY{o}{*}\PY{n}{np}\PY{o}{.}\PY{n}{log10}\PY{p}{(}\PY{n}{calAtten}\PY{p}{)}\PY{p}{)}
             \PY{n+nb}{print}\PY{p}{(}\PY{n}{err}\PY{p}{)}
         \PY{c+c1}{\PYZsh{} Less than 4\PYZpc{} error}
\end{Verbatim}

    \subsubsection{Step 5.5: Pull Filter
Shapes}\label{step-5.5-pull-filter-shapes}

    \begin{Verbatim}[commandchars=\\\{\}]
{\color{incolor}In [{\color{incolor}38}]:} \PY{k+kn}{from} \PY{n+nn}{lightlab}\PY{n+nn}{.}\PY{n+nn}{util}\PY{n+nn}{.}\PY{n+nn}{data} \PY{k}{import} \PY{n}{Spectrum}
         
         \PY{c+c1}{\PYZsh{} Pull Filter Shapes}
         \PY{n}{spect} \PY{o}{=} \PY{n}{spctAssist}\PY{o}{.}\PY{n}{fgSpect}\PY{p}{(}\PY{n}{avgCnt}\PY{o}{=}\PY{l+m+mi}{5}\PY{p}{,} \PY{n}{bgType}\PY{o}{=}\PY{l+s+s1}{\PYZsq{}}\PY{l+s+s1}{tuned}\PY{l+s+s1}{\PYZsq{}}\PY{p}{)}
         \PY{n}{axonCurves} \PY{o}{=} \PY{n+nb}{list}\PY{p}{(}\PY{p}{)}
         \PY{n}{dendriteCurves} \PY{o}{=} \PY{n+nb}{list}\PY{p}{(}\PY{p}{)}
         \PY{k}{for} \PY{n}{i}\PY{p}{,} \PY{n}{r} \PY{o+ow}{in} \PY{n+nb}{enumerate}\PY{p}{(}\PY{n}{spctAssist}\PY{o}{.}\PY{n}{resonances}\PY{p}{(}\PY{n}{spect}\PY{p}{)}\PY{p}{)}\PY{p}{:}
             \PY{n}{relWindow} \PY{o}{=} \PY{l+m+mi}{8} \PY{o}{*} \PY{n}{r}\PY{o}{.}\PY{n}{fwhm} \PY{o}{*} \PY{n}{np}\PY{o}{.}\PY{n}{array}\PY{p}{(}\PY{p}{[}\PY{o}{\PYZhy{}}\PY{l+m+mi}{1}\PY{p}{,}\PY{l+m+mi}{1}\PY{p}{]}\PY{p}{)}\PY{o}{/}\PY{l+m+mi}{2}
             \PY{n}{proximitySpect} \PY{o}{=} \PY{n}{spect}\PY{o}{.}\PY{n}{shift}\PY{p}{(}\PY{o}{\PYZhy{}}\PY{n}{r}\PY{o}{.}\PY{n}{lam}\PY{p}{)}\PY{o}{.}\PY{n}{crop}\PY{p}{(}\PY{n}{relWindow}\PY{p}{)}
             \PY{n}{nm}\PY{p}{,} \PY{n}{dbm} \PY{o}{=} \PY{n}{proximitySpect}\PY{o}{.}\PY{n}{shift}\PY{p}{(}\PY{n}{r}\PY{o}{.}\PY{n}{lam}\PY{p}{)}\PY{o}{.}\PY{n}{getData}\PY{p}{(}\PY{p}{)}
             \PY{n}{plt}\PY{o}{.}\PY{n}{plot}\PY{p}{(}\PY{n}{nm}\PY{p}{,} \PY{n}{dbm}\PY{p}{)}
             \PY{n}{lin} \PY{o}{=} \PY{n}{np}\PY{o}{.}\PY{n}{clip}\PY{p}{(}\PY{n}{dbm2lin}\PY{p}{(}\PY{n}{dbm}\PY{p}{)}\PY{p}{,} \PY{l+m+mf}{0.0}\PY{p}{,} \PY{l+m+mf}{1.0}\PY{p}{)}
             \PY{k}{if} \PY{n}{pMap}\PY{p}{[}\PY{n}{i}\PY{p}{]} \PY{o+ow}{in} \PY{n}{calAxonChan}\PY{p}{:}
                 \PY{n}{axonCurves}\PY{o}{.}\PY{n}{append}\PY{p}{(}\PY{p}{(}\PY{n}{nm}\PY{p}{,}\PY{n}{lin}\PY{p}{)}\PY{p}{)}
             \PY{k}{else}\PY{p}{:}
                 \PY{n}{dendriteCurves}\PY{o}{.}\PY{n}{append}\PY{p}{(}\PY{p}{(}\PY{n}{nm}\PY{p}{,} \PY{n}{lin}\PY{p}{)}\PY{p}{)}
                 
         \PY{k}{assert} \PY{n+nb}{len}\PY{p}{(}\PY{n}{axonCurves}\PY{p}{)} \PY{o}{==} \PY{n}{numRings}
         \PY{k}{assert} \PY{n+nb}{len}\PY{p}{(}\PY{n}{dendriteCurves}\PY{p}{)} \PY{o}{==} \PY{n}{numRings}
         \PY{n}{plt}\PY{o}{.}\PY{n}{show}\PY{p}{(}\PY{p}{)}
         
         \PY{c+c1}{\PYZsh{} Don\PYZsq{}t add them until AFTER bias calculation}
\end{Verbatim}

    
    \subsubsection{Step 6: Determine Heat Bias}\label{step-6-determine-heat-bias}

\begin{enumerate}
\def\labelenumi{\arabic{enumi}.}
\tightlist
\item
  Put peaks on AD order (should already be done manually).
\item
  Move to within FWHM of targets
\item
  Focus on each target, track 2 peaks together.
\end{enumerate}

    \begin{Verbatim}[commandchars=\\\{\}]
{\color{incolor}In [{\color{incolor}39}]:} \PY{c+c1}{\PYZsh{} Set Targets}
         \PY{n}{maxfwhm} \PY{o}{=} \PY{n}{np}\PY{o}{.}\PY{n}{max}\PY{p}{(}\PY{n}{np}\PY{o}{.}\PY{n}{array}\PY{p}{(}\PY{p}{[}\PY{n}{r}\PY{o}{.}\PY{n}{fwhm} 
            \PY{k}{for} \PY{n}{r} \PY{o+ow}{in} \PY{n}{spctAssist}\PY{o}{.}\PY{n}{resonances}\PY{p}{(}\PY{p}{)}\PY{p}{]}\PY{p}{)}\PY{p}{)}
         \PY{n}{targets} \PY{o}{=} \PY{n+nb}{list}\PY{p}{(}\PY{p}{)}
         \PY{k}{for} \PY{n}{wl} \PY{o+ow}{in} \PY{n}{wlChannels}\PY{p}{:}
             \PY{n}{targets}\PY{o}{.}\PY{n}{append}\PY{p}{(}\PY{n}{wl} \PY{o}{\PYZhy{}} \PY{n}{maxfwhm}\PY{p}{)}
             \PY{n}{targets}\PY{o}{.}\PY{n}{append}\PY{p}{(}\PY{n}{wl} \PY{o}{+} \PY{n}{maxfwhm}\PY{p}{)}
             
         \PY{n}{targets} \PY{o}{=} \PY{n}{np}\PY{o}{.}\PY{n}{array}\PY{p}{(}\PY{n}{targets}\PY{p}{)}
\end{Verbatim}

    \begin{Verbatim}[commandchars=\\\{\}]
{\color{incolor}In [{\color{incolor}40}]:} \PY{c+c1}{\PYZsh{}\PYZsh{}\PYZsh{} Tracking to targets}
         
         \PY{n}{precision} \PY{o}{=} \PY{l+m+mf}{0.005} \PY{c+c1}{\PYZsh{} Threshold}
         \PY{n}{propCoef} \PY{o}{=} \PY{l+m+mf}{0.5} \PY{c+c1}{\PYZsh{} kP}
         \PY{n}{avgCnt} \PY{o}{=} \PY{l+m+mi}{4}
         \PY{n}{nowTune} \PY{o}{=} \PY{n}{baseTune}\PY{o}{.}\PY{n}{copy}\PY{p}{(}\PY{p}{)}
         \PY{n}{appxThrmCoefs} \PY{o}{=} \PY{n}{tEstMap}
         \PY{n}{inst}\PY{o}{.}\PY{n}{setCurrentChanTuning}\PY{p}{(}\PY{n}{baseTune}\PY{p}{,} \PY{n}{token}\PY{p}{,} \PY{n}{CurrentUnit}\PY{o}{.}\PY{n}{mW}\PY{p}{)}
\end{Verbatim}

    \begin{Verbatim}[commandchars=\\\{\}]
{\color{incolor}In [{\color{incolor}41}]:} \PY{k}{for} \PY{n}{i} \PY{o+ow}{in} \PY{n+nb}{range}\PY{p}{(}\PY{l+m+mi}{100}\PY{p}{)}\PY{p}{:}
             \PY{n}{spect} \PY{o}{=} \PY{n}{spctAssist}\PY{o}{.}\PY{n}{fgSpect}\PY{p}{(}\PY{n}{bgType}\PY{o}{=}\PY{l+s+s1}{\PYZsq{}}\PY{l+s+s1}{tuned}\PY{l+s+s1}{\PYZsq{}}\PY{p}{,} \PY{n}{avgCnt} \PY{o}{=} \PY{n}{avgCnt}\PY{p}{)}
             \PY{n}{actualPeaks} \PY{o}{=} \PY{n}{np}\PY{o}{.}\PY{n}{array}\PY{p}{(}\PY{p}{[}\PY{n}{r}\PY{o}{.}\PY{n}{lam} 
                \PY{k}{for} \PY{n}{r} \PY{o+ow}{in} \PY{n}{spctAssist}\PY{o}{.}\PY{n}{resonances}\PY{p}{(}\PY{n}{spect}\PY{p}{)}\PY{p}{]}\PY{p}{)}
             \PY{n}{errs} \PY{o}{=} \PY{n}{targets} \PY{o}{\PYZhy{}} \PY{n}{actualPeaks}
             \PY{k}{if} \PY{n+nb}{max}\PY{p}{(}\PY{n+nb}{abs}\PY{p}{(}\PY{n}{errs}\PY{p}{)}\PY{p}{)} \PY{o}{\PYZlt{}} \PY{n}{precision}\PY{p}{:}
                 \PY{n+nb}{print}\PY{p}{(}\PY{l+s+s1}{\PYZsq{}}\PY{l+s+se}{\PYZbs{}n}\PY{l+s+s1}{Tracking complete}\PY{l+s+s1}{\PYZsq{}}\PY{p}{)}
                 \PY{k}{break}
             \PY{c+c1}{\PYZsh{} recenter wlRange to avoid other FSR resonances}
             \PY{n}{wlRangeTight} \PY{o}{=} \PY{n}{np}\PY{o}{.}\PY{n}{array}\PY{p}{(}\PY{p}{[}\PY{n+nb}{min}\PY{p}{(}\PY{n+nb}{min}\PY{p}{(}\PY{n}{actualPeaks}\PY{p}{)}\PY{p}{,} \PY{n+nb}{min}\PY{p}{(}\PY{n}{targets}\PY{p}{)}\PY{p}{)}\PY{p}{,}
                \PY{n+nb}{max}\PY{p}{(}\PY{n+nb}{max}\PY{p}{(}\PY{n}{actualPeaks}\PY{p}{)}\PY{p}{,} \PY{n+nb}{max}\PY{p}{(}\PY{n}{targets}\PY{p}{)}\PY{p}{)}\PY{p}{]}\PY{p}{)}
             \PY{n}{newwlRange} \PY{o}{=} \PY{n}{np}\PY{o}{.}\PY{n}{mean}\PY{p}{(}\PY{n}{wlRangeTight}\PY{p}{)} \PY{o}{+}
                \PY{n}{np}\PY{o}{.}\PY{n}{diff}\PY{p}{(}\PY{n}{wlRangeTight}\PY{p}{)} \PY{o}{*} \PY{n}{np}\PY{o}{.}\PY{n}{array}\PY{p}{(}\PY{p}{[}\PY{o}{\PYZhy{}}\PY{l+m+mi}{1}\PY{p}{,} \PY{l+m+mi}{1}\PY{p}{]}\PY{p}{)} \PY{o}{/} \PY{l+m+mi}{2} \PY{o}{*} \PY{l+m+mi}{2}
             \PY{n}{spctAssist}\PY{o}{.}\PY{n}{wlRange} \PY{o}{=} \PY{n}{newwlRange}
             \PY{k}{try}\PY{p}{:}
                 \PY{k}{for} \PY{n}{p}\PY{p}{,} \PY{n}{ch} \PY{o+ow}{in} \PY{n}{pMap}\PY{o}{.}\PY{n}{items}\PY{p}{(}\PY{p}{)}\PY{p}{:}
                     \PY{n}{nowTune}\PY{p}{[}\PY{n}{ch}\PY{p}{]} \PY{o}{=} \PY{n}{nowTune}\PY{p}{[}\PY{n}{ch}\PY{p}{]} \PY{o}{+}
                        \PY{n}{propCoef}\PY{o}{*}\PY{n}{errs}\PY{p}{[}\PY{n}{p}\PY{p}{]} \PY{o}{/} \PY{n}{appxThrmCoefs}\PY{p}{[}\PY{n}{ch}\PY{p}{]}
                 \PY{n}{inst}\PY{o}{.}\PY{n}{setCurrentChanTuning}\PY{p}{(}\PY{n}{nowTune}\PY{p}{,} \PY{n}{token}\PY{p}{,} \PY{n}{CurrentUnit}\PY{o}{.}\PY{n}{mW}\PY{p}{)}
             \PY{k}{except} \PY{n}{io}\PY{o}{.}\PY{n}{RangeError} \PY{k}{as} \PY{n}{err}\PY{p}{:}
                 \PY{n+nb}{print}\PY{p}{(}\PY{l+s+s1}{\PYZsq{}}\PY{l+s+s1}{Out of range during tracking. See plot}\PY{l+s+s1}{\PYZsq{}}\PY{p}{)}
                 \PY{n}{spctAssist}\PY{o}{.}\PY{n}{fgResPlot}\PY{p}{(}\PY{p}{)}
                 \PY{k}{raise} \PY{n}{err}
                 
         \PY{n}{spctAssist}\PY{o}{.}\PY{n}{fgResPlot}\PY{p}{(}\PY{p}{)}
         \PY{n}{plt}\PY{o}{.}\PY{n}{show}\PY{p}{(}\PY{p}{)}
\end{Verbatim}

    \begin{Verbatim}[commandchars=\\\{\}]

Tracking complete

    \end{Verbatim}

    
    \begin{Verbatim}[commandchars=\\\{\}]
{\color{incolor}In [{\color{incolor}42}]:} \PY{k+kn}{import} \PY{n+nn}{copy}
         \PY{k}{def} \PY{n+nf}{mergePeaks}\PY{p}{(}\PY{n}{wlTarget}\PY{p}{,} \PY{n}{newPMap}\PY{p}{,} \PY{n}{spctAssist}\PY{p}{,} 
            \PY{n}{currentTune}\PY{p}{,} \PY{n}{fwhmThresh}\PY{p}{,} \PY{n}{tEstMap}\PY{p}{,} \PY{n}{skipStart}\PY{p}{)}\PY{p}{:}
             \PY{n}{targets} \PY{o}{=} \PY{n}{np}\PY{o}{.}\PY{n}{array}\PY{p}{(}\PY{p}{[}\PY{n}{wlTarget}\PY{p}{,} \PY{n}{wlTarget}\PY{p}{]}\PY{p}{)}
             \PY{n}{newTune} \PY{o}{=} \PY{n}{currentTune}\PY{o}{.}\PY{n}{copy}\PY{p}{(}\PY{p}{)}
             \PY{n}{newSp} \PY{o}{=} \PY{n}{copy}\PY{o}{.}\PY{n}{deepcopy}\PY{p}{(}\PY{n}{spctAssist}\PY{p}{)}
             \PY{n}{newSp}\PY{o}{.}\PY{n}{wlRange} \PY{o}{=} \PY{p}{[}\PY{n}{wlTarget}\PY{o}{\PYZhy{}}\PY{l+m+mi}{2}\PY{o}{*}\PY{n}{fwhmThresh}\PY{p}{,} \PY{n}{wlTarget}\PY{o}{+}\PY{l+m+mi}{2}\PY{o}{*}\PY{n}{fwhmThresh}\PY{p}{]}
             \PY{n}{newSp}\PY{o}{.}\PY{n}{nChan} \PY{o}{=} \PY{l+m+mi}{2}
             
             \PY{c+c1}{\PYZsh{}\PYZsh{}\PYZsh{} Tracking to single detected peak}
         
             \PY{n}{precision} \PY{o}{=} \PY{l+m+mf}{0.005} \PY{c+c1}{\PYZsh{} Threshold}
             \PY{n}{propCoef} \PY{o}{=} \PY{l+m+mf}{0.1} \PY{c+c1}{\PYZsh{} kP}
             \PY{n}{avgCnt} \PY{o}{=} \PY{l+m+mi}{4}
             \PY{n}{appxThrmCoefs} \PY{o}{=} \PY{n}{tEstMap}
             \PY{n}{inst}\PY{o}{.}\PY{n}{setCurrentChanTuning}\PY{p}{(}\PY{n}{newTune}\PY{p}{,} \PY{n}{token}\PY{p}{,} \PY{n}{CurrentUnit}\PY{o}{.}\PY{n}{mW}\PY{p}{)}
             
             \PY{k}{for} \PY{n}{i} \PY{o+ow}{in} \PY{n+nb}{range}\PY{p}{(}\PY{l+m+mi}{100}\PY{p}{)}\PY{p}{:}
                 \PY{k}{if} \PY{n}{skipStart}\PY{p}{:}
                     \PY{k}{break}
                     
                 \PY{n}{spect} \PY{o}{=} \PY{n}{newSp}\PY{o}{.}\PY{n}{fgSpect}\PY{p}{(}\PY{n}{bgType}\PY{o}{=}\PY{l+s+s1}{\PYZsq{}}\PY{l+s+s1}{tuned}\PY{l+s+s1}{\PYZsq{}}\PY{p}{,} \PY{n}{avgCnt} \PY{o}{=} \PY{n}{avgCnt}\PY{p}{)}
                 \PY{k}{try}\PY{p}{:}
                     \PY{n}{actualPeaks} \PY{o}{=} \PY{n}{np}\PY{o}{.}\PY{n}{array}\PY{p}{(}\PY{p}{[}\PY{n}{r}\PY{o}{.}\PY{n}{lam} 
                        \PY{k}{for} \PY{n}{r} \PY{o+ow}{in} \PY{n}{newSp}\PY{o}{.}\PY{n}{resonances}\PY{p}{(}\PY{n}{spect}\PY{p}{)}\PY{p}{]}\PY{p}{)}
                     \PY{n}{actualfwhms} \PY{o}{=} \PY{n}{np}\PY{o}{.}\PY{n}{array}\PY{p}{(}\PY{p}{[}\PY{n}{r}\PY{o}{.}\PY{n}{fwhm} 
                        \PY{k}{for} \PY{n}{r} \PY{o+ow}{in} \PY{n}{newSp}\PY{o}{.}\PY{n}{resonances}\PY{p}{(}\PY{n}{spect}\PY{p}{)}\PY{p}{]}\PY{p}{)}
                 \PY{k}{except}\PY{p}{:}
                     \PY{n+nb}{print}\PY{p}{(}\PY{l+s+s2}{\PYZdq{}}\PY{l+s+s2}{Peaks have merged! (exception)}\PY{l+s+s2}{\PYZdq{}}\PY{p}{)}
                     \PY{k}{break}
         
                 \PY{c+c1}{\PYZsh{} Check for merged peaks}
                 \PY{n}{flag} \PY{o}{=} \PY{k+kc}{False}
                 \PY{k}{for} \PY{n}{f} \PY{o+ow}{in} \PY{n}{actualfwhms}\PY{p}{:}
                         \PY{k}{if} \PY{n}{f} \PY{o}{\PYZgt{}} \PY{l+m+mi}{2}\PY{o}{*}\PY{n}{maxfwhm}\PY{p}{:}
                             \PY{n+nb}{print}\PY{p}{(}\PY{l+s+s2}{\PYZdq{}}\PY{l+s+s2}{Peaks have merged! (fwhm)}\PY{l+s+s2}{\PYZdq{}}\PY{p}{)}
                             \PY{n}{flag} \PY{o}{=} \PY{k+kc}{True}
                 \PY{k}{if} \PY{n}{flag}\PY{p}{:}
                     \PY{k}{break}
         
                 \PY{c+c1}{\PYZsh{} Run PID}
                 \PY{n}{errs} \PY{o}{=} \PY{n}{targets} \PY{o}{\PYZhy{}} \PY{n}{actualPeaks}
                 \PY{k}{if} \PY{n+nb}{max}\PY{p}{(}\PY{n+nb}{abs}\PY{p}{(}\PY{n}{errs}\PY{p}{)}\PY{p}{)} \PY{o}{\PYZlt{}} \PY{n}{precision}\PY{p}{:}
                     \PY{n+nb}{print}\PY{p}{(}\PY{l+s+s1}{\PYZsq{}}\PY{l+s+se}{\PYZbs{}n}\PY{l+s+s1}{Tracking complete}\PY{l+s+s1}{\PYZsq{}}\PY{p}{)}
                     \PY{k}{break}
                 \PY{k}{try}\PY{p}{:}
                     \PY{k}{for} \PY{n}{p}\PY{p}{,} \PY{n}{ch} \PY{o+ow}{in} \PY{n}{newPMap}\PY{o}{.}\PY{n}{items}\PY{p}{(}\PY{p}{)}\PY{p}{:}
                         \PY{n}{newTune}\PY{p}{[}\PY{n}{ch}\PY{p}{]} \PY{o}{=} \PY{n}{newTune}\PY{p}{[}\PY{n}{ch}\PY{p}{]} \PY{o}{+} \PY{n}{propCoef}\PY{o}{*}\PY{n}{errs}\PY{p}{[}\PY{n}{p}\PY{p}{]} \PY{o}{/} 
                            \PY{n}{appxThrmCoefs}\PY{p}{[}\PY{n}{ch}\PY{p}{]}
                     \PY{n}{inst}\PY{o}{.}\PY{n}{setCurrentChanTuning}\PY{p}{(}\PY{n}{newTune}\PY{p}{,} \PY{n}{token}\PY{p}{,}
                        \PY{n}{CurrentUnit}\PY{o}{.}\PY{n}{mW}\PY{p}{)}
                 \PY{k}{except} \PY{n}{io}\PY{o}{.}\PY{n}{RangeError} \PY{k}{as} \PY{n}{err}\PY{p}{:}
                     \PY{n+nb}{print}\PY{p}{(}\PY{l+s+s1}{\PYZsq{}}\PY{l+s+s1}{Out of range during tracking. See plot}\PY{l+s+s1}{\PYZsq{}}\PY{p}{)}
                     \PY{n}{newSp}\PY{o}{.}\PY{n}{fgResPlot}\PY{p}{(}\PY{p}{)}
                     \PY{k}{raise} \PY{n}{err}
             \PY{c+c1}{\PYZsh{}Peaks should have merged}
             \PY{n}{newSp}\PY{o}{.}\PY{n}{nChan} \PY{o}{=} \PY{l+m+mi}{1}
             
             \PY{c+c1}{\PYZsh{}\PYZsh{}\PYZsh{} Tracking to minimum FWHM (actually 1 peak)}
             
             \PY{n}{leftCh} \PY{o}{=} \PY{n}{newPMap}\PY{p}{[}\PY{l+m+mi}{0}\PY{p}{]}
             \PY{n}{rightCh} \PY{o}{=} \PY{n}{newPMap}\PY{p}{[}\PY{l+m+mi}{1}\PY{p}{]}
             \PY{n}{errfwhm} \PY{o}{=} \PY{l+m+mi}{10000}
             \PY{n}{propCoef} \PY{o}{=} \PY{l+m+mf}{0.11}
             \PY{n}{prevTune} \PY{o}{=} \PY{n}{newTune}\PY{o}{.}\PY{n}{copy}\PY{p}{(}\PY{p}{)}
             \PY{n}{strike} \PY{o}{=} \PY{k+kc}{False}
             
             \PY{k}{for} \PY{n}{i} \PY{o+ow}{in} \PY{n+nb}{range}\PY{p}{(}\PY{l+m+mi}{100}\PY{p}{)}\PY{p}{:}
                 \PY{n}{spect} \PY{o}{=} \PY{n}{newSp}\PY{o}{.}\PY{n}{fgSpect}\PY{p}{(}\PY{n}{bgType}\PY{o}{=}\PY{l+s+s1}{\PYZsq{}}\PY{l+s+s1}{tuned}\PY{l+s+s1}{\PYZsq{}}\PY{p}{,} \PY{n}{avgCnt} \PY{o}{=} \PY{n}{avgCnt}\PY{p}{)}
         
                 \PY{n}{aCenter} \PY{o}{=} \PY{n}{np}\PY{o}{.}\PY{n}{array}\PY{p}{(}\PY{p}{[}\PY{n}{r}\PY{o}{.}\PY{n}{lam} 
                    \PY{k}{for} \PY{n}{r} \PY{o+ow}{in} \PY{n}{newSp}\PY{o}{.}\PY{n}{resonances}\PY{p}{(}\PY{n}{spect}\PY{p}{)}\PY{p}{]}\PY{p}{)}\PY{p}{[}\PY{l+m+mi}{0}\PY{p}{]}
                 \PY{n}{aFwhm} \PY{o}{=} \PY{n}{np}\PY{o}{.}\PY{n}{array}\PY{p}{(}\PY{p}{[}\PY{n}{r}\PY{o}{.}\PY{n}{fwhm} 
                    \PY{k}{for} \PY{n}{r} \PY{o+ow}{in} \PY{n}{newSp}\PY{o}{.}\PY{n}{resonances}\PY{p}{(}\PY{n}{spect}\PY{p}{)}\PY{p}{]}\PY{p}{)}\PY{p}{[}\PY{l+m+mi}{0}\PY{p}{]}
         
                 \PY{c+c1}{\PYZsh{} Run PID}
                 \PY{n}{prevErr} \PY{o}{=} \PY{n}{errfwhm}
                 \PY{n}{errfwhm} \PY{o}{=} \PY{n}{aFwhm} \PY{o}{\PYZhy{}} \PY{n}{maxfwhm}
                 \PY{n}{errCenter} \PY{o}{=} \PY{n}{aCenter} \PY{o}{\PYZhy{}} \PY{n}{targets}\PY{p}{[}\PY{l+m+mi}{0}\PY{p}{]}
         
                 \PY{k}{if} \PY{n}{errfwhm} \PY{o}{\PYZgt{}} \PY{n}{prevErr}\PY{p}{:}
                     \PY{n}{newTune} \PY{o}{=} \PY{n}{prevTune}
                     \PY{n}{inst}\PY{o}{.}\PY{n}{setCurrentChanTuning}\PY{p}{(}\PY{n}{newTune}\PY{p}{,} \PY{n}{token}\PY{p}{,}
                        \PY{n}{CurrentUnit}\PY{o}{.}\PY{n}{mW}\PY{p}{)}
                     \PY{k}{if} \PY{n}{strike}\PY{p}{:}
                         \PY{k}{break}
                     \PY{k}{else}\PY{p}{:}
                         \PY{c+c1}{\PYZsh{} Maybe the two peaks have crossed}
                         \PY{n}{strike} \PY{o}{=} \PY{k+kc}{True}
                         \PY{n}{tmp} \PY{o}{=} \PY{n}{leftCh}
                         \PY{n}{leftCh} \PY{o}{=} \PY{n}{rightCh}
                         \PY{n}{rightCh} \PY{o}{=} \PY{n}{tmp}
         
                 \PY{k}{if} \PY{n+nb}{abs}\PY{p}{(}\PY{n}{errfwhm}\PY{p}{)} \PY{o}{\PYZlt{}} \PY{n}{precision}\PY{p}{:}
                     \PY{n+nb}{print}\PY{p}{(}\PY{l+s+s1}{\PYZsq{}}\PY{l+s+se}{\PYZbs{}n}\PY{l+s+s1}{Tracking complete}\PY{l+s+s1}{\PYZsq{}}\PY{p}{)}
                     \PY{k}{break}
                 \PY{k}{try}\PY{p}{:}
                     \PY{n}{prevTune} \PY{o}{=} \PY{n}{newTune}\PY{o}{.}\PY{n}{copy}\PY{p}{(}\PY{p}{)}
                     \PY{n}{newTune}\PY{p}{[}\PY{n}{leftCh}\PY{p}{]} \PY{o}{=} \PY{n}{newTune}\PY{p}{[}\PY{n}{leftCh}\PY{p}{]} \PY{o}{+}
                        \PY{n}{propCoef}\PY{o}{*}\PY{p}{(}\PY{n}{errfwhm}\PY{o}{\PYZhy{}}\PY{n}{errCenter}\PY{p}{)} \PY{o}{/}
                            \PY{n}{appxThrmCoefs}\PY{p}{[}\PY{n}{leftCh}\PY{p}{]}
                     \PY{n}{newTune}\PY{p}{[}\PY{n}{rightCh}\PY{p}{]} \PY{o}{=} \PY{n}{newTune}\PY{p}{[}\PY{n}{rightCh}\PY{p}{]} \PY{o}{\PYZhy{}}
                        \PY{n}{propCoef}\PY{o}{*}\PY{p}{(}\PY{n}{errfwhm}\PY{o}{\PYZhy{}}\PY{n}{errCenter}\PY{p}{)} \PY{o}{/}
                            \PY{n}{appxThrmCoefs}\PY{p}{[}\PY{n}{rightCh}\PY{p}{]}
                     \PY{n}{inst}\PY{o}{.}\PY{n}{setCurrentChanTuning}\PY{p}{(}\PY{n}{newTune}\PY{p}{,} \PY{n}{token}\PY{p}{,}
                        \PY{n}{CurrentUnit}\PY{o}{.}\PY{n}{mW}\PY{p}{)}
                 \PY{k}{except} \PY{n}{io}\PY{o}{.}\PY{n}{RangeError} \PY{k}{as} \PY{n}{err}\PY{p}{:}
                     \PY{n+nb}{print}\PY{p}{(}\PY{l+s+s1}{\PYZsq{}}\PY{l+s+s1}{Out of range during tracking. See plot}\PY{l+s+s1}{\PYZsq{}}\PY{p}{)}
                     \PY{n}{newSp}\PY{o}{.}\PY{n}{fgResPlot}\PY{p}{(}\PY{p}{)}
                     \PY{k}{raise} \PY{n}{err}
                     
             \PY{c+c1}{\PYZsh{} Peaks should really have merged}
             \PY{n}{newSp}\PY{o}{.}\PY{n}{fgResPlot}\PY{p}{(}\PY{p}{)}
             \PY{k}{return} \PY{n}{newTune}
\end{Verbatim}

    \begin{Verbatim}[commandchars=\\\{\}]
{\color{incolor}In [{\color{incolor}43}]:} \PY{c+c1}{\PYZsh{}\PYZsh{}\PYZsh{} MAY HAVE TO RUN A FEW TIMES TO GET GOOD RESULTS}
         
         \PY{n}{inst}\PY{o}{.}\PY{n}{setCurrentChanTuning}\PY{p}{(}\PY{n}{nowTune}\PY{p}{,} \PY{n}{token}\PY{p}{,} \PY{n}{CurrentUnit}\PY{o}{.}\PY{n}{mW}\PY{p}{)}
         \PY{n}{newTune} \PY{o}{=} \PY{n}{nowTune}\PY{o}{.}\PY{n}{copy}\PY{p}{(}\PY{p}{)}
         
         \PY{c+c1}{\PYZsh{} Initial Run\PYZhy{}thru}
         \PY{k}{for} \PY{n}{i}\PY{p}{,} \PY{n}{wl} \PY{o+ow}{in} \PY{n+nb}{enumerate}\PY{p}{(}\PY{n}{wlChannels}\PY{p}{)}\PY{p}{:}
             \PY{n}{newPMap} \PY{o}{=} \PY{n+nb}{dict}\PY{p}{(}\PY{p}{)}
             \PY{n}{newPMap}\PY{p}{[}\PY{l+m+mi}{0}\PY{p}{]} \PY{o}{=} \PY{n}{pMap}\PY{p}{[}\PY{l+m+mi}{2}\PY{o}{*}\PY{n}{i}\PY{p}{]}
             \PY{n}{newPMap}\PY{p}{[}\PY{l+m+mi}{1}\PY{p}{]} \PY{o}{=} \PY{n}{pMap}\PY{p}{[}\PY{l+m+mi}{2}\PY{o}{*}\PY{n}{i}\PY{o}{+}\PY{l+m+mi}{1}\PY{p}{]}
             \PY{n}{newTune} \PY{o}{=} \PY{n}{mergePeaks}\PY{p}{(}\PY{n}{wl}\PY{p}{,} \PY{n}{newPMap}\PY{p}{,} \PY{n}{spctAssist}\PY{p}{,} 
                \PY{n}{newTune}\PY{p}{,} \PY{n}{maxfwhm}\PY{p}{,} \PY{n}{tEstMap}\PY{p}{,} \PY{k+kc}{False}\PY{p}{)}
             \PY{n}{plt}\PY{o}{.}\PY{n}{show}\PY{p}{(}\PY{p}{)}
             \PY{n}{sleep}\PY{p}{(}\PY{l+m+mi}{1}\PY{p}{)}
             
         \PY{c+c1}{\PYZsh{} Another run\PYZhy{}thru to smooth out any errant cross\PYZhy{}talk}
         \PY{k}{for} \PY{n}{i}\PY{p}{,} \PY{n}{wl} \PY{o+ow}{in} \PY{n+nb}{enumerate}\PY{p}{(}\PY{n}{wlChannels}\PY{p}{)}\PY{p}{:}
             \PY{n}{newPMap} \PY{o}{=} \PY{n+nb}{dict}\PY{p}{(}\PY{p}{)}
             \PY{n}{newPMap}\PY{p}{[}\PY{l+m+mi}{0}\PY{p}{]} \PY{o}{=} \PY{n}{pMap}\PY{p}{[}\PY{l+m+mi}{2}\PY{o}{*}\PY{n}{i}\PY{p}{]}
             \PY{n}{newPMap}\PY{p}{[}\PY{l+m+mi}{1}\PY{p}{]} \PY{o}{=} \PY{n}{pMap}\PY{p}{[}\PY{l+m+mi}{2}\PY{o}{*}\PY{n}{i}\PY{o}{+}\PY{l+m+mi}{1}\PY{p}{]}
             \PY{n}{newTune} \PY{o}{=} \PY{n}{mergePeaks}\PY{p}{(}\PY{n}{wl}\PY{p}{,} \PY{n}{newPMap}\PY{p}{,} \PY{n}{spctAssist}\PY{p}{,} 
                \PY{n}{newTune}\PY{p}{,} \PY{n}{maxfwhm}\PY{p}{,} \PY{n}{tEstMap}\PY{p}{,} \PY{k+kc}{True}\PY{p}{)}
             \PY{n}{plt}\PY{o}{.}\PY{n}{show}\PY{p}{(}\PY{p}{)}
             \PY{n}{sleep}\PY{p}{(}\PY{l+m+mi}{1}\PY{p}{)}
\end{Verbatim}

    \begin{Verbatim}[commandchars=\\\{\}]
Peaks have merged! (fwhm)

    \end{Verbatim}

    
    \begin{Verbatim}[commandchars=\\\{\}]

    \end{Verbatim}

    
    \begin{Verbatim}[commandchars=\\\{\}]

    \end{Verbatim}

    
    \begin{Verbatim}[commandchars=\\\{\}]
{\color{incolor}In [{\color{incolor}44}]:} \PY{c+c1}{\PYZsh{} There should now be two peaks}
         \PY{n}{spctAssist}\PY{o}{.}\PY{n}{nChan} \PY{o}{=} \PY{n+nb}{len}\PY{p}{(}\PY{n}{wlChannels}\PY{p}{)}
         \PY{n}{spctAssist}\PY{o}{.}\PY{n}{fgResPlot}\PY{p}{(}\PY{p}{)}
         \PY{n}{plt}\PY{o}{.}\PY{n}{show}\PY{p}{(}\PY{p}{)}
\end{Verbatim}

    
    \begin{Verbatim}[commandchars=\\\{\}]
{\color{incolor}In [{\color{incolor}45}]:} \PY{n}{calTherm}\PY{o}{.}\PY{n}{setHeatBias}\PY{p}{(}\PY{n}{newTune}\PY{p}{,} \PY{n}{CurrentUnit}\PY{o}{.}\PY{n}{mW}\PY{p}{)}
\end{Verbatim}

    \begin{Verbatim}[commandchars=\\\{\}]
{\color{incolor}In [{\color{incolor}47}]:} \PY{c+c1}{\PYZsh{} Validation}
         \PY{n}{threshold} \PY{o}{=} \PY{l+m+mf}{0.01}
         
         \PY{n}{calBias} \PY{o}{=} \PY{n}{np}\PY{o}{.}\PY{n}{array}\PY{p}{(}\PY{n+nb}{sorted}\PY{p}{(}\PY{p}{[}\PY{n}{CurrentUnit}\PY{o}{.}\PY{n}{toVolt}\PY{p}{(}\PY{n}{b}\PY{p}{,} \PY{n}{CurrentUnit}\PY{o}{.}\PY{n}{mW}\PY{p}{)}
            \PY{k}{for} \PY{n}{b} \PY{o+ow}{in} \PY{n}{calTherm}\PY{o}{.}\PY{n}{heatBias}\PY{p}{]}\PY{p}{)}\PY{p}{)}
         \PY{n}{virtBias} \PY{o}{=} \PY{n}{np}\PY{o}{.}\PY{n}{array}\PY{p}{(}\PY{n+nb}{sorted}\PY{p}{(}\PY{n+nb}{list}\PY{p}{(}\PY{n}{heatBias}\PY{o}{.}\PY{n}{values}\PY{p}{(}\PY{p}{)}\PY{p}{)}\PY{p}{)}\PY{p}{)}
         \PY{n}{diff} \PY{o}{=} \PY{n}{np}\PY{o}{.}\PY{n}{absolute}\PY{p}{(}\PY{n}{calBias} \PY{o}{\PYZhy{}} \PY{n}{virtBias}\PY{p}{)}
         \PY{n+nb}{print}\PY{p}{(}\PY{n}{diff}\PY{p}{)}
         \PY{k}{for} \PY{n}{d} \PY{o+ow}{in} \PY{n}{diff}\PY{p}{:}
             \PY{k}{assert} \PY{n}{d} \PY{o}{\PYZlt{}} \PY{n}{threshold}
\end{Verbatim}

    \begin{Verbatim}[commandchars=\\\{\}]
[ 0.0028115   0.00295138  0.00819335  0.00469781  0.0002451   0.0007584 ]

    \end{Verbatim}

    \begin{Verbatim}[commandchars=\\\{\}]
{\color{incolor}In [{\color{incolor}48}]:} \PY{n}{calAxon}\PY{o}{.}\PY{n}{setBiasParams}\PY{p}{(}\PY{p}{[}\PY{n}{r}\PY{o}{.}\PY{n}{lam} \PY{k}{for} \PY{n}{r} \PY{o+ow}{in} \PY{n}{spctAssist}\PY{o}{.}\PY{n}{resonances}\PY{p}{(}\PY{p}{)}\PY{p}{]}\PY{p}{)}
         \PY{n}{calFb}\PY{o}{.}\PY{n}{setBiasParams}\PY{p}{(}\PY{p}{[}\PY{n}{r}\PY{o}{.}\PY{n}{lam} 
            \PY{k}{for} \PY{n}{r} \PY{o+ow}{in} \PY{n}{spctAssist}\PY{o}{.}\PY{n}{resonances}\PY{p}{(}\PY{p}{)}\PY{p}{]}\PY{p}{)}
\end{Verbatim}

    \begin{Verbatim}[commandchars=\\\{\}]
{\color{incolor}In [{\color{incolor}49}]:} \PY{c+c1}{\PYZsh{}\PYZsh{} Validation}
         \PY{n}{threshold} \PY{o}{=} \PY{l+m+mf}{0.01}
         
         \PY{n}{calBias} \PY{o}{=} \PY{p}{[}\PY{n}{r}\PY{o}{.}\PY{n}{lam} 
            \PY{k}{for} \PY{n}{r} \PY{o+ow}{in} \PY{n}{spctAssist}\PY{o}{.}\PY{n}{resonances}\PY{p}{(}\PY{p}{)}\PY{p}{]}
         \PY{n}{virtBias} \PY{o}{=} \PY{n}{np}\PY{o}{.}\PY{n}{array}\PY{p}{(}\PY{n}{wlChannels}\PY{p}{)}
         \PY{n}{diff} \PY{o}{=} \PY{n}{np}\PY{o}{.}\PY{n}{absolute}\PY{p}{(}\PY{n}{calBias} \PY{o}{\PYZhy{}} \PY{n}{virtBias}\PY{p}{)}
         \PY{n+nb}{print}\PY{p}{(}\PY{n}{diff}\PY{p}{)}
         \PY{k}{for} \PY{n}{d} \PY{o+ow}{in} \PY{n}{diff}\PY{p}{:}
             \PY{k}{assert} \PY{n}{d} \PY{o}{\PYZlt{}} \PY{n}{threshold}
\end{Verbatim}

    \begin{Verbatim}[commandchars=\\\{\}]
[ 0.00308762  0.00493711  0.00198601]

    \end{Verbatim}

    \begin{Verbatim}[commandchars=\\\{\}]
{\color{incolor}In [{\color{incolor}50}]:} \PY{c+c1}{\PYZsh{} NOW Add filter shapes, after shifting to new bias}
         \PY{k}{for} \PY{n}{i} \PY{o+ow}{in} \PY{n+nb}{range}\PY{p}{(}\PY{n+nb}{len}\PY{p}{(}\PY{n}{axonCurves}\PY{p}{)}\PY{p}{)}\PY{p}{:}
             \PY{n}{oldCurve} \PY{o}{=} \PY{n}{axonCurves}\PY{p}{[}\PY{n}{i}\PY{p}{]}
             \PY{n}{shift} \PY{o}{=} \PY{n}{calBias}\PY{p}{[}\PY{n}{i}\PY{p}{]} \PY{o}{\PYZhy{}} \PY{n}{np}\PY{o}{.}\PY{n}{mean}\PY{p}{(}\PY{n}{oldCurve}\PY{p}{[}\PY{l+m+mi}{0}\PY{p}{]}\PY{p}{)}
             \PY{n}{newCurve} \PY{o}{=} \PY{p}{(}\PY{n}{np}\PY{o}{.}\PY{n}{add}\PY{p}{(}\PY{n}{oldCurve}\PY{p}{[}\PY{l+m+mi}{0}\PY{p}{]}\PY{p}{,} \PY{n}{shift}\PY{p}{)}\PY{p}{,} \PY{n}{oldCurve}\PY{p}{[}\PY{l+m+mi}{1}\PY{p}{]}\PY{p}{)}
             \PY{n}{axonCurves}\PY{p}{[}\PY{n}{i}\PY{p}{]} \PY{o}{=} \PY{n}{newCurve}
             
         \PY{k}{for} \PY{n}{i} \PY{o+ow}{in} \PY{n+nb}{range}\PY{p}{(}\PY{n+nb}{len}\PY{p}{(}\PY{n}{dendriteCurves}\PY{p}{)}\PY{p}{)}\PY{p}{:}
             \PY{n}{oldCurve} \PY{o}{=} \PY{n}{dendriteCurves}\PY{p}{[}\PY{n}{i}\PY{p}{]}
             \PY{n}{shift} \PY{o}{=} \PY{n}{calBias}\PY{p}{[}\PY{n}{i}\PY{p}{]} \PY{o}{\PYZhy{}} \PY{n}{np}\PY{o}{.}\PY{n}{mean}\PY{p}{(}\PY{n}{oldCurve}\PY{p}{[}\PY{l+m+mi}{0}\PY{p}{]}\PY{p}{)}
             \PY{n}{newCurve} \PY{o}{=} \PY{p}{(}\PY{n}{np}\PY{o}{.}\PY{n}{add}\PY{p}{(}\PY{n}{oldCurve}\PY{p}{[}\PY{l+m+mi}{0}\PY{p}{]}\PY{p}{,} \PY{n}{shift}\PY{p}{)}\PY{p}{,} \PY{n}{oldCurve}\PY{p}{[}\PY{l+m+mi}{1}\PY{p}{]}\PY{p}{)}
             \PY{n}{dendriteCurves}\PY{p}{[}\PY{n}{i}\PY{p}{]} \PY{o}{=} \PY{n}{newCurve}
             
         \PY{n}{calAxon}\PY{o}{.}\PY{n}{setCurve}\PY{p}{(}\PY{n}{axonCurves}\PY{p}{,} \PY{n}{MrrOut}\PY{o}{.}\PY{n}{kThru}\PY{p}{)}
         \PY{n}{calFb}\PY{o}{.}\PY{n}{setCurve}\PY{p}{(}\PY{n}{dendriteCurves}\PY{p}{,} \PY{n}{MrrOut}\PY{o}{.}\PY{n}{kThru}\PY{p}{)}
\end{Verbatim}

    \begin{Verbatim}[commandchars=\\\{\}]
{\color{incolor}In [{\color{incolor}51}]:} \PY{c+c1}{\PYZsh{} Hot\PYZhy{}Swap in Calibration Module and make sure things look good}
         \PY{n}{inst}\PY{o}{.}\PY{n}{lockPhony}\PY{p}{(}\PY{n}{calCsc}\PY{p}{)}
         \PY{n}{calCsc}\PY{o}{.}\PY{n}{setOsaOut}\PY{p}{(}\PY{n}{MrrOut}\PY{o}{.}\PY{n}{kThru}\PY{o}{.}\PY{n}{value}\PY{p}{)}
         \PY{n}{spctAssist}\PY{o}{.}\PY{n}{fgResPlot}\PY{p}{(}\PY{p}{)}
         \PY{n}{plt}\PY{o}{.}\PY{n}{show}\PY{p}{(}\PY{p}{)}
         \PY{n}{inst}\PY{o}{.}\PY{n}{releasePhony}\PY{p}{(}\PY{p}{)}
\end{Verbatim}

    
    \subsubsection{Step 7: Fill K Matrix}\label{step-7-fill-k-matrix}

K is the coefficients between mW and deltaWL.

    \begin{Verbatim}[commandchars=\\\{\}]
{\color{incolor}In [{\color{incolor}52}]:} \PY{n}{biasTune} \PY{o}{=} \PY{n}{newTune}\PY{o}{.}\PY{n}{copy}\PY{p}{(}\PY{p}{)}
\end{Verbatim}

    \begin{Verbatim}[commandchars=\\\{\}]
{\color{incolor}In [{\color{incolor}75}]:} \PY{k}{def} \PY{n+nf}{partialK}\PY{p}{(}\PY{n}{wlTarget}\PY{p}{,} \PY{n}{primNum}\PY{p}{,} \PY{n}{axonChan}\PY{p}{,} \PY{n}{dendChan}\PY{p}{,} 
    \PY{n}{tEstMap}\PY{p}{,} \PY{n}{biasTune}\PY{p}{,} \PY{n}{spctAssist}\PY{p}{,} \PY{n}{avgCnt}\PY{p}{,} \PY{n}{nPts}\PY{p}{)}\PY{p}{:}
             \PY{l+s+sd}{\PYZsq{}\PYZsq{}\PYZsq{}}
         \PY{l+s+sd}{    For each wlTarget, retuns 2 rows of K.}
         \PY{l+s+sd}{    rows: axon, dendrite}
         \PY{l+s+sd}{    cols: a\PYZhy{}d alternating in order provided by}
         \PY{l+s+sd}{    axonChan and dendChan}
         \PY{l+s+sd}{    \PYZsq{}\PYZsq{}\PYZsq{}}
             \PY{n}{newSp} \PY{o}{=} \PY{n}{copy}\PY{o}{.}\PY{n}{deepcopy}\PY{p}{(}\PY{n}{spctAssist}\PY{p}{)}
             \PY{n}{newSp}\PY{o}{.}\PY{n}{wlRange} \PY{o}{=} \PY{p}{[}\PY{n}{wlTarget}\PY{o}{\PYZhy{}}\PY{l+m+mi}{1}\PY{p}{,} \PY{n}{wlTarget}\PY{o}{+}\PY{l+m+mi}{1}\PY{p}{]}
             \PY{n}{newSp}\PY{o}{.}\PY{n}{nChan} \PY{o}{=} \PY{l+m+mi}{2}
             \PY{n}{nowTune} \PY{o}{=} \PY{n}{biasTune}\PY{o}{.}\PY{n}{copy}\PY{p}{(}\PY{p}{)}
             \PY{n}{biasWL} \PY{o}{=} \PY{n}{np}\PY{o}{.}\PY{n}{multiply}\PY{p}{(}\PY{n}{wlTarget}\PY{p}{,} \PY{n}{np}\PY{o}{.}\PY{n}{ones}\PY{p}{(}\PY{l+m+mi}{2}\PY{p}{)}\PY{p}{)}
             \PY{n}{KList} \PY{o}{=} \PY{p}{[}\PY{k+kc}{None}\PY{p}{]} \PY{o}{*} \PY{n+nb}{len}\PY{p}{(}\PY{n}{axonChan}\PY{p}{)}
             \PY{k}{assert} \PY{n+nb}{len}\PY{p}{(}\PY{n}{axonChan}\PY{p}{)} \PY{o}{==} \PY{n+nb}{len}\PY{p}{(}\PY{n}{dendChan}\PY{p}{)}
             
             \PY{c+c1}{\PYZsh{} Step 1: Do primary square}
             \PY{n}{square} \PY{o}{=} \PY{n}{np}\PY{o}{.}\PY{n}{zeros}\PY{p}{(}\PY{p}{(}\PY{l+m+mi}{2}\PY{p}{,} \PY{l+m+mi}{2}\PY{p}{)}\PY{p}{)}
             \PY{k}{for} \PY{n}{ich}\PY{p}{,} \PY{n}{ch} \PY{o+ow}{in} \PY{n+nb}{enumerate}\PY{p}{(}\PY{p}{[}\PY{n}{axonChan}\PY{p}{[}\PY{n}{primNum}\PY{p}{]}\PY{p}{,}
                \PY{n}{dendChan}\PY{p}{[}\PY{n}{primNum}\PY{p}{]}\PY{p}{]}\PY{p}{)}\PY{p}{:}
                 \PY{c+c1}{\PYZsh{} Shift in a 0.75nm range around bias WL}
                 \PY{n}{dB} \PY{o}{=} \PY{l+m+mf}{0.75} \PY{o}{/} \PY{n}{tEstMap}\PY{p}{[}\PY{n}{ch}\PY{p}{]}
                 \PY{n}{x} \PY{o}{=} \PY{n}{np}\PY{o}{.}\PY{n}{linspace}\PY{p}{(}\PY{n+nb}{max}\PY{p}{(}\PY{l+m+mf}{0.0}\PY{p}{,}
                    \PY{p}{(}\PY{n}{biasTune}\PY{p}{[}\PY{n}{ch}\PY{p}{]}\PY{o}{\PYZhy{}}\PY{n}{dB}\PY{p}{)}\PY{p}{)}\PY{p}{,} \PY{n}{biasTune}\PY{p}{[}\PY{n}{ch}\PY{p}{]}\PY{o}{+}\PY{n}{dB}\PY{p}{,} \PY{n}{nPts}\PY{p}{)}
                 \PY{n}{y} \PY{o}{=} \PY{n}{np}\PY{o}{.}\PY{n}{zeros}\PY{p}{(}\PY{p}{(}\PY{l+m+mi}{2}\PY{p}{,} \PY{n}{nPts}\PY{p}{)}\PY{p}{)}
                 \PY{k}{for} \PY{n}{ipt}\PY{p}{,} \PY{n}{pt} \PY{o+ow}{in} \PY{n+nb}{enumerate}\PY{p}{(}\PY{n}{x}\PY{p}{)}\PY{p}{:}
                     \PY{k}{if} \PY{n}{pt} \PY{o}{==} \PY{n}{biasTune}\PY{p}{[}\PY{n}{ch}\PY{p}{]}\PY{p}{:}
                         \PY{k}{continue}
                     \PY{n}{nowTune}\PY{p}{[}\PY{n}{ch}\PY{p}{]} \PY{o}{=} \PY{n}{pt}
                     \PY{n}{inst}\PY{o}{.}\PY{n}{setCurrentChanTuning}\PY{p}{(}\PY{n}{nowTune}\PY{p}{,} \PY{n}{token}\PY{p}{,}
                        \PY{n}{CurrentUnit}\PY{o}{.}\PY{n}{mW}\PY{p}{)}
                     \PY{n}{nowWL} \PY{o}{=} \PY{n}{np}\PY{o}{.}\PY{n}{array}\PY{p}{(}\PY{p}{[}\PY{n}{r}\PY{o}{.}\PY{n}{lam}
                        \PY{k}{for} \PY{n}{r} \PY{o+ow}{in} \PY{n}{newSp}\PY{o}{.}\PY{n}{resonances}\PY{p}{(}\PY{p}{)}\PY{p}{]}\PY{p}{)}
                     \PY{n}{diff} \PY{o}{=} \PY{n}{np}\PY{o}{.}\PY{n}{sort}\PY{p}{(}\PY{n}{nowWL} \PY{o}{\PYZhy{}} \PY{n}{biasWL}\PY{p}{)}
                     \PY{c+c1}{\PYZsh{} Catch Negative Case}
                     \PY{k}{if} \PY{n}{np}\PY{o}{.}\PY{n}{absolute}\PY{p}{(}\PY{n}{diff}\PY{p}{[}\PY{l+m+mi}{1}\PY{p}{]}\PY{p}{)} \PY{o}{\PYZlt{}} \PY{n}{np}\PY{o}{.}\PY{n}{absolute}\PY{p}{(}\PY{n}{diff}\PY{p}{[}\PY{l+m+mi}{0}\PY{p}{]}\PY{p}{)}\PY{p}{:}
                         \PY{n}{tmp} \PY{o}{=} \PY{n}{diff}\PY{p}{[}\PY{l+m+mi}{1}\PY{p}{]}
                         \PY{n}{diff}\PY{p}{[}\PY{l+m+mi}{1}\PY{p}{]} \PY{o}{=} \PY{n}{diff}\PY{p}{[}\PY{l+m+mi}{0}\PY{p}{]}
                         \PY{n}{diff}\PY{p}{[}\PY{l+m+mi}{0}\PY{p}{]} \PY{o}{=} \PY{n}{tmp}
                     \PY{n}{y}\PY{p}{[}\PY{n}{ich}\PY{p}{,} \PY{n}{ipt}\PY{p}{]} \PY{o}{=} \PY{n}{diff}\PY{p}{[}\PY{l+m+mi}{1}\PY{p}{]}
                     \PY{n}{y}\PY{p}{[}\PY{l+m+mi}{1}\PY{o}{\PYZhy{}}\PY{n}{ich}\PY{p}{,} \PY{n}{ipt}\PY{p}{]} \PY{o}{=} \PY{n}{diff}\PY{p}{[}\PY{l+m+mi}{0}\PY{p}{]}
                     
                 \PY{c+c1}{\PYZsh{} Make Linear Fit}
             
                 \PY{k}{for} \PY{n}{r} \PY{o+ow}{in} \PY{n+nb}{range}\PY{p}{(}\PY{l+m+mi}{2}\PY{p}{)}\PY{p}{:}
                     \PY{n}{yr} \PY{o}{=} \PY{n}{y}\PY{p}{[}\PY{n}{r}\PY{p}{,} \PY{p}{:}\PY{p}{]}
                     \PY{n}{p} \PY{o}{=} \PY{n}{np}\PY{o}{.}\PY{n}{polyfit}\PY{p}{(}\PY{n}{x}\PY{p}{,} \PY{n}{yr}\PY{p}{,} \PY{l+m+mi}{1}\PY{p}{)}
                     \PY{n}{square}\PY{p}{[}\PY{n}{r}\PY{p}{,} \PY{n}{ich}\PY{p}{]} \PY{o}{=} \PY{n+nb}{max}\PY{p}{(}\PY{n}{p}\PY{p}{[}\PY{l+m+mi}{0}\PY{p}{]}\PY{p}{,} \PY{l+m+mf}{0.0}\PY{p}{)}
                 
                 \PY{k}{if} \PY{n}{ich} \PY{o}{==} \PY{l+m+mi}{1}\PY{p}{:}
                     \PY{c+c1}{\PYZsh{} Change bias to separate two peaks}
                     \PY{n}{biasTune} \PY{o}{=} \PY{n}{nowTune}\PY{o}{.}\PY{n}{copy}\PY{p}{(}\PY{p}{)}
                     \PY{n}{biasWL} \PY{o}{=} \PY{n}{nowWL}
                     
                 \PY{n}{nowTune}\PY{p}{[}\PY{n}{ch}\PY{p}{]} \PY{o}{=} \PY{n}{biasTune}\PY{p}{[}\PY{n}{ch}\PY{p}{]}

             \PY{n}{KList}\PY{p}{[}\PY{n}{primNum}\PY{p}{]} \PY{o}{=} \PY{n}{square}
             
             \PY{c+c1}{\PYZsh{} Step 2: Do all Other Squares}
             \PY{n}{nPts} \PY{o}{=} \PY{n}{nPts} \PY{o}{+} \PY{l+m+mi}{6}
             
             \PY{k}{for} \PY{n}{num} \PY{o+ow}{in} \PY{n+nb}{range}\PY{p}{(}\PY{n+nb}{len}\PY{p}{(}\PY{n}{KList}\PY{p}{)}\PY{p}{)}\PY{p}{:}
                 \PY{k}{if} \PY{n}{num} \PY{o}{==} \PY{n}{primNum}\PY{p}{:}
                     \PY{k}{continue}
                 \PY{n}{square} \PY{o}{=} \PY{n}{np}\PY{o}{.}\PY{n}{zeros}\PY{p}{(}\PY{p}{(}\PY{l+m+mi}{2}\PY{p}{,} \PY{l+m+mi}{2}\PY{p}{)}\PY{p}{)}
                 \PY{k}{for} \PY{n}{ich}\PY{p}{,} \PY{n}{ch} \PY{o+ow}{in} \PY{n+nb}{enumerate}\PY{p}{(}\PY{p}{[}\PY{n}{axonChan}\PY{p}{[}\PY{n}{num}\PY{p}{]}\PY{p}{,} \PY{n}{dendChan}\PY{p}{[}\PY{n}{num}\PY{p}{]}\PY{p}{]}\PY{p}{)}\PY{p}{:}
                     \PY{c+c1}{\PYZsh{} Shift in a 1.0nm range around bias WL}
                     \PY{n}{dB} \PY{o}{=} \PY{l+m+mf}{1.25} \PY{o}{/} \PY{n}{tEstMap}\PY{p}{[}\PY{n}{ch}\PY{p}{]}
                     \PY{n}{x} \PY{o}{=} \PY{n}{np}\PY{o}{.}\PY{n}{linspace}\PY{p}{(}\PY{n+nb}{max}\PY{p}{(}\PY{l+m+mf}{0.0}\PY{p}{,}
                        \PY{p}{(}\PY{n}{biasTune}\PY{p}{[}\PY{n}{ch}\PY{p}{]}\PY{o}{\PYZhy{}}\PY{n}{dB}\PY{p}{)}\PY{p}{)}\PY{p}{,} \PY{n}{biasTune}\PY{p}{[}\PY{n}{ch}\PY{p}{]}\PY{o}{+}\PY{n}{dB}\PY{p}{,} \PY{n}{nPts}\PY{p}{)}
                     \PY{n}{y} \PY{o}{=} \PY{n}{np}\PY{o}{.}\PY{n}{zeros}\PY{p}{(}\PY{p}{(}\PY{l+m+mi}{2}\PY{p}{,} \PY{n}{nPts}\PY{p}{)}\PY{p}{)}
                     \PY{k}{for} \PY{n}{ipt}\PY{p}{,} \PY{n}{pt} \PY{o+ow}{in} \PY{n+nb}{enumerate}\PY{p}{(}\PY{n}{x}\PY{p}{)}\PY{p}{:}
                         \PY{k}{if} \PY{n}{pt} \PY{o}{==} \PY{n}{biasTune}\PY{p}{[}\PY{n}{ch}\PY{p}{]}\PY{p}{:}
                             \PY{k}{continue}
                         \PY{n}{nowTune}\PY{p}{[}\PY{n}{ch}\PY{p}{]} \PY{o}{=} \PY{n}{pt}
                         \PY{n}{inst}\PY{o}{.}\PY{n}{setCurrentChanTuning}\PY{p}{(}\PY{n}{nowTune}\PY{p}{,} \PY{n}{token}\PY{p}{,}
                            \PY{n}{CurrentUnit}\PY{o}{.}\PY{n}{mW}\PY{p}{)}
                         \PY{n}{nowWL} \PY{o}{=} \PY{n}{np}\PY{o}{.}\PY{n}{array}\PY{p}{(}\PY{p}{[}\PY{n}{r}\PY{o}{.}\PY{n}{lam}
                            \PY{k}{for} \PY{n}{r} \PY{o+ow}{in} \PY{n}{newSp}\PY{o}{.}\PY{n}{resonances}\PY{p}{(}\PY{p}{)}\PY{p}{]}\PY{p}{)}
                         \PY{n}{diff} \PY{o}{=} \PY{n}{nowWL} \PY{o}{\PYZhy{}} \PY{n}{biasWL}
                         \PY{n}{y}\PY{p}{[}\PY{p}{:}\PY{p}{,} \PY{n}{ipt}\PY{p}{]} \PY{o}{=} \PY{n}{diff}
         
                     \PY{c+c1}{\PYZsh{} Make Linear Fit}
         
                     \PY{k}{for} \PY{n}{r} \PY{o+ow}{in} \PY{n+nb}{range}\PY{p}{(}\PY{l+m+mi}{2}\PY{p}{)}\PY{p}{:}
                         \PY{n}{yr} \PY{o}{=} \PY{n}{y}\PY{p}{[}\PY{n}{r}\PY{p}{,} \PY{p}{:}\PY{p}{]}
                         \PY{n}{p} \PY{o}{=} \PY{n}{np}\PY{o}{.}\PY{n}{polyfit}\PY{p}{(}\PY{n}{x}\PY{p}{,} \PY{n}{yr}\PY{p}{,} \PY{l+m+mi}{1}\PY{p}{)}
                         \PY{n}{square}\PY{p}{[}\PY{n}{r}\PY{p}{,} \PY{n}{ich}\PY{p}{]} \PY{o}{=} \PY{n+nb}{max}\PY{p}{(}\PY{n}{p}\PY{p}{[}\PY{l+m+mi}{0}\PY{p}{]}\PY{p}{,} \PY{l+m+mf}{0.0}\PY{p}{)}
         
                     \PY{k}{if} \PY{n}{ich} \PY{o}{==} \PY{l+m+mi}{1}\PY{p}{:}
                         \PY{c+c1}{\PYZsh{} Change bias to separate two peaks}
                         \PY{n}{biasTune} \PY{o}{=} \PY{n}{nowTune}\PY{o}{.}\PY{n}{copy}\PY{p}{(}\PY{p}{)}
         
                     \PY{n}{nowTune}\PY{p}{[}\PY{n}{ch}\PY{p}{]} \PY{o}{=} \PY{n}{biasTune}\PY{p}{[}\PY{n}{ch}\PY{p}{]}
                     
                 \PY{n}{KList}\PY{p}{[}\PY{n}{num}\PY{p}{]} \PY{o}{=} \PY{n}{square}
                 
             \PY{k}{return} \PY{n}{np}\PY{o}{.}\PY{n}{hstack}\PY{p}{(}\PY{n}{KList}\PY{p}{)}
\end{Verbatim}

    \begin{Verbatim}[commandchars=\\\{\}]
{\color{incolor}In [{\color{incolor}76}]:} \PY{n}{rows} \PY{o}{=} \PY{n+nb}{list}\PY{p}{(}\PY{p}{)}
         \PY{k}{for} \PY{n}{j}\PY{p}{,} \PY{n}{wl} \PY{o+ow}{in} \PY{n+nb}{enumerate}\PY{p}{(}\PY{n}{wlChannels}\PY{p}{)}\PY{p}{:}
             \PY{n}{retRow} \PY{o}{=} \PY{n}{partialK}\PY{p}{(}\PY{n}{wl}\PY{p}{,} \PY{n}{j}\PY{p}{,} \PY{n}{calAxonChan}\PY{p}{,} \PY{n}{calDendriteChan}\PY{p}{,}
                \PY{n}{tEstMap}\PY{p}{,} \PY{n}{biasTune}\PY{p}{,} \PY{n}{spctAssist}\PY{p}{,} \PY{n}{avgCnt}\PY{o}{=}\PY{l+m+mi}{5}\PY{p}{,} \PY{n}{nPts}\PY{o}{=}\PY{l+m+mi}{7}\PY{p}{)}
             \PY{n}{rows}\PY{o}{.}\PY{n}{append}\PY{p}{(}\PY{n}{retRow}\PY{p}{)}
         \PY{n}{newK} \PY{o}{=} \PY{n}{np}\PY{o}{.}\PY{n}{vstack}\PY{p}{(}\PY{n}{rows}\PY{p}{)}
         
         \PY{k}{def} \PY{n+nf}{swapK}\PY{p}{(}\PY{n}{i}\PY{p}{,} \PY{n}{j}\PY{p}{)}\PY{p}{:}
             \PY{n}{newK}\PY{p}{[}\PY{p}{:}\PY{p}{,}\PY{p}{[}\PY{n}{i}\PY{p}{,} \PY{n}{j}\PY{p}{]}\PY{p}{]} \PY{o}{=} \PY{n}{newK}\PY{p}{[}\PY{p}{:}\PY{p}{,}\PY{p}{[}\PY{n}{j}\PY{p}{,} \PY{n}{i}\PY{p}{]}\PY{p}{]}
             \PY{n}{newK}\PY{p}{[}\PY{p}{[}\PY{n}{i}\PY{p}{,} \PY{n}{j}\PY{p}{]}\PY{p}{,} \PY{p}{:}\PY{p}{]} \PY{o}{=} \PY{n}{newK}\PY{p}{[}\PY{p}{[}\PY{n}{j}\PY{p}{,} \PY{n}{i}\PY{p}{]}\PY{p}{,} \PY{p}{:}\PY{p}{]}
         
         \PY{c+c1}{\PYZsh{} At ADADAD}
         \PY{c+c1}{\PYZsh{} Want AAADDD}
         \PY{c+c1}{\PYZsh{} Swap rows/cols 1 and 2 (AADDAD)}
         \PY{n}{swapK}\PY{p}{(}\PY{l+m+mi}{1}\PY{p}{,} \PY{l+m+mi}{2}\PY{p}{)}
         \PY{c+c1}{\PYZsh{} Swap rows/cols 3 and 4 (AADADD)}
         \PY{n}{swapK}\PY{p}{(}\PY{l+m+mi}{3}\PY{p}{,} \PY{l+m+mi}{4}\PY{p}{)}
         \PY{c+c1}{\PYZsh{} Swap rows/cols 2 and 3 (AAADDD)}
         \PY{n}{swapK}\PY{p}{(}\PY{l+m+mi}{2}\PY{p}{,} \PY{l+m+mi}{3}\PY{p}{)}
         
         \PY{c+c1}{\PYZsh{} Swap Middle Rows and Columns to order axons then dendrites}
         
         \PY{c+c1}{\PYZsh{} Add to thermalgroup}
         \PY{n}{calTherm}\PY{o}{.}\PY{n}{setK}\PY{p}{(}\PY{n}{newK}\PY{p}{)}
\end{Verbatim}

    \begin{Verbatim}[commandchars=\\\{\}]
{\color{incolor}In [{\color{incolor}77}]:} \PY{n+nb}{print}\PY{p}{(}\PY{l+s+s2}{\PYZdq{}}\PY{l+s+s2}{K:}\PY{l+s+s2}{\PYZdq{}}\PY{p}{)}
         \PY{n+nb}{print}\PY{p}{(}\PY{n}{np}\PY{o}{.}\PY{n}{round}\PY{p}{(}\PY{n}{K}\PY{p}{)}\PY{p}{)}
         \PY{n+nb}{print}\PY{p}{(}\PY{p}{)}
         
         \PY{n+nb}{print}\PY{p}{(}\PY{l+s+s2}{\PYZdq{}}\PY{l+s+s2}{New K:}\PY{l+s+s2}{\PYZdq{}}\PY{p}{)}
         \PY{n+nb}{print}\PY{p}{(}\PY{n}{np}\PY{o}{.}\PY{n}{round}\PY{p}{(}\PY{n}{newK}\PY{p}{)}\PY{p}{)}
         \PY{n+nb}{print}\PY{p}{(}\PY{p}{)}

         \PY{n+nb}{print}\PY{p}{(}\PY{l+s+s2}{\PYZdq{}}\PY{l+s+s2}{Round Error:}\PY{l+s+s2}{\PYZdq{}}\PY{p}{)}
         \PY{n+nb}{print}\PY{p}{(}\PY{n}{np}\PY{o}{.}\PY{n}{round}\PY{p}{(}\PY{n}{np}\PY{o}{.}\PY{n}{absolute}\PY{p}{(}\PY{n}{newK} \PY{o}{\PYZhy{}} \PY{n}{K}\PY{p}{)}\PY{p}{)}\PY{p}{)}
         \PY{n+nb}{print}\PY{p}{(}\PY{p}{)}
         
         \PY{n+nb}{print}\PY{p}{(}\PY{l+s+s2}{\PYZdq{}}\PY{l+s+s2}{Absolute Error:}\PY{l+s+s2}{\PYZdq{}}\PY{p}{)}
         \PY{n+nb}{print}\PY{p}{(}\PY{n}{np}\PY{o}{.}\PY{n}{absolute}\PY{p}{(}\PY{n}{newK}\PY{o}{\PYZhy{}}\PY{n}{K}\PY{p}{)}\PY{p}{)}
\end{Verbatim}

    \begin{Verbatim}[commandchars=\\\{\}]
K:
[[ 19.   1.   1.   1.   0.   0.]
 [  1.  21.   2.   1.   0.   0.]
 [  0.   2.  20.   0.   1.   0.]
 [  0.   1.   1.  20.   1.   0.]
 [  1.   1.   1.   0.  20.   1.]
 [  0.   1.   0.   0.   0.  20.]]

New K:
[[ 19.   1.   1.   1.   0.   0.]
 [  1.  21.   2.   0.   0.   0.]
 [  0.   2.  20.   0.   0.   0.]
 [  0.   1.   1.  20.   1.   0.]
 [  1.   1.   0.   0.  20.   1.]
 [  0.   1.   0.   0.   0.  20.]]

Round Error:
[[ 0.  0.  0.  0.  0.  0.]
 [ 0.  0.  0.  1.  0.  0.]
 [ 0.  0.  0.  0.  1.  0.]
 [ 0.  0.  0.  0.  0.  0.]
 [ 0.  0.  1.  0.  0.  0.]
 [ 0.  0.  0.  0.  0.  0.]]

Absolute Error:
[[ 0.02004703  0.00661128  0.00038421  0.0148798   0.00423898  0.02573439]
 [ 0.00783713  0.02285229  0.11606766  0.60466732  0.05026481  0.44121404]
 [ 0.00027619  0.00252188  0.04469797  0.01197408  1.06470847  0.02240962]
 [ 0.00851921  0.18221379  0.01600566  0.03894927  0.00990297  0.01298121]
 [ 0.0043928   0.02412916  1.11338314  0.05033487  0.04016287  0.47670709]
 [ 0.00217652  0.00233603  0.02640214  0.02192149  0.20880726  0.00609685]]

    \end{Verbatim}

    \textbf{And calCsc is a calibrated device!}

Validation was done at the following points: * Ascription is guaranteed
to be correct. * HeatBias is within 0.01mW * Wavelength bias is within
0.01nm * Check K cross-terms to make sure error is 1 or 0, ESPECIALLY
along diagonals

Note that these validation constraints \textbf{degrade} as the cross
terms between axons and dendrites increase. Fortunately, those
cross-terms will not be significant on the actual chip.

    \begin{Verbatim}[commandchars=\\\{\}]
{\color{incolor}In [{\color{incolor} }]:} 
\end{Verbatim}

%% file: 2-3-1_FFNet_Backprop.tex
    \subsubsection{Step 1: Initialization}\label{step-1-initializations}

\begin{itemize}
\tightlist
\item
  Import anything necessary.
\item
  Pull Neural Network measured thermal parameters (thK and Ib).
\item
  Define activation function and load it into tensorflow.
\end{itemize}

    \begin{Verbatim}[commandchars=\\\{\}]
{\color{incolor}In [{\color{incolor}43}]:} \PY{k+kn}{import} \PY{n+nn}{tensorflow} \PY{k}{as} \PY{n+nn}{tf}
         \PY{k+kn}{import} \PY{n+nn}{numpy} \PY{k}{as} \PY{n+nn}{np}
         \PY{k+kn}{import} \PY{n+nn}{matplotlib}\PY{n+nn}{.}\PY{n+nn}{pyplot} \PY{k}{as} \PY{n+nn}{plt}
         \PY{k+kn}{from} \PY{n+nn}{lightlab}\PY{n+nn}{.}\PY{n+nn}{util}\PY{n+nn}{.}\PY{n+nn}{modeling} \PY{k}{import} \PY{n}{lorentz}
         \PY{k+kn}{from} \PY{n+nn}{tensorflow}\PY{n+nn}{.}\PY{n+nn}{python}\PY{n+nn}{.}\PY{n+nn}{framework} \PY{k}{import} \PY{n}{ops}
\end{Verbatim}

    \begin{Verbatim}[commandchars=\\\{\}]
{\color{incolor}In [{\color{incolor}34}]:} \PY{c+c1}{\PYZsh{} Step 1: Required Parameters}
         
         \PY{c+c1}{\PYZsh{} K Diagonals}
         \PY{n}{thK} \PY{o}{=} \PY{l+m+mf}{20.0}
         \PY{c+c1}{\PYZsh{} 0 Current Bias (mA)}
         \PY{n}{Ib} \PY{o}{=} \PY{l+m+mf}{6.0}
\end{Verbatim}

    \begin{Verbatim}[commandchars=\\\{\}]
{\color{incolor}In [{\color{incolor}42}]:} \PY{c+c1}{\PYZsh{} Step 2: Transfer Function, approximate as lorentzian}
         \PY{n}{nm}\PY{p}{,} \PY{n}{lin} \PY{o}{=} \PY{n}{lorentz}\PY{p}{(}\PY{p}{[}\PY{o}{\PYZhy{}}\PY{l+m+mf}{0.5}\PY{p}{,} \PY{l+m+mf}{0.5}\PY{p}{]}\PY{p}{,} \PY{l+m+mi}{0}\PY{p}{,} \PY{l+m+mf}{0.1}\PY{p}{,} \PY{l+m+mf}{0.98}\PY{p}{)}
         \PY{n}{lin} \PY{o}{=} \PY{l+m+mf}{1.0}\PY{o}{\PYZhy{}}\PY{n}{lin}
         
         \PY{k}{def} \PY{n+nf}{h}\PY{p}{(}\PY{n}{x}\PY{p}{)}\PY{p}{:}
             \PY{k}{return} \PY{n}{np}\PY{o}{.}\PY{n}{interp}\PY{p}{(}\PY{n}{x}\PY{p}{,} \PY{n}{nm}\PY{p}{,} \PY{n}{lin}\PY{p}{)}
         
         \PY{k}{def} \PY{n+nf}{dh}\PY{p}{(}\PY{n}{x}\PY{p}{)}\PY{p}{:}
             \PY{k}{return} \PY{n}{np}\PY{o}{.}\PY{n}{interp}\PY{p}{(}\PY{n}{x}\PY{p}{,} \PY{n}{nm}\PY{p}{,} \PY{n}{np}\PY{o}{.}\PY{n}{gradient}\PY{p}{(}\PY{n}{lin}\PY{p}{)}\PY{p}{)}
         
         \PY{k}{def} \PY{n+nf}{g}\PY{p}{(}\PY{n}{x}\PY{p}{)}\PY{p}{:}
             \PY{k}{return} \PY{n}{np}\PY{o}{.}\PY{n}{divide}\PY{p}{(}\PY{n}{thK}\PY{o}{*}\PY{n}{np}\PY{o}{.}\PY{n}{add}\PY{p}{(}\PY{n}{np}\PY{o}{.}\PY{n}{power}\PY{p}{(}\PY{n}{x}\PY{p}{,} \PY{l+m+mf}{2.0}\PY{p}{)}\PY{p}{,} \PY{l+m+mi}{2}\PY{o}{*}\PY{n}{Ib}\PY{o}{*}\PY{n}{x}\PY{p}{)}\PY{p}{,} \PY{l+m+mi}{1000}\PY{p}{)}
         
         \PY{k}{def} \PY{n+nf}{dg}\PY{p}{(}\PY{n}{x}\PY{p}{)}\PY{p}{:}
             \PY{k}{return} \PY{n}{np}\PY{o}{.}\PY{n}{divide}\PY{p}{(}\PY{n}{np}\PY{o}{.}\PY{n}{add}\PY{p}{(}\PY{l+m+mi}{2}\PY{o}{*}\PY{n}{thK}\PY{o}{*}\PY{n}{x}\PY{p}{,} \PY{l+m+mi}{2}\PY{o}{*}\PY{n}{thK}\PY{o}{*}\PY{n}{Ib}\PY{p}{)}\PY{p}{,} \PY{l+m+mi}{1000}\PY{p}{)}
         
         \PY{k}{def} \PY{n+nf}{f}\PY{p}{(}\PY{n}{x}\PY{p}{)}\PY{p}{:}
             \PY{k}{return} \PY{n}{h}\PY{p}{(}\PY{n}{g}\PY{p}{(}\PY{n}{x}\PY{p}{)}\PY{p}{)}
         
         \PY{k}{def} \PY{n+nf}{df}\PY{p}{(}\PY{n}{x}\PY{p}{)}\PY{p}{:}
             \PY{k}{return} \PY{n}{np}\PY{o}{.}\PY{n}{multiply}\PY{p}{(}\PY{n}{dh}\PY{p}{(}\PY{n}{g}\PY{p}{(}\PY{n}{x}\PY{p}{)}\PY{p}{)}\PY{p}{,} \PY{n}{dg}\PY{p}{(}\PY{n}{x}\PY{p}{)}\PY{p}{)}
         
         \PY{n}{x} \PY{o}{=} \PY{n}{np}\PY{o}{.}\PY{n}{linspace}\PY{p}{(}\PY{o}{\PYZhy{}}\PY{l+m+mi}{1}\PY{p}{,} \PY{l+m+mi}{1}\PY{p}{,} \PY{l+m+mi}{200}\PY{p}{)}
         \PY{n}{plt}\PY{o}{.}\PY{n}{plot}\PY{p}{(}\PY{n}{x}\PY{p}{,} \PY{n}{f}\PY{p}{(}\PY{n}{x}\PY{p}{)}\PY{p}{,} \PY{l+s+s1}{\PYZsq{}}\PY{l+s+s1}{b}\PY{l+s+s1}{\PYZsq{}}\PY{p}{)}
         \PY{n}{plt}\PY{o}{.}\PY{n}{xlim}\PY{p}{(}\PY{p}{[}\PY{o}{\PYZhy{}}\PY{l+m+mi}{1}\PY{p}{,} \PY{l+m+mi}{1}\PY{p}{]}\PY{p}{)}
         \PY{n}{plt}\PY{o}{.}\PY{n}{show}\PY{p}{(}\PY{p}{)}
         \PY{n}{plt}\PY{o}{.}\PY{n}{plot}\PY{p}{(}\PY{n}{x}\PY{p}{,} \PY{n}{df}\PY{p}{(}\PY{n}{x}\PY{p}{)}\PY{p}{,} \PY{l+s+s1}{\PYZsq{}}\PY{l+s+s1}{b}\PY{l+s+s1}{\PYZsq{}}\PY{p}{)}
         \PY{n}{plt}\PY{o}{.}\PY{n}{show}\PY{p}{(}\PY{p}{)}
\end{Verbatim}

    
    
    \begin{Verbatim}[commandchars=\\\{\}]
{\color{incolor}In [{\color{incolor}50}]:} \PY{c+c1}{\PYZsh{} Step 3: Add to TensorFlow}
         
         \PY{c+c1}{\PYZsh{} Py\PYZus{}Func Hack}
         \PY{c+c1}{(http://stackoverflow.com/questions/39921607}
         \PY{k}{def} \PY{n+nf}{py\PYZus{}func}\PY{p}{(}\PY{n}{func}\PY{p}{,} \PY{n}{inp}\PY{p}{,} \PY{n}{Tout}\PY{p}{,} \PY{n}{stateful}\PY{o}{=}\PY{k+kc}{True}\PY{p}{,} \PY{n}{name}\PY{o}{=}\PY{k+kc}{None}\PY{p}{,} \PY{n}{grad}\PY{o}{=}\PY{k+kc}{None}\PY{p}{)}\PY{p}{:}
         
             \PY{c+c1}{\PYZsh{} Need to generate a unique name to avoid duplicates:}
             \PY{n}{rnd\PYZus{}name} \PY{o}{=} \PY{l+s+s1}{\PYZsq{}}\PY{l+s+s1}{PyFuncGrad}\PY{l+s+s1}{\PYZsq{}} \PY{o}{+} \PY{n+nb}{str}\PY{p}{(}\PY{n}{np}\PY{o}{.}\PY{n}{random}\PY{o}{.}\PY{n}{randint}\PY{p}{(}\PY{l+m+mi}{0}\PY{p}{,} \PY{l+m+mf}{1E+8}\PY{p}{)}\PY{p}{)}
         
             \PY{n}{tf}\PY{o}{.}\PY{n}{RegisterGradient}\PY{p}{(}\PY{n}{rnd\PYZus{}name}\PY{p}{)}\PY{p}{(}\PY{n}{grad}\PY{p}{)}
             \PY{n}{g} \PY{o}{=} \PY{n}{tf}\PY{o}{.}\PY{n}{get\PYZus{}default\PYZus{}graph}\PY{p}{(}\PY{p}{)}
             \PY{k}{with} \PY{n}{g}\PY{o}{.}\PY{n}{gradient\PYZus{}override\PYZus{}map}\PY{p}{(}\PY{p}{\PYZob{}}\PY{l+s+s2}{\PYZdq{}}\PY{l+s+s2}{PyFunc}\PY{l+s+s2}{\PYZdq{}}\PY{p}{:} \PY{n}{rnd\PYZus{}name}\PY{p}{\PYZcb{}}\PY{p}{)}\PY{p}{:}
                 \PY{k}{return} \PY{n}{tf}\PY{o}{.}\PY{n}{py\PYZus{}func}\PY{p}{(}\PY{n}{func}\PY{p}{,}
                    \PY{n}{inp}\PY{p}{,} \PY{n}{Tout}\PY{p}{,} \PY{n}{stateful}\PY{o}{=}\PY{n}{stateful}\PY{p}{,} \PY{n}{name}\PY{o}{=}\PY{n}{name}\PY{p}{)}
             
         \PY{n}{np\PYZus{}df} \PY{o}{=} \PY{n}{np}\PY{o}{.}\PY{n}{vectorize}\PY{p}{(}\PY{n}{df}\PY{p}{)}
             
         \PY{n}{np\PYZus{}df\PYZus{}32} \PY{o}{=} \PY{k}{lambda} \PY{n}{x}\PY{p}{:} \PY{n}{np\PYZus{}df}\PY{p}{(}\PY{n}{x}\PY{p}{)}\PY{o}{.}\PY{n}{astype}\PY{p}{(}\PY{n}{np}\PY{o}{.}\PY{n}{float32}\PY{p}{)}
         
         \PY{k}{def} \PY{n+nf}{tf\PYZus{}df}\PY{p}{(}\PY{n}{x}\PY{p}{,}\PY{n}{name}\PY{o}{=}\PY{k+kc}{None}\PY{p}{)}\PY{p}{:}
             \PY{k}{with} \PY{n}{ops}\PY{o}{.}\PY{n}{op\PYZus{}scope}\PY{p}{(}\PY{p}{[}\PY{n}{x}\PY{p}{]}\PY{p}{,} \PY{n}{name}\PY{p}{,} \PY{l+s+s2}{\PYZdq{}}\PY{l+s+s2}{df}\PY{l+s+s2}{\PYZdq{}}\PY{p}{)} \PY{k}{as} \PY{n}{name}\PY{p}{:}
                 \PY{n}{y} \PY{o}{=} \PY{n}{tf}\PY{o}{.}\PY{n}{py\PYZus{}func}\PY{p}{(}\PY{n}{np\PYZus{}df\PYZus{}32}\PY{p}{,}
                                 \PY{p}{[}\PY{n}{x}\PY{p}{]}\PY{p}{,}
                                 \PY{p}{[}\PY{n}{tf}\PY{o}{.}\PY{n}{float32}\PY{p}{]}\PY{p}{,}
                                 \PY{n}{name}\PY{o}{=}\PY{n}{name}\PY{p}{,}
                                 \PY{n}{stateful}\PY{o}{=}\PY{k+kc}{False}\PY{p}{)}
                 \PY{k}{return} \PY{n}{y}\PY{p}{[}\PY{l+m+mi}{0}\PY{p}{]}
         
         \PY{k}{def} \PY{n+nf}{fgrad}\PY{p}{(}\PY{n}{op}\PY{p}{,} \PY{n}{grad}\PY{p}{)}\PY{p}{:}
             \PY{n}{x} \PY{o}{=} \PY{n}{op}\PY{o}{.}\PY{n}{inputs}\PY{p}{[}\PY{l+m+mi}{0}\PY{p}{]}
         
             \PY{n}{n\PYZus{}gr} \PY{o}{=} \PY{n}{tf\PYZus{}df}\PY{p}{(}\PY{n}{x}\PY{p}{)}
             \PY{k}{return} \PY{n}{grad} \PY{o}{*} \PY{n}{n\PYZus{}gr}
         
         \PY{n}{np\PYZus{}f} \PY{o}{=} \PY{n}{np}\PY{o}{.}\PY{n}{vectorize}\PY{p}{(}\PY{n}{f}\PY{p}{)}
         
         \PY{n}{np\PYZus{}f\PYZus{}32} \PY{o}{=} \PY{k}{lambda} \PY{n}{x}\PY{p}{:} \PY{n}{np\PYZus{}f}\PY{p}{(}\PY{n}{x}\PY{p}{)}\PY{o}{.}\PY{n}{astype}\PY{p}{(}\PY{n}{np}\PY{o}{.}\PY{n}{float32}\PY{p}{)}
         
         \PY{k}{def} \PY{n+nf}{tf\PYZus{}f}\PY{p}{(}\PY{n}{x}\PY{p}{,} \PY{n}{name}\PY{o}{=}\PY{k+kc}{None}\PY{p}{)}\PY{p}{:}
         
             \PY{k}{with} \PY{n}{ops}\PY{o}{.}\PY{n}{op\PYZus{}scope}\PY{p}{(}\PY{p}{[}\PY{n}{x}\PY{p}{]}\PY{p}{,} \PY{n}{name}\PY{p}{,} \PY{l+s+s2}{\PYZdq{}}\PY{l+s+s2}{f}\PY{l+s+s2}{\PYZdq{}}\PY{p}{)} \PY{k}{as} \PY{n}{name}\PY{p}{:}
                 \PY{n}{y} \PY{o}{=} \PY{n}{py\PYZus{}func}\PY{p}{(}\PY{n}{np\PYZus{}f\PYZus{}32}\PY{p}{,}
                                 \PY{p}{[}\PY{n}{x}\PY{p}{]}\PY{p}{,}
                                 \PY{p}{[}\PY{n}{tf}\PY{o}{.}\PY{n}{float32}\PY{p}{]}\PY{p}{,}
                                 \PY{n}{name}\PY{o}{=}\PY{n}{name}\PY{p}{,}
                                 \PY{n}{grad}\PY{o}{=}\PY{n}{fgrad}\PY{p}{)}
                 \PY{k}{return} \PY{n}{y}\PY{p}{[}\PY{l+m+mi}{0}\PY{p}{]}
\end{Verbatim}

    \begin{Verbatim}[commandchars=\\\{\}]
{\color{incolor}In [{\color{incolor}51}]:} \PY{k}{with} \PY{n}{tf}\PY{o}{.}\PY{n}{Session}\PY{p}{(}\PY{p}{)} \PY{k}{as} \PY{n}{sess}\PY{p}{:}
         
             \PY{n}{x} \PY{o}{=} \PY{n}{tf}\PY{o}{.}\PY{n}{constant}\PY{p}{(}\PY{p}{[}\PY{o}{\PYZhy{}}\PY{l+m+mf}{0.4}\PY{p}{,}\PY{o}{\PYZhy{}}\PY{l+m+mf}{0.2}\PY{p}{,}\PY{l+m+mf}{0.0}\PY{p}{,}\PY{l+m+mf}{0.2}\PY{p}{,}\PY{l+m+mf}{0.4}\PY{p}{]}\PY{p}{)}
             \PY{n}{y} \PY{o}{=} \PY{n}{tf\PYZus{}f}\PY{p}{(}\PY{n}{x}\PY{p}{)}
             \PY{n}{tf}\PY{o}{.}\PY{n}{initialize\PYZus{}all\PYZus{}variables}\PY{p}{(}\PY{p}{)}\PY{o}{.}\PY{n}{run}\PY{p}{(}\PY{p}{)}
         
             \PY{n+nb}{print}\PY{p}{(}\PY{n}{x}\PY{o}{.}\PY{n}{eval}\PY{p}{(}\PY{p}{)}\PY{p}{,} \PY{n}{y}\PY{o}{.}\PY{n}{eval}\PY{p}{(}\PY{p}{)}\PY{p}{,} \PY{n}{tf}\PY{o}{.}\PY{n}{gradients}\PY{p}{(}\PY{n}{y}\PY{p}{,} \PY{p}{[}\PY{n}{x}\PY{p}{]}\PY{p}{)}\PY{p}{[}\PY{l+m+mi}{0}\PY{p}{]}\PY{o}{.}\PY{n}{eval}\PY{p}{(}\PY{p}{)}\PY{p}{)}
\end{Verbatim}

    \subsubsection{Step 2: Set Up Neural
Network}\label{step-2-set-up-neural-network}

\begin{itemize}
\tightlist
\item
  Create random XOR Data
\item
  Create tensorflow model
\end{itemize}

    \begin{Verbatim}[commandchars=\\\{\}]
{\color{incolor}In [{\color{incolor}273}]:} \PY{c+c1}{\PYZsh{} Generate XOR Data}
          \PY{n}{tData} \PY{o}{=} \PY{n}{np}\PY{o}{.}\PY{n}{random}\PY{o}{.}\PY{n}{uniform}\PY{p}{(}\PY{o}{\PYZhy{}}\PY{l+m+mf}{1.0}\PY{p}{,} \PY{l+m+mf}{1.0}\PY{p}{,} \PY{p}{(}\PY{l+m+mi}{400}\PY{p}{,} \PY{l+m+mi}{2}\PY{p}{)}\PY{p}{)}
          \PY{n}{tData}\PY{p}{[}\PY{l+m+mi}{0}\PY{p}{:}\PY{l+m+mi}{100}\PY{p}{,} \PY{p}{:}\PY{p}{]} \PY{o}{=} \PY{n}{tData}\PY{p}{[}\PY{l+m+mi}{0}\PY{p}{:}\PY{l+m+mi}{100}\PY{p}{,} \PY{p}{:}\PY{p}{]}\PY{o}{*}\PY{l+m+mf}{0.2} \PY{o}{+} \PY{l+m+mf}{0.2}
          \PY{n}{tData}\PY{p}{[}\PY{l+m+mi}{100}\PY{p}{:}\PY{l+m+mi}{200}\PY{p}{,} \PY{p}{:}\PY{p}{]} \PY{o}{=} \PY{n}{tData}\PY{p}{[}\PY{l+m+mi}{100}\PY{p}{:}\PY{l+m+mi}{200}\PY{p}{,} \PY{p}{:}\PY{p}{]}\PY{o}{*}\PY{l+m+mf}{0.2} \PY{o}{+} \PY{l+m+mf}{0.6}
          \PY{n}{tData}\PY{p}{[}\PY{l+m+mi}{200}\PY{p}{:}\PY{l+m+mi}{300}\PY{p}{,} \PY{l+m+mi}{0}\PY{p}{]} \PY{o}{=} \PY{n}{tData}\PY{p}{[}\PY{l+m+mi}{200}\PY{p}{:}\PY{l+m+mi}{300}\PY{p}{,} \PY{l+m+mi}{0}\PY{p}{]}\PY{o}{*}\PY{l+m+mf}{0.2} \PY{o}{+} \PY{l+m+mf}{0.2}
          \PY{n}{tData}\PY{p}{[}\PY{l+m+mi}{200}\PY{p}{:}\PY{l+m+mi}{300}\PY{p}{,} \PY{l+m+mi}{1}\PY{p}{]} \PY{o}{=} \PY{n}{tData}\PY{p}{[}\PY{l+m+mi}{200}\PY{p}{:}\PY{l+m+mi}{300}\PY{p}{,} \PY{l+m+mi}{1}\PY{p}{]}\PY{o}{*}\PY{l+m+mf}{0.2} \PY{o}{+} \PY{l+m+mf}{0.6}
          \PY{n}{tData}\PY{p}{[}\PY{l+m+mi}{300}\PY{p}{:}\PY{l+m+mi}{400}\PY{p}{,} \PY{l+m+mi}{1}\PY{p}{]} \PY{o}{=} \PY{n}{tData}\PY{p}{[}\PY{l+m+mi}{300}\PY{p}{:}\PY{l+m+mi}{400}\PY{p}{,} \PY{l+m+mi}{1}\PY{p}{]}\PY{o}{*}\PY{l+m+mf}{0.2} \PY{o}{+} \PY{l+m+mf}{0.2}
          \PY{n}{tData}\PY{p}{[}\PY{l+m+mi}{300}\PY{p}{:}\PY{l+m+mi}{400}\PY{p}{,} \PY{l+m+mi}{0}\PY{p}{]} \PY{o}{=} \PY{n}{tData}\PY{p}{[}\PY{l+m+mi}{300}\PY{p}{:}\PY{l+m+mi}{400}\PY{p}{,} \PY{l+m+mi}{0}\PY{p}{]}\PY{o}{*}\PY{l+m+mf}{0.2} \PY{o}{+} \PY{l+m+mf}{0.6}
          \PY{n}{tData} \PY{o}{=} \PY{n}{np}\PY{o}{.}\PY{n}{clip}\PY{p}{(}\PY{n}{tData}\PY{p}{,} \PY{l+m+mf}{0.0}\PY{p}{,} \PY{l+m+mf}{0.8}\PY{p}{)}
          \PY{n}{tLabels} \PY{o}{=} \PY{n}{np}\PY{o}{.}\PY{n}{zeros}\PY{p}{(}\PY{p}{(}\PY{l+m+mi}{400}\PY{p}{,} \PY{l+m+mi}{2}\PY{p}{)}\PY{p}{)}
          \PY{n}{tLabels}\PY{p}{[}\PY{l+m+mi}{0}\PY{p}{:}\PY{l+m+mi}{200}\PY{p}{,} \PY{l+m+mi}{0}\PY{p}{]} \PY{o}{=} \PY{n}{np}\PY{o}{.}\PY{n}{ones}\PY{p}{(}\PY{l+m+mi}{200}\PY{p}{)}
          \PY{n}{tLabels}\PY{p}{[}\PY{l+m+mi}{200}\PY{p}{:}\PY{l+m+mi}{400}\PY{p}{,} \PY{l+m+mi}{1}\PY{p}{]} \PY{o}{=} \PY{n}{np}\PY{o}{.}\PY{n}{ones}\PY{p}{(}\PY{l+m+mi}{200}\PY{p}{)}
          \PY{n}{plt}\PY{o}{.}\PY{n}{plot}\PY{p}{(}\PY{n}{tData}\PY{p}{[}\PY{l+m+mi}{0}\PY{p}{:}\PY{l+m+mi}{200}\PY{p}{,} \PY{l+m+mi}{0}\PY{p}{]}\PY{p}{,} \PY{n}{tData}\PY{p}{[}\PY{l+m+mi}{0}\PY{p}{:}\PY{l+m+mi}{200}\PY{p}{,} \PY{l+m+mi}{1}\PY{p}{]}\PY{p}{,} \PY{l+s+s1}{\PYZsq{}}\PY{l+s+s1}{r*}\PY{l+s+s1}{\PYZsq{}}\PY{p}{)}
          \PY{n}{plt}\PY{o}{.}\PY{n}{plot}\PY{p}{(}\PY{n}{tData}\PY{p}{[}\PY{l+m+mi}{200}\PY{p}{:}\PY{l+m+mi}{400}\PY{p}{,} \PY{l+m+mi}{0}\PY{p}{]}\PY{p}{,} \PY{n}{tData}\PY{p}{[}\PY{l+m+mi}{200}\PY{p}{:}\PY{l+m+mi}{400}\PY{p}{,} \PY{l+m+mi}{1}\PY{p}{]}\PY{p}{,} \PY{l+s+s1}{\PYZsq{}}\PY{l+s+s1}{bs}\PY{l+s+s1}{\PYZsq{}}\PY{p}{)}
          \PY{n}{plt}\PY{o}{.}\PY{n}{show}\PY{p}{(}\PY{p}{)}
\end{Verbatim}

    
    \begin{Verbatim}[commandchars=\\\{\}]
{\color{incolor}In [{\color{incolor}274}]:} \PY{c+c1}{\PYZsh{} Parameters}
          \PY{n}{learning\PYZus{}rate} \PY{o}{=} \PY{l+m+mf}{0.01}
          \PY{n}{training\PYZus{}epochs} \PY{o}{=} \PY{l+m+mi}{1000}
          \PY{n}{display\PYZus{}step} \PY{o}{=} \PY{l+m+mi}{50}
\end{Verbatim}

    \begin{Verbatim}[commandchars=\\\{\}]
{\color{incolor}In [{\color{incolor}275}]:} \PY{c+c1}{\PYZsh{} Network Parameters}
          \PY{n}{n\PYZus{}hidden} \PY{o}{=} \PY{l+m+mi}{3} \PY{c+c1}{\PYZsh{} 1st layer number of features}
          \PY{n}{n\PYZus{}input} \PY{o}{=} \PY{l+m+mi}{2} \PY{c+c1}{\PYZsh{} XOR data input}
          \PY{n}{n\PYZus{}classes} \PY{o}{=} \PY{l+m+mi}{2} \PY{c+c1}{\PYZsh{} total classes (0 or 1)}
          
          \PY{c+c1}{\PYZsh{} tf Graph input}
          \PY{n}{x} \PY{o}{=} \PY{n}{tf}\PY{o}{.}\PY{n}{placeholder}\PY{p}{(}\PY{l+s+s2}{\PYZdq{}}\PY{l+s+s2}{float}\PY{l+s+s2}{\PYZdq{}}\PY{p}{,} \PY{p}{[}\PY{k+kc}{None}\PY{p}{,} \PY{n}{n\PYZus{}input}\PY{p}{]}\PY{p}{)}
          \PY{n}{y} \PY{o}{=} \PY{n}{tf}\PY{o}{.}\PY{n}{placeholder}\PY{p}{(}\PY{l+s+s2}{\PYZdq{}}\PY{l+s+s2}{float}\PY{l+s+s2}{\PYZdq{}}\PY{p}{,} \PY{p}{[}\PY{k+kc}{None}\PY{p}{,} \PY{n}{n\PYZus{}classes}\PY{p}{]}\PY{p}{)}
\end{Verbatim}

    \begin{Verbatim}[commandchars=\\\{\}]
{\color{incolor}In [{\color{incolor}276}]:} \PY{c+c1}{\PYZsh{} Create model}
          \PY{k}{def} \PY{n+nf}{multilayer\PYZus{}perceptron}\PY{p}{(}\PY{n}{x}\PY{p}{,} \PY{n}{weights}\PY{p}{,} \PY{n}{biases}\PY{p}{)}\PY{p}{:}
              \PY{c+c1}{\PYZsh{} Hidden layer with Photonic Axon}
              \PY{n}{hidden} \PY{o}{=} \PY{n}{tf}\PY{o}{.}\PY{n}{add}\PY{p}{(}\PY{n}{tf}\PY{o}{.}\PY{n}{matmul}\PY{p}{(}\PY{n}{x}\PY{p}{,} \PY{n}{weights}\PY{p}{[}\PY{l+s+s1}{\PYZsq{}}\PY{l+s+s1}{w0}\PY{l+s+s1}{\PYZsq{}}\PY{p}{]}\PY{p}{)}\PY{p}{,} \PY{n}{biases}\PY{p}{[}\PY{l+s+s1}{\PYZsq{}}\PY{l+s+s1}{b0}\PY{l+s+s1}{\PYZsq{}}\PY{p}{]}\PY{p}{)}
              \PY{n}{hidden} \PY{o}{=} \PY{n}{tf\PYZus{}f}\PY{p}{(}\PY{n}{hidden}\PY{p}{)}
              \PY{c+c1}{\PYZsh{} Output layer with linear activation}
              \PY{n}{out\PYZus{}layer} \PY{o}{=} \PY{n}{tf}\PY{o}{.}\PY{n}{matmul}\PY{p}{(}\PY{n}{hidden}\PY{p}{,} \PY{n}{weights}\PY{p}{[}\PY{l+s+s1}{\PYZsq{}}\PY{l+s+s1}{w1}\PY{l+s+s1}{\PYZsq{}}\PY{p}{]}\PY{p}{)} \PY{o}{+} \PY{n}{biases}\PY{p}{[}\PY{l+s+s1}{\PYZsq{}}\PY{l+s+s1}{b1}\PY{l+s+s1}{\PYZsq{}}\PY{p}{]}
              \PY{k}{return} \PY{n}{out\PYZus{}layer}
          
          \PY{c+c1}{\PYZsh{} Store layers weight \PYZam{} bias}
          \PY{n}{weights} \PY{o}{=} \PY{p}{\PYZob{}}
              \PY{l+s+s1}{\PYZsq{}}\PY{l+s+s1}{w0}\PY{l+s+s1}{\PYZsq{}}\PY{p}{:} \PY{n}{tf}\PY{o}{.}\PY{n}{Variable}\PY{p}{(}\PY{n}{tf}\PY{o}{.}\PY{n}{random\PYZus{}normal}\PY{p}{(}\PY{p}{[}\PY{n}{n\PYZus{}input}\PY{p}{,} \PY{n}{n\PYZus{}hidden}\PY{p}{]}\PY{p}{)}\PY{p}{)}\PY{p}{,}
              \PY{l+s+s1}{\PYZsq{}}\PY{l+s+s1}{w1}\PY{l+s+s1}{\PYZsq{}}\PY{p}{:} \PY{n}{tf}\PY{o}{.}\PY{n}{Variable}\PY{p}{(}\PY{n}{tf}\PY{o}{.}\PY{n}{random\PYZus{}normal}\PY{p}{(}\PY{p}{[}\PY{n}{n\PYZus{}hidden}\PY{p}{,} \PY{n}{n\PYZus{}classes}\PY{p}{]}\PY{p}{)}\PY{p}{)}
          \PY{p}{\PYZcb{}}
          \PY{n}{biases} \PY{o}{=} \PY{p}{\PYZob{}}
              \PY{l+s+s1}{\PYZsq{}}\PY{l+s+s1}{b0}\PY{l+s+s1}{\PYZsq{}}\PY{p}{:} \PY{n}{tf}\PY{o}{.}\PY{n}{Variable}\PY{p}{(}\PY{n}{tf}\PY{o}{.}\PY{n}{random\PYZus{}normal}\PY{p}{(}\PY{p}{[}\PY{n}{n\PYZus{}hidden}\PY{p}{]}\PY{p}{)}\PY{p}{)}\PY{p}{,}
              \PY{l+s+s1}{\PYZsq{}}\PY{l+s+s1}{b1}\PY{l+s+s1}{\PYZsq{}}\PY{p}{:} \PY{n}{tf}\PY{o}{.}\PY{n}{Variable}\PY{p}{(}\PY{n}{tf}\PY{o}{.}\PY{n}{random\PYZus{}normal}\PY{p}{(}\PY{p}{[}\PY{n}{n\PYZus{}classes}\PY{p}{]}\PY{p}{)}\PY{p}{)}
          \PY{p}{\PYZcb{}}
\end{Verbatim}

    \begin{Verbatim}[commandchars=\\\{\}]
{\color{incolor}In [{\color{incolor}277}]:} \PY{c+c1}{\PYZsh{} Construct model}
          \PY{n}{pred} \PY{o}{=} \PY{n}{multilayer\PYZus{}perceptron}\PY{p}{(}\PY{n}{x}\PY{p}{,} \PY{n}{weights}\PY{p}{,} \PY{n}{biases}\PY{p}{)}
          
          \PY{c+c1}{\PYZsh{} Define loss and optimizer}
          \PY{n}{cost} \PY{o}{=} \PY{n}{tf}\PY{o}{.}\PY{n}{reduce\PYZus{}mean}\PY{p}{(}\PY{n}{tf}\PY{o}{.}\PY{n}{nn}\PY{o}{.}\PY{n}{softmax\PYZus{}cross\PYZus{}entropy\PYZus{}with\PYZus{}logits}\PY{p}{(}
            \PY{n}{logits}\PY{o}{=}\PY{n}{pred}\PY{p}{,} \PY{n}{labels}\PY{o}{=}\PY{n}{y}\PY{p}{)}\PY{p}{)}
          \PY{n}{optimizer} \PY{o}{=} \PY{n}{tf}\PY{o}{.}\PY{n}{train}\PY{o}{.}\PY{n}{AdamOptimizer}\PY{p}{(}
            \PY{n}{learning\PYZus{}rate}\PY{o}{=}\PY{n}{learning\PYZus{}rate}\PY{p}{)}\PY{o}{.}\PY{n}{minimize}\PY{p}{(}\PY{n}{cost}\PY{p}{)}
          
          \PY{c+c1}{\PYZsh{} Initializing the variables}
          \PY{n}{init} \PY{o}{=} \PY{n}{tf}\PY{o}{.}\PY{n}{global\PYZus{}variables\PYZus{}initializer}\PY{p}{(}\PY{p}{)}
\end{Verbatim}

    \subsubsection{Step 3: Train}\label{step-3-train}

\begin{itemize}
\tightlist
\item
  Train the network.
\item
  Check accuracy and see the 2D sweep.
\item
  Pull weights to pass back into the network simulation.
\end{itemize}

    \begin{Verbatim}[commandchars=\\\{\}]
{\color{incolor}In [{\color{incolor}303}]:} \PY{c+c1}{\PYZsh{} Launch the graph}
          
          \PY{c+c1}{\PYZsh{}sess = tf.Session()}
          \PY{c+c1}{\PYZsh{}sess.run(init)}
          
          \PY{c+c1}{\PYZsh{} Training cycle}
          \PY{k}{try}\PY{p}{:}
              \PY{k}{for} \PY{n}{epoch} \PY{o+ow}{in} \PY{n+nb}{range}\PY{p}{(}\PY{n}{training\PYZus{}epochs}\PY{p}{)}\PY{p}{:}
                  \PY{c+c1}{\PYZsh{} Run backprop and cost func}
                  \PY{n}{\PYZus{}}\PY{p}{,} \PY{n}{c} \PY{o}{=} \PY{n}{sess}\PY{o}{.}\PY{n}{run}\PY{p}{(}\PY{p}{[}\PY{n}{optimizer}\PY{p}{,} \PY{n}{cost}\PY{p}{]}\PY{p}{,} \PY{n}{feed\PYZus{}dict}\PY{o}{=}\PY{p}{\PYZob{}}\PY{n}{x}\PY{p}{:} \PY{n}{tData}\PY{p}{,}
                        \PY{n}{y}\PY{p}{:} \PY{n}{tLabels}\PY{p}{\PYZcb{}}\PY{p}{)}
                  \PY{c+c1}{\PYZsh{} Display logs per epoch step}
                  \PY{k}{if} \PY{n}{epoch} \PY{o}{\PYZpc{}} \PY{n}{display\PYZus{}step} \PY{o}{==} \PY{l+m+mi}{0}\PY{p}{:}
                      \PY{n+nb}{print}\PY{p}{(}\PY{l+s+s2}{\PYZdq{}}\PY{l+s+s2}{Epoch:}\PY{l+s+s2}{\PYZdq{}}\PY{p}{,} \PY{l+s+s1}{\PYZsq{}}\PY{l+s+si}{\PYZpc{}04d}\PY{l+s+s1}{\PYZsq{}} \PY{o}{\PYZpc{}} \PY{p}{(}\PY{n}{epoch}\PY{o}{+}\PY{l+m+mi}{1}\PY{p}{)}\PY{p}{,} \PY{l+s+s2}{\PYZdq{}}\PY{l+s+s2}{cost=}\PY{l+s+s2}{\PYZdq{}}\PY{p}{,} \PYZbs{}
                          \PY{l+s+s2}{\PYZdq{}}\PY{l+s+si}{\PYZob{}:.9f\PYZcb{}}\PY{l+s+s2}{\PYZdq{}}\PY{o}{.}\PY{n}{format}\PY{p}{(}\PY{n}{c}\PY{p}{)}\PY{p}{)}
              \PY{n+nb}{print}\PY{p}{(}\PY{l+s+s2}{\PYZdq{}}\PY{l+s+s2}{Optimization Finished!}\PY{l+s+s2}{\PYZdq{}}\PY{p}{)}
          \PY{k}{except}\PY{p}{:}
              \PY{n+nb}{print}\PY{p}{(}\PY{l+s+s2}{\PYZdq{}}\PY{l+s+s2}{Training Interrupted}\PY{l+s+s2}{\PYZdq{}}\PY{p}{)}
\end{Verbatim}

    \begin{Verbatim}[commandchars=\\\{\}]
Epoch: 0001 cost= 0.059564065
Epoch: 0051 cost= 0.058935978
Training Interrupted

    \end{Verbatim}

    \begin{Verbatim}[commandchars=\\\{\}]
{\color{incolor}In [{\color{incolor}304}]:} \PY{k}{with} \PY{n}{sess}\PY{o}{.}\PY{n}{as\PYZus{}default}\PY{p}{(}\PY{p}{)}\PY{p}{:}
              \PY{c+c1}{\PYZsh{} Test model}
              \PY{n}{correct\PYZus{}prediction} \PY{o}{=}
                \PY{n}{tf}\PY{o}{.}\PY{n}{equal}\PY{p}{(}\PY{n}{tf}\PY{o}{.}\PY{n}{argmax}\PY{p}{(}\PY{n}{pred}\PY{p}{,} \PY{l+m+mi}{1}\PY{p}{)}\PY{p}{,} \PY{n}{tf}\PY{o}{.}\PY{n}{argmax}\PY{p}{(}\PY{n}{y}\PY{p}{,} \PY{l+m+mi}{1}\PY{p}{)}\PY{p}{)}
              \PY{c+c1}{\PYZsh{} Calculate accuracy}
              \PY{n}{accuracy} \PY{o}{=} \PY{n}{tf}\PY{o}{.}\PY{n}{reduce\PYZus{}mean}\PY{p}{(}
                \PY{n}{tf}\PY{o}{.}\PY{n}{cast}\PY{p}{(}\PY{n}{correct\PYZus{}prediction}\PY{p}{,} \PY{l+s+s2}{\PYZdq{}}\PY{l+s+s2}{float}\PY{l+s+s2}{\PYZdq{}}\PY{p}{)}\PY{p}{)}
              \PY{n+nb}{print}\PY{p}{(}\PY{l+s+s2}{\PYZdq{}}\PY{l+s+s2}{Accuracy:}\PY{l+s+s2}{\PYZdq{}}\PY{p}{,} \PY{n}{accuracy}\PY{o}{.}\PY{n}{eval}\PY{p}{(}\PY{p}{\PYZob{}}\PY{n}{x}\PY{p}{:} \PY{n}{tData}\PY{p}{,} \PY{n}{y}\PY{p}{:} \PY{n}{tLabels}\PY{p}{\PYZcb{}}\PY{p}{)}\PY{p}{)}
\end{Verbatim}

    \begin{Verbatim}[commandchars=\\\{\}]
Accuracy: 0.9875

    \end{Verbatim}

    \begin{Verbatim}[commandchars=\\\{\}]
{\color{incolor}In [{\color{incolor}296}]:} \PY{n}{cList} \PY{o}{=} \PY{n}{np}\PY{o}{.}\PY{n}{linspace}\PY{p}{(}\PY{l+m+mf}{0.0}\PY{p}{,} \PY{l+m+mf}{0.8}\PY{p}{,} \PY{l+m+mi}{100}\PY{p}{)}
          \PY{n}{yMat} \PY{o}{=} \PY{n}{np}\PY{o}{.}\PY{n}{zeros}\PY{p}{(}\PY{p}{(}\PY{l+m+mi}{100}\PY{p}{,} \PY{l+m+mi}{100}\PY{p}{,} \PY{l+m+mi}{2}\PY{p}{)}\PY{p}{)}
          
          \PY{k}{with} \PY{n}{sess}\PY{o}{.}\PY{n}{as\PYZus{}default}\PY{p}{(}\PY{p}{)}\PY{p}{:}
              \PY{k}{for} \PY{n}{i}\PY{p}{,} \PY{n}{c1} \PY{o+ow}{in} \PY{n+nb}{enumerate}\PY{p}{(}\PY{n}{cList}\PY{p}{)}\PY{p}{:}
                  \PY{k}{for} \PY{n}{j}\PY{p}{,} \PY{n}{c2} \PY{o+ow}{in} \PY{n+nb}{enumerate}\PY{p}{(}\PY{n}{cList}\PY{p}{)}\PY{p}{:}
                      \PY{n}{cIn} \PY{o}{=} \PY{n}{np}\PY{o}{.}\PY{n}{zeros}\PY{p}{(}\PY{p}{(}\PY{l+m+mi}{1}\PY{p}{,} \PY{l+m+mi}{2}\PY{p}{)}\PY{p}{)}
                      \PY{n}{cIn}\PY{p}{[}\PY{l+m+mi}{0}\PY{p}{,} \PY{l+m+mi}{0}\PY{p}{]} \PY{o}{=} \PY{n}{c1}
                      \PY{n}{cIn}\PY{p}{[}\PY{l+m+mi}{0}\PY{p}{,} \PY{l+m+mi}{1}\PY{p}{]} \PY{o}{=} \PY{n}{c2}
                      \PY{n}{yMat}\PY{p}{[}\PY{n}{i}\PY{p}{,} \PY{n}{j}\PY{p}{,} \PY{p}{:}\PY{p}{]} \PY{o}{=} \PY{n}{pred}\PY{o}{.}\PY{n}{eval}\PY{p}{(}\PY{p}{\PYZob{}}\PY{n}{x}\PY{p}{:} \PY{n}{cIn}\PY{p}{\PYZcb{}}\PY{p}{)}
\end{Verbatim}

    \begin{Verbatim}[commandchars=\\\{\}]
{\color{incolor}In [{\color{incolor}297}]:} \PY{c+c1}{\PYZsh{} Plot Normalized Output}
          \PY{n}{im} \PY{o}{=} \PY{n}{plt}\PY{o}{.}\PY{n}{imshow}\PY{p}{(}\PY{n}{yMat}\PY{p}{[}\PY{p}{:}\PY{p}{,} \PY{p}{:}\PY{p}{,} \PY{l+m+mi}{0}\PY{p}{]}\PY{p}{,} 
            \PY{n}{interpolation}\PY{o}{=}\PY{l+s+s1}{\PYZsq{}}\PY{l+s+s1}{bilinear}\PY{l+s+s1}{\PYZsq{}}\PY{p}{,} \PY{n}{cmap}\PY{o}{=}\PY{n}{cm}\PY{o}{.}\PY{n}{RdYlGn}\PY{p}{,}
                              \PY{n}{origin}\PY{o}{=}\PY{l+s+s1}{\PYZsq{}}\PY{l+s+s1}{lower}\PY{l+s+s1}{\PYZsq{}}\PY{p}{,} \PY{n}{extent}\PY{o}{=}\PY{p}{[}\PY{l+m+mi}{0}\PY{p}{,} \PY{l+m+mf}{0.8}\PY{p}{,} \PY{l+m+mi}{0}\PY{p}{,} \PY{l+m+mf}{0.8}\PY{p}{]}\PY{p}{,}
                              \PY{n}{vmax}\PY{o}{=}\PY{n+nb}{abs}\PY{p}{(}\PY{n}{yMat}\PY{p}{)}\PY{o}{.}\PY{n}{max}\PY{p}{(}\PY{p}{)}\PY{p}{,} \PY{n}{vmin}\PY{o}{=}\PY{o}{\PYZhy{}}\PY{n+nb}{abs}\PY{p}{(}\PY{n}{yMat}\PY{p}{)}\PY{o}{.}\PY{n}{max}\PY{p}{(}\PY{p}{)}\PY{p}{)}
          \PY{n}{plt}\PY{o}{.}\PY{n}{colorbar}\PY{p}{(}\PY{p}{)}
          \PY{n}{plt}\PY{o}{.}\PY{n}{show}\PY{p}{(}\PY{p}{)}
          \PY{c+c1}{\PYZsh{} Classification}
          \PY{n+nb+bp}{cls} \PY{o}{=} \PY{n}{np}\PY{o}{.}\PY{n}{sign}\PY{p}{(}\PY{n}{yMat}\PY{p}{[}\PY{p}{:}\PY{p}{,} \PY{p}{:}\PY{p}{,} \PY{l+m+mi}{0}\PY{p}{]}\PY{p}{)}
          \PY{n}{im} \PY{o}{=} \PY{n}{plt}\PY{o}{.}\PY{n}{imshow}\PY{p}{(}\PY{n+nb+bp}{cls}\PY{p}{,} \PY{n}{interpolation}\PY{o}{=}\PY{l+s+s1}{\PYZsq{}}\PY{l+s+s1}{bilinear}\PY{l+s+s1}{\PYZsq{}}\PY{p}{,} \PY{n}{cmap}\PY{o}{=}\PY{n}{cm}\PY{o}{.}\PY{n}{RdYlGn}\PY{p}{,}
                              \PY{n}{origin}\PY{o}{=}\PY{l+s+s1}{\PYZsq{}}\PY{l+s+s1}{lower}\PY{l+s+s1}{\PYZsq{}}\PY{p}{,} \PY{n}{extent}\PY{o}{=}\PY{p}{[}\PY{l+m+mi}{0}\PY{p}{,} \PY{l+m+mf}{0.8}\PY{p}{,} \PY{l+m+mi}{0}\PY{p}{,} \PY{l+m+mf}{0.8}\PY{p}{]}\PY{p}{,}
                              \PY{n}{vmax}\PY{o}{=}\PY{l+m+mi}{1}\PY{p}{,} \PY{n}{vmin}\PY{o}{=}\PY{o}{\PYZhy{}}\PY{l+m+mi}{1}\PY{p}{)}
          
          \PY{n}{plt}\PY{o}{.}\PY{n}{show}\PY{p}{(}\PY{p}{)}
\end{Verbatim}

    
    
    \begin{Verbatim}[commandchars=\\\{\}]
{\color{incolor}In [{\color{incolor}301}]:} \PY{c+c1}{\PYZsh{} Pull Weights}
          \PY{k}{with} \PY{n}{sess}\PY{o}{.}\PY{n}{as\PYZus{}default}\PY{p}{(}\PY{p}{)}\PY{p}{:}
              \PY{n}{allweights} \PY{o}{=} \PY{n}{sess}\PY{o}{.}\PY{n}{run}\PY{p}{(}\PY{n}{weights}\PY{p}{)}
              \PY{n}{w0} \PY{o}{=} \PY{n}{allweights}\PY{p}{[}\PY{l+s+s1}{\PYZsq{}}\PY{l+s+s1}{w0}\PY{l+s+s1}{\PYZsq{}}\PY{p}{]}\PY{o}{.}\PY{n}{T}
              \PY{n}{w1} \PY{o}{=} \PY{n}{allweights}\PY{p}{[}\PY{l+s+s1}{\PYZsq{}}\PY{l+s+s1}{w1}\PY{l+s+s1}{\PYZsq{}}\PY{p}{]}\PY{o}{.}\PY{n}{T}\PY{p}{[}\PY{l+m+mi}{0}\PY{p}{,} \PY{p}{:}\PY{p}{]}
              \PY{n+nb}{print}\PY{p}{(}\PY{n}{w0}\PY{p}{)}
              \PY{n+nb}{print}\PY{p}{(}\PY{n}{w1}\PY{p}{)}
\end{Verbatim}

    \begin{Verbatim}[commandchars=\\\{\}]
[[ 1.30109513  0.90975827]
 [-0.79916418 -0.85981083]
 [ 0.9742766  -1.02000749]]
[ -4.16287088  11.24845219  -9.0137167 ]

    \end{Verbatim}

    \begin{Verbatim}[commandchars=\\\{\}]
{\color{incolor}In [{\color{incolor}302}]:} \PY{c+c1}{\PYZsh{} Pull Biases}
          \PY{k}{with} \PY{n}{sess}\PY{o}{.}\PY{n}{as\PYZus{}default}\PY{p}{(}\PY{p}{)}\PY{p}{:}
              \PY{n}{allbiases} \PY{o}{=} \PY{n}{sess}\PY{o}{.}\PY{n}{run}\PY{p}{(}\PY{n}{biases}\PY{p}{)}
              \PY{n}{b0} \PY{o}{=} \PY{n}{allbiases}\PY{p}{[}\PY{l+s+s1}{\PYZsq{}}\PY{l+s+s1}{b0}\PY{l+s+s1}{\PYZsq{}}\PY{p}{]}
              \PY{n}{b1} \PY{o}{=} \PY{n}{allbiases}\PY{p}{[}\PY{l+s+s1}{\PYZsq{}}\PY{l+s+s1}{b1}\PY{l+s+s1}{\PYZsq{}}\PY{p}{]}\PY{o}{.}\PY{n}{T}\PY{p}{[}\PY{l+m+mi}{0}\PY{p}{]}
              \PY{n+nb}{print}\PY{p}{(}\PY{n}{b0}\PY{p}{)}
              \PY{n+nb}{print}\PY{p}{(}\PY{n}{b1}\PY{p}{)}
\end{Verbatim}

    \begin{Verbatim}[commandchars=\\\{\}]
[ 1.0609535   0.65707952  0.01618425]
2.17837

    \end{Verbatim}

    \begin{Verbatim}[commandchars=\\\{\}]
{\color{incolor}In [{\color{incolor} }]:} 
\end{Verbatim}

%% file: 2-3-1_FFNet_Simulation.tex
    
    \subsubsection{Step 1: Initialize 2-3-1 Virtual Network
(Calibrated)}\label{step-1-initialize-2-3-1-virtual-network-calibrated}

    \begin{Verbatim}[commandchars=\\\{\}]
{\color{incolor}In [{\color{incolor}1}]:} \PY{k+kn}{import} \PY{n+nn}{numpy} \PY{k}{as} \PY{n+nn}{np}
        \PY{k+kn}{import} \PY{n+nn}{matplotlib}\PY{n+nn}{.}\PY{n+nn}{pyplot} \PY{k}{as} \PY{n+nn}{plt}
        \PY{k+kn}{import} \PY{n+nn}{lightlab}\PY{n+nn}{.}\PY{n+nn}{instruments} \PY{k}{as} \PY{n+nn}{inst}
        \PY{k+kn}{import} \PY{n+nn}{lightlab}\PY{n+nn}{.}\PY{n+nn}{model} \PY{k}{as} \PY{n+nn}{m}
        \PY{k+kn}{from} \PY{n+nn}{lightlab}\PY{n+nn}{.}\PY{n+nn}{util}\PY{n+nn}{.}\PY{n+nn}{modeling} \PY{k}{import} \PY{n}{kOSAPwr}\PY{p}{,} \PY{n}{dbm2lin}\PY{p}{,} \PY{n}{CurrentUnit}
        \PY{k+kn}{from} \PY{n+nn}{lightlab}\PY{n+nn}{.}\PY{n+nn}{util}\PY{n+nn}{.}\PY{n+nn}{calibrating} \PY{k}{import} \PY{n}{SpectrumMeasurementAssistant}
        \PY{k+kn}{from} \PY{n+nn}{scipy} \PY{k}{import} \PY{n}{interpolate}
        \PY{k+kn}{import} \PY{n+nn}{matplotlib}\PY{n+nn}{.}\PY{n+nn}{cm} \PY{k}{as} \PY{n+nn}{cm}
\end{Verbatim}

    \begin{Verbatim}[commandchars=\\\{\}]
{\color{incolor}In [{\color{incolor}2}]:} \PY{c+c1}{\PYZsh{} Thermal Group, Uses 14 current channels}
        \PY{n}{numCurrents} \PY{o}{=} \PY{l+m+mi}{14}
        \PY{n}{currentChan} \PY{o}{=} \PY{n+nb}{range}\PY{p}{(}\PY{n}{numCurrents}\PY{p}{)}
        \PY{n}{therm} \PY{o}{=} \PY{n}{m}\PY{o}{.}\PY{n}{ThermalGroup}\PY{p}{(}\PY{n}{currentChan}\PY{p}{)}
        
        \PY{c+c1}{\PYZsh{} Axons first, two branches, 2 in 1st branch, 3 in second branch}
        \PY{n}{axonChannels} \PY{o}{=} \PY{p}{[}\PY{p}{[}\PY{l+m+mi}{0}\PY{p}{,} \PY{l+m+mi}{1}\PY{p}{]}\PY{p}{,} \PY{p}{[}\PY{l+m+mi}{2}\PY{p}{,} \PY{l+m+mi}{3}\PY{p}{,} \PY{l+m+mi}{4}\PY{p}{]}\PY{p}{]}
        
        \PY{c+c1}{\PYZsh{} Master Variables}
        \PY{n}{numBranches} \PY{o}{=} \PY{l+m+mi}{2}
        \PY{n}{numRings} \PY{o}{=} \PY{p}{[}\PY{l+m+mi}{2}\PY{p}{,} \PY{l+m+mi}{3}\PY{p}{]}
        \PY{n}{numBanks} \PY{o}{=} \PY{p}{[}\PY{l+m+mi}{3}\PY{p}{,} \PY{l+m+mi}{1}\PY{p}{]}
        \PY{n}{wlChannels} \PY{o}{=} \PY{n}{np}\PY{o}{.}\PY{n}{array}\PY{p}{(}\PY{p}{[}\PY{l+m+mi}{1550}\PY{p}{,} \PY{l+m+mi}{1552}\PY{p}{,} \PY{l+m+mi}{1554}\PY{p}{]}\PY{p}{)}
        \PY{n}{lorentzAtten} \PY{o}{=} \PY{l+m+mf}{0.99}
        \PY{n}{lorentzfwhm} \PY{o}{=} \PY{l+m+mf}{0.1}
        \PY{n}{fullAtten} \PY{o}{=} \PY{l+m+mf}{0.01}
        
        \PY{c+c1}{\PYZsh{} Make Axons}
        \PY{n}{axons} \PY{o}{=} \PY{n+nb}{list}\PY{p}{(}\PY{p}{)}
        \PY{k}{for} \PY{n}{i} \PY{o+ow}{in} \PY{n+nb}{range}\PY{p}{(}\PY{n}{numBranches}\PY{p}{)}\PY{p}{:}
            \PY{n}{axon} \PY{o}{=} \PY{n}{m}\PY{o}{.}\PY{n}{FilterBank}\PY{p}{(}\PY{n}{therm}\PY{p}{,} \PY{n}{numRings}\PY{p}{[}\PY{n}{i}\PY{p}{]}\PY{p}{)}
            \PY{n}{axon}\PY{o}{.}\PY{n}{setAxon}\PY{p}{(}\PY{p}{)}
            \PY{n}{axon}\PY{o}{.}\PY{n}{setBiasParams}\PY{p}{(}\PY{p}{[}\PY{n}{wlChannels}\PY{p}{[}\PY{n}{j}\PY{p}{]} \PY{k}{for} \PY{n}{j} \PY{o+ow}{in} \PY{n+nb}{range}\PY{p}{(}\PY{n}{numRings}\PY{p}{[}\PY{n}{i}\PY{p}{]}\PY{p}{)}\PY{p}{]}\PY{p}{,}
            	\PY{n}{lorentzAtten}\PY{o}{*}\PY{n}{np}\PY{o}{.}\PY{n}{ones}\PY{p}{(}\PY{n}{numRings}\PY{p}{[}\PY{n}{i}\PY{p}{]}\PY{p}{)}\PY{p}{,} 
            	\PY{n}{lorentzfwhm} \PY{o}{*} \PY{n}{np}\PY{o}{.}\PY{n}{ones}\PY{p}{(}\PY{n}{numRings}\PY{p}{[}\PY{n}{i}\PY{p}{]}\PY{p}{)}\PY{p}{)}
            \PY{n}{axons}\PY{o}{.}\PY{n}{append}\PY{p}{(}\PY{n}{axon}\PY{p}{)}
        
        \PY{c+c1}{\PYZsh{} Make Dendrites}
        \PY{n}{fbs} \PY{o}{=} \PY{p}{[}\PY{k+kc}{None}\PY{p}{]} \PY{o}{*} \PY{n}{numBranches}
        \PY{k}{for} \PY{n}{i} \PY{o+ow}{in} \PY{n+nb}{range}\PY{p}{(}\PY{n}{numBranches}\PY{p}{)}\PY{p}{:}
            \PY{n}{fbs}\PY{p}{[}\PY{n}{i}\PY{p}{]} \PY{o}{=} \PY{n+nb}{list}\PY{p}{(}\PY{p}{)}
            \PY{k}{for} \PY{n}{j} \PY{o+ow}{in} \PY{n+nb}{range}\PY{p}{(}\PY{n}{numBanks}\PY{p}{[}\PY{n}{i}\PY{p}{]}\PY{p}{)}\PY{p}{:}
                \PY{n}{fb} \PY{o}{=} \PY{n}{m}\PY{o}{.}\PY{n}{FilterBank}\PY{p}{(}\PY{n}{therm}\PY{p}{,} \PY{n}{numRings}\PY{p}{[}\PY{n}{i}\PY{p}{]}\PY{p}{)}
                \PY{n}{fb}\PY{o}{.}\PY{n}{setBiasParams}\PY{p}{(}\PY{p}{[}\PY{n}{wlChannels}\PY{p}{[}\PY{n}{k}\PY{p}{]} \PY{k}{for} \PY{n}{k} \PY{o+ow}{in} \PY{n+nb}{range}\PY{p}{(}
                    \PY{n}{numRings}\PY{p}{[}\PY{n}{i}\PY{p}{]}\PY{p}{)}\PY{p}{]}\PY{p}{,} \PY{n}{lorentzAtten}\PY{o}{*}\PY{n}{np}\PY{o}{.}\PY{n}{ones}\PY{p}{(}\PY{n}{numRings}\PY{p}{[}\PY{n}{i}\PY{p}{]}\PY{p}{)}\PY{p}{,} 
                    \PY{n}{lorentzfwhm} \PY{o}{*} \PY{n}{np}\PY{o}{.}\PY{n}{ones}\PY{p}{(}\PY{n}{numRings}\PY{p}{[}\PY{n}{i}\PY{p}{]}\PY{p}{)}\PY{p}{)}
                \PY{n}{fbs}\PY{p}{[}\PY{n}{i}\PY{p}{]}\PY{o}{.}\PY{n}{append}\PY{p}{(}\PY{n}{fb}\PY{p}{)}
        
        \PY{c+c1}{\PYZsh{} Connect Components}
        \PY{n}{cscs} \PY{o}{=} \PY{n+nb}{list}\PY{p}{(}\PY{p}{)}
        \PY{k}{for} \PY{n}{i} \PY{o+ow}{in} \PY{n+nb}{range}\PY{p}{(}\PY{n}{numBranches}\PY{p}{)}\PY{p}{:}
            \PY{n}{cscs}\PY{o}{.}\PY{n}{append}\PY{p}{(}\PY{n}{m}\PY{o}{.}\PY{n}{Cascade}\PY{p}{(}\PY{n}{axons}\PY{p}{[}\PY{n}{i}\PY{p}{]}\PY{p}{,} \PY{n}{m}\PY{o}{.}\PY{n}{Splitter}\PY{p}{(}\PY{n}{fbs}\PY{p}{[}\PY{n}{i}\PY{p}{]}\PY{p}{)}\PY{p}{)}\PY{p}{)}
        \PY{n}{network} \PY{o}{=} \PY{n}{m}\PY{o}{.}\PY{n}{Splitter}\PY{p}{(}\PY{n}{cscs}\PY{p}{)}
        \PY{n}{network}\PY{o}{.}\PY{n}{setAttenuation}\PY{p}{(}\PY{n}{fullAtten}\PY{p}{)}

        \PY{c+c1}{\PYZsh{} Set K and bias of hidden phony module}
        \PY{n}{K} \PY{o}{=} \PY{l+m+mf}{20.0}\PY{o}{*}\PY{n}{np}\PY{o}{.}\PY{n}{eye}\PY{p}{(}\PY{n}{numCurrents}\PY{p}{)}
        \PY{n}{therm}\PY{o}{.}\PY{n}{setK}\PY{p}{(}\PY{n}{K}\PY{p}{)}
        
        \PY{n}{heatBias} \PY{o}{=} \PY{n+nb}{dict}\PY{p}{(}\PY{p}{)}
        \PY{k}{for} \PY{n}{c} \PY{o+ow}{in} \PY{n+nb}{range}\PY{p}{(}\PY{n}{numCurrents}\PY{p}{)}\PY{p}{:}
            \PY{n}{heatBias}\PY{p}{[}\PY{n}{c}\PY{p}{]} \PY{o}{=} \PY{n}{c}\PY{o}{*}\PY{l+m+mf}{0.2} \PY{o}{+} \PY{l+m+mi}{1}  \PY{c+c1}{\PYZsh{} In Volts}
            
        \PY{c+c1}{\PYZsh{}heatBias[2] = heatBias[3] = 1.5}
        \PY{c+c1}{\PYZsh{}heatBias[4] = heatBias[3]}
        
        \PY{n}{therm}\PY{o}{.}\PY{n}{setHeatBias}\PY{p}{(}\PY{n}{heatBias}\PY{p}{,} \PY{n}{CurrentUnit}\PY{o}{.}\PY{n}{V}\PY{p}{)}
        
        \PY{c+c1}{\PYZsh{} Register Network}
        \PY{n}{inst}\PY{o}{.}\PY{n}{togglePhony}\PY{p}{(}\PY{k+kc}{True}\PY{p}{,} \PY{n}{network}\PY{p}{)}
        \PY{n}{token} \PY{o}{=} \PY{n}{inst}\PY{o}{.}\PY{n}{reserveCurrentChan}\PY{p}{(}\PY{n}{currentChan}\PY{p}{)}
        
        \PY{n}{wlRange} \PY{o}{=} \PY{p}{[}\PY{l+m+mi}{1547}\PY{p}{,} \PY{l+m+mi}{1555}\PY{p}{]}
\end{Verbatim}

    \subsubsection{Step 2: Network Simulation
Function}\label{step-2-network-simulation-function}

    \begin{Verbatim}[commandchars=\\\{\}]
{\color{incolor}In [{\color{incolor}3}]:} \PY{c+c1}{\PYZsh{} Global Parameters}
        \PY{n}{Resp} \PY{o}{=} \PY{l+m+mf}{0.9}
        \PY{n}{Rt} \PY{o}{=} \PY{l+m+mi}{15000} \PY{c+c1}{\PYZsh{} (15MOhm)}
        \PY{n}{Rs} \PY{o}{=} \PY{l+m+mf}{1.5} \PY{c+c1}{\PYZsh{}(1kOhm)}
        \PY{n}{bv} \PY{o}{=} \PY{l+m+mi}{4} \PY{c+c1}{\PYZsh{} 4V}
\end{Verbatim}

    \begin{Verbatim}[commandchars=\\\{\}]
{\color{incolor}In [{\color{incolor}4}]:} \PY{k}{def} \PY{n+nf}{getWeightsBias}\PY{p}{(}\PY{n}{weight23}\PY{p}{,} \PY{n}{weight31}\PY{p}{)}\PY{p}{:}
            \PY{c+c1}{\PYZsh{} Generate transmission matrices}
            \PY{c+c1}{\PYZsh{} Note: Thru port is positive}
            \PY{c+c1}{\PYZsh{} t\PYZus{}thru = (w+1)/2}
            
            \PY{c+c1}{\PYZsh{} Start with all axons 0 deltaLambda}
            \PY{n}{dummyAxonBias} \PY{o}{=} \PY{p}{[}\PY{n}{np}\PY{o}{.}\PY{n}{array}\PY{p}{(}\PY{p}{[}\PY{l+m+mi}{0}\PY{p}{,} \PY{l+m+mi}{0}\PY{p}{]}\PY{p}{)}\PY{p}{,} \PY{n}{np}\PY{o}{.}\PY{n}{array}\PY{p}{(}\PY{p}{[}\PY{l+m+mi}{0}\PY{p}{,}\PY{l+m+mi}{0}\PY{p}{,}\PY{l+m+mi}{0}\PY{p}{]}\PY{p}{)}\PY{p}{]}
            
            \PY{c+c1}{\PYZsh{} First Branch}
            \PY{n}{tM1} \PY{o}{=} \PY{n}{np}\PY{o}{.}\PY{n}{ones}\PY{p}{(}\PY{p}{(}\PY{l+m+mi}{6}\PY{p}{,} \PY{l+m+mi}{3}\PY{p}{)}\PY{p}{)} \PY{c+c1}{\PYZsh{} Third Channel keep 1, acts as bias}
            \PY{k}{for} \PY{n}{bank} \PY{o+ow}{in} \PY{n+nb}{range}\PY{p}{(}\PY{l+m+mi}{3}\PY{p}{)}\PY{p}{:}
                \PY{k}{for} \PY{n}{channel} \PY{o+ow}{in} \PY{n+nb}{range}\PY{p}{(}\PY{l+m+mi}{2}\PY{p}{)}\PY{p}{:}
                    \PY{n}{tM1}\PY{p}{[}\PY{l+m+mi}{2}\PY{o}{*}\PY{n}{bank}\PY{p}{,} \PY{n}{channel}\PY{p}{]} \PY{o}{=}
                        \PY{n}{np}\PY{o}{.}\PY{n}{divide}\PY{p}{(}\PY{n}{weight23}\PY{p}{[}\PY{n}{bank}\PY{p}{,} \PY{n}{channel}\PY{p}{]} \PY{o}{+} \PY{l+m+mi}{1}\PY{p}{,} \PY{l+m+mi}{2}\PY{p}{)}
                \PY{n}{tM1}\PY{p}{[}\PY{l+m+mi}{2}\PY{o}{*}\PY{n}{bank}\PY{o}{+}\PY{l+m+mi}{1}\PY{p}{,} \PY{p}{:}\PY{p}{]} \PY{o}{=} \PY{n}{np}\PY{o}{.}\PY{n}{subtract}\PY{p}{(}\PY{l+m+mf}{1.0}\PY{p}{,} \PY{n}{tM1}\PY{p}{[}\PY{l+m+mi}{2}\PY{o}{*}\PY{n}{bank}\PY{p}{,} \PY{p}{:}\PY{p}{]}\PY{p}{)}
            
            \PY{c+c1}{\PYZsh{} Second Branch}
            \PY{n}{tM2} \PY{o}{=} \PY{n}{np}\PY{o}{.}\PY{n}{ones}\PY{p}{(}\PY{p}{(}\PY{l+m+mi}{2}\PY{p}{,} \PY{l+m+mi}{3}\PY{p}{)}\PY{p}{)}
            \PY{n}{tM2}\PY{p}{[}\PY{l+m+mi}{0}\PY{p}{,} \PY{p}{:}\PY{p}{]} \PY{o}{=} \PY{n}{np}\PY{o}{.}\PY{n}{divide}\PY{p}{(}\PY{n}{np}\PY{o}{.}\PY{n}{add}\PY{p}{(}\PY{n}{weight31}\PY{p}{,} \PY{l+m+mf}{1.0}\PY{p}{)}\PY{p}{,} \PY{l+m+mf}{2.0}\PY{p}{)}
            \PY{n}{tM2}\PY{p}{[}\PY{l+m+mi}{1}\PY{p}{,} \PY{p}{:}\PY{p}{]} \PY{o}{=} \PY{n}{np}\PY{o}{.}\PY{n}{subtract}\PY{p}{(}\PY{l+m+mf}{1.0}\PY{p}{,} \PY{n}{tM2}\PY{p}{[}\PY{l+m+mi}{0}\PY{p}{,} \PY{p}{:}\PY{p}{]}\PY{p}{)}
            
            \PY{n}{tM} \PY{o}{=} \PY{n+nb}{list}\PY{p}{(}\PY{p}{)}
            \PY{n}{tM}\PY{o}{.}\PY{n}{append}\PY{p}{(}\PY{n}{np}\PY{o}{.}\PY{n}{divide}\PY{p}{(}\PY{n}{tM1}\PY{p}{,} \PY{l+m+mf}{3.0}\PY{p}{)}\PY{p}{)} \PY{c+c1}{\PYZsh{} Don\PYZsq{}t forget splitter.}
            \PY{n}{tM}\PY{o}{.}\PY{n}{append}\PY{p}{(}\PY{n}{tM2}\PY{p}{)}
            
            \PY{k}{return} \PY{n}{network}\PY{o}{.}\PY{n}{control}\PY{p}{(}\PY{n}{dummyAxonBias}\PY{p}{,} \PY{n}{tM}\PY{p}{,} \PY{n}{wlChannels}\PY{p}{)}
\end{Verbatim}

    \begin{Verbatim}[commandchars=\\\{\}]
{\color{incolor}In [{\color{incolor}5}]:} \PY{k+kn}{from} \PY{n+nn}{time} \PY{k}{import} \PY{n}{sleep}
        \PY{k}{def} \PY{n+nf}{netSim}\PY{p}{(}\PY{n}{axon2}\PY{p}{,} \PY{n}{axon3}\PY{p}{,} \PY{n}{weightBias}\PY{p}{,} \PY{n}{bOut}\PY{p}{,} \PY{n}{debug} \PY{o}{=} \PY{k+kc}{False}\PY{p}{)}\PY{p}{:}
            \PY{l+s+sd}{\PYZsq{}\PYZsq{}\PYZsq{}For a given set of weights and axon biases,}
            \PY{l+s+sd}{simulate the neural network.}
        \PY{l+s+sd}{    \PYZsq{}\PYZsq{}\PYZsq{}}
            
            \PY{n}{biasCurrent} \PY{o}{=} \PY{n}{weightBias}\PY{o}{.}\PY{n}{copy}\PY{p}{(}\PY{p}{)}
            
            \PY{c+c1}{\PYZsh{} Add In Axon Biases}
            \PY{k}{for} \PY{n}{i}\PY{p}{,} \PY{n}{ch} \PY{o+ow}{in} \PY{n+nb}{enumerate}\PY{p}{(}\PY{n}{axonChannels}\PY{p}{[}\PY{l+m+mi}{0}\PY{p}{]}\PY{p}{)}\PY{p}{:}
                \PY{n}{biasCurrent}\PY{p}{[}\PY{n}{ch}\PY{p}{]} \PY{o}{+}\PY{o}{=} \PY{n}{axon2}\PY{p}{[}\PY{n}{i}\PY{p}{]}
                
            \PY{c+c1}{\PYZsh{} Convert to mA}
            \PY{n}{nowCurrent} \PY{o}{=} \PY{n+nb}{dict}\PY{p}{(}\PY{p}{)}
            \PY{k}{for} \PY{n}{ch}\PY{p}{,} \PY{n}{v} \PY{o+ow}{in} \PY{n}{biasCurrent}\PY{o}{.}\PY{n}{items}\PY{p}{(}\PY{p}{)}\PY{p}{:}
                \PY{n}{nowCurrent}\PY{p}{[}\PY{n}{ch}\PY{p}{]} \PY{o}{=} \PY{n}{CurrentUnit}\PY{o}{.}\PY{n}{voltTo}\PY{p}{(}\PY{n}{biasCurrent}\PY{p}{[}\PY{n}{ch}\PY{p}{]}\PY{p}{,} 
                    \PY{n}{CurrentUnit}\PY{o}{.}\PY{n}{mA}\PY{p}{)}
                
            \PY{c+c1}{\PYZsh{} Write Current to instruments}
            \PY{n}{inst}\PY{o}{.}\PY{n}{setCurrentChanTuning}\PY{p}{(}\PY{n}{biasCurrent}\PY{p}{,} \PY{n}{token}\PY{p}{)}
            
            \PY{c+c1}{\PYZsh{} See what our normalized input is:}
            \PY{n}{x0} \PY{o}{=} \PY{n}{np}\PY{o}{.}\PY{n}{zeros}\PY{p}{(}\PY{l+m+mi}{2}\PY{p}{)}
            
            \PY{n}{inst}\PY{o}{.}\PY{n}{lockPhony}\PY{p}{(}\PY{n}{axons}\PY{p}{[}\PY{l+m+mi}{0}\PY{p}{]}\PY{p}{)}
            \PY{n}{axons}\PY{p}{[}\PY{l+m+mi}{0}\PY{p}{]}\PY{o}{.}\PY{n}{setAttenuation}\PY{p}{(}\PY{n}{fullAtten}\PY{p}{)}
            \PY{n}{nm}\PY{p}{,} \PY{n}{dbm} \PY{o}{=} \PY{n}{inst}\PY{o}{.}\PY{n}{spectrum}\PY{p}{(}\PY{n}{wlRange}\PY{p}{)}
            \PY{n}{inst}\PY{o}{.}\PY{n}{releasePhony}\PY{p}{(}\PY{p}{)}
            
            \PY{n}{lin} \PY{o}{=} \PY{n}{dbm2lin}\PY{p}{(}\PY{n}{dbm}\PY{p}{)} \PY{o}{/} \PY{p}{(}\PY{n}{kOSAPwr} \PY{o}{*} \PY{n}{fullAtten}\PY{p}{)}
            \PY{n}{x0} \PY{o}{=} \PY{n}{np}\PY{o}{.}\PY{n}{clip}\PY{p}{(}\PY{n}{np}\PY{o}{.}\PY{n}{interp}\PY{p}{(}\PY{n}{wlChannels}\PY{p}{[}\PY{p}{:}\PY{l+m+mi}{2}\PY{p}{]}\PY{p}{,} \PY{n}{nm}\PY{p}{,} \PY{n}{lin}\PY{p}{)}\PY{p}{,} \PY{l+m+mf}{0.0}\PY{p}{,} \PY{l+m+mf}{1.0}\PY{p}{)}
            \PY{c+c1}{\PYZsh{} Store for Return}

            \PY{c+c1}{\PYZsh{} Get three Currents in mA}
            \PY{n}{c0} \PY{o}{=} \PY{n}{np}\PY{o}{.}\PY{n}{zeros}\PY{p}{(}\PY{l+m+mi}{3}\PY{p}{)}
            \PY{k}{for} \PY{n}{i} \PY{o+ow}{in} \PY{n+nb}{range}\PY{p}{(}\PY{l+m+mi}{3}\PY{p}{)}\PY{p}{:}
                \PY{n}{network}\PY{o}{.}\PY{n}{setOsaOut}\PY{p}{(}\PY{l+m+mi}{2}\PY{o}{*}\PY{n}{i}\PY{p}{)}
                \PY{n}{nm}\PY{p}{,} \PY{n}{dbm} \PY{o}{=} \PY{n}{inst}\PY{o}{.}\PY{n}{spectrum}\PY{p}{(}\PY{n}{wlRange}\PY{p}{,} \PY{n}{avgCnt}\PY{o}{=}\PY{l+m+mi}{10}\PY{p}{)}
                \PY{n}{thru} \PY{o}{=} \PY{n}{np}\PY{o}{.}\PY{n}{sum}\PY{p}{(}\PY{n}{dbm2lin}\PY{p}{(}\PY{n}{np}\PY{o}{.}\PY{n}{interp}\PY{p}{(}\PY{n}{wlChannels}\PY{p}{,} \PY{n}{nm}\PY{p}{,} \PY{n}{dbm}\PY{p}{)}\PY{p}{)}\PY{p}{)}

                \PY{n}{network}\PY{o}{.}\PY{n}{setOsaOut}\PY{p}{(}\PY{l+m+mi}{2}\PY{o}{*}\PY{n}{i} \PY{o}{+} \PY{l+m+mi}{1}\PY{p}{)}
                \PY{n}{nm}\PY{p}{,} \PY{n}{dbm} \PY{o}{=} \PY{n}{inst}\PY{o}{.}\PY{n}{spectrum}\PY{p}{(}\PY{n}{wlRange}\PY{p}{,} \PY{n}{avgCnt}\PY{o}{=}\PY{l+m+mi}{10}\PY{p}{)}
                \PY{n}{drop} \PY{o}{=} \PY{n}{np}\PY{o}{.}\PY{n}{sum}\PY{p}{(}\PY{n}{dbm2lin}\PY{p}{(}\PY{n}{np}\PY{o}{.}\PY{n}{interp}\PY{p}{(}\PY{n}{wlChannels}\PY{p}{,} \PY{n}{nm}\PY{p}{,} \PY{n}{dbm}\PY{p}{)}\PY{p}{)}\PY{p}{)}
                
                \PY{n}{c0}\PY{p}{[}\PY{n}{i}\PY{p}{]} \PY{o}{=} \PY{n}{Resp}\PY{o}{*}\PY{p}{(}\PY{n}{thru} \PY{o}{\PYZhy{}} \PY{n}{drop}\PY{p}{)}
            
            \PY{c+c1}{\PYZsh{} Pre\PYZhy{}Amplification}
            \PY{c+c1}{\PYZsh{}if debug:}
            \PY{c+c1}{\PYZsh{}    return x0, c0}
            
            \PY{c+c1}{\PYZsh{} Send through amplifiers}
            \PY{n}{c0} \PY{o}{=} \PY{p}{(}\PY{n}{Rt}\PY{o}{*}\PY{n}{c0} \PY{o}{+} \PY{n}{bv}\PY{p}{)}\PY{o}{/}\PY{n}{Rs} \PY{o}{+} \PY{n}{axon3}
        
            \PY{k}{if} \PY{n}{debug}\PY{p}{:}
                \PY{c+c1}{\PYZsh{} Give Effective Weight Matrix}
                \PY{k}{return} \PY{n}{x0}\PY{p}{,} \PY{n}{c0}

            \PY{c+c1}{\PYZsh{} Send to axons}
            \PY{c+c1}{\PYZsh{}print(c0)}
            \PY{k}{for} \PY{n}{i}\PY{p}{,} \PY{n}{ch} \PY{o+ow}{in} \PY{n+nb}{enumerate}\PY{p}{(}\PY{n}{axonChannels}\PY{p}{[}\PY{l+m+mi}{1}\PY{p}{]}\PY{p}{)}\PY{p}{:}
                \PY{n}{nowCurrent}\PY{p}{[}\PY{n}{ch}\PY{p}{]} \PY{o}{+}\PY{o}{=} \PY{n}{c0}\PY{p}{[}\PY{n}{i}\PY{p}{]}
                
            \PY{n}{inst}\PY{o}{.}\PY{n}{setCurrentChanTuning}\PY{p}{(}\PY{n}{nowCurrent}\PY{p}{,} \PY{n}{token}\PY{p}{,} \PY{n}{CurrentUnit}\PY{o}{.}\PY{n}{mA}\PY{p}{)}
            
            \PY{c+c1}{\PYZsh{} See what our normalized hidden layer is:}
            \PY{n}{x1} \PY{o}{=} \PY{n}{np}\PY{o}{.}\PY{n}{zeros}\PY{p}{(}\PY{l+m+mi}{3}\PY{p}{)}
            
            \PY{n}{inst}\PY{o}{.}\PY{n}{lockPhony}\PY{p}{(}\PY{n}{axons}\PY{p}{[}\PY{l+m+mi}{1}\PY{p}{]}\PY{p}{)}
            \PY{n}{axons}\PY{p}{[}\PY{l+m+mi}{1}\PY{p}{]}\PY{o}{.}\PY{n}{setAttenuation}\PY{p}{(}\PY{n}{fullAtten}\PY{p}{)}
            \PY{n}{nm}\PY{p}{,} \PY{n}{dbm} \PY{o}{=} \PY{n}{inst}\PY{o}{.}\PY{n}{spectrum}\PY{p}{(}\PY{n}{wlRange}\PY{p}{)}
            \PY{n}{inst}\PY{o}{.}\PY{n}{releasePhony}\PY{p}{(}\PY{p}{)}
            
            \PY{n}{lin} \PY{o}{=} \PY{n}{dbm2lin}\PY{p}{(}\PY{n}{dbm}\PY{p}{)} \PY{o}{/} \PY{p}{(}\PY{n}{kOSAPwr} \PY{o}{*} \PY{n}{fullAtten}\PY{p}{)}
            \PY{n}{x1} \PY{o}{=} \PY{n}{np}\PY{o}{.}\PY{n}{clip}\PY{p}{(}\PY{n}{np}\PY{o}{.}\PY{n}{interp}\PY{p}{(}\PY{n}{wlChannels}\PY{p}{,} \PY{n}{nm}\PY{p}{,} \PY{n}{lin}\PY{p}{)}\PY{p}{,} \PY{l+m+mf}{0.0}\PY{p}{,} \PY{l+m+mf}{1.0}\PY{p}{)}
            \PY{c+c1}{\PYZsh{} Store for Return}
            
            \PY{c+c1}{\PYZsh{}return x0, x1}
            
            \PY{c+c1}{\PYZsh{}\PYZsq{}\PYZsq{}\PYZsq{}}
            \PY{c+c1}{\PYZsh{} Get current in mA}
            \PY{n}{network}\PY{o}{.}\PY{n}{setOsaOut}\PY{p}{(}\PY{l+m+mi}{6}\PY{p}{)}
            \PY{n}{nm}\PY{p}{,} \PY{n}{dbm} \PY{o}{=} \PY{n}{inst}\PY{o}{.}\PY{n}{spectrum}\PY{p}{(}\PY{n}{wlRange}\PY{p}{)}
            \PY{n}{thru} \PY{o}{=} \PY{n}{np}\PY{o}{.}\PY{n}{sum}\PY{p}{(}\PY{n}{dbm2lin}\PY{p}{(}\PY{n}{np}\PY{o}{.}\PY{n}{interp}\PY{p}{(}\PY{n}{wlChannels}\PY{p}{,} \PY{n}{nm}\PY{p}{,} \PY{n}{dbm}\PY{p}{)}\PY{p}{)}\PY{p}{)}
                
            \PY{n}{network}\PY{o}{.}\PY{n}{setOsaOut}\PY{p}{(}\PY{l+m+mi}{7}\PY{p}{)}
            \PY{n}{nm}\PY{p}{,} \PY{n}{dbm} \PY{o}{=} \PY{n}{inst}\PY{o}{.}\PY{n}{spectrum}\PY{p}{(}\PY{n}{wlRange}\PY{p}{)}
            \PY{n}{drop} \PY{o}{=} \PY{n}{np}\PY{o}{.}\PY{n}{sum}\PY{p}{(}\PY{n}{dbm2lin}\PY{p}{(}\PY{n}{np}\PY{o}{.}\PY{n}{interp}\PY{p}{(}\PY{n}{wlChannels}\PY{p}{,} \PY{n}{nm}\PY{p}{,} \PY{n}{dbm}\PY{p}{)}\PY{p}{)}\PY{p}{)}
                
            \PY{n}{c1} \PY{o}{=} \PY{n}{Resp}\PY{o}{*}\PY{p}{(}\PY{n}{thru} \PY{o}{\PYZhy{}} \PY{n}{drop}\PY{p}{)}
            
            \PY{n}{y} \PY{o}{=} \PY{n}{Rt2}\PY{o}{*}\PY{n}{c1} \PY{o}{+} \PY{n}{bOut}
            
            \PY{k}{return} \PY{n}{x0}\PY{p}{,} \PY{n}{y}
\end{Verbatim}

    \begin{Verbatim}[commandchars=\\\{\}]
{\color{incolor}In [{\color{incolor}6}]:} \PY{c+c1}{\PYZsh{}\PYZsh{}\PYZsh{} Function to Plot Sweep of Network}
        
        \PY{k}{def} \PY{n+nf}{sweepNetwork}\PY{p}{(}\PY{n}{weight23}\PY{p}{,} \PY{n}{weight31}\PY{p}{,} \PY{n}{axonBias}\PY{p}{,} \PY{n}{outBias}\PY{p}{,} \PY{n}{sweepNum}\PY{p}{)}\PY{p}{:}
        
            \PY{c+c1}{\PYZsh{} Calculate currents that give said weights}
            \PY{n}{weightBias} \PY{o}{=} \PY{n}{getWeightsBias}\PY{p}{(}\PY{n}{weight23}\PY{p}{,} \PY{n}{weight31}\PY{p}{)}

            \PY{c+c1}{\PYZsh{} Run Sweep}
            \PY{n}{cList} \PY{o}{=} \PY{n}{np}\PY{o}{.}\PY{n}{linspace}\PY{p}{(}\PY{l+m+mi}{0}\PY{p}{,} \PY{l+m+mf}{0.15}\PY{p}{,} \PY{n}{sweepNum}\PY{p}{)}
            \PY{n}{z} \PY{o}{=} \PY{n}{np}\PY{o}{.}\PY{n}{zeros}\PY{p}{(}\PY{p}{(}\PY{n}{sweepNum}\PY{p}{,} \PY{n}{sweepNum}\PY{p}{)}\PY{p}{)}
            \PY{n}{x} \PY{o}{=} \PY{n}{np}\PY{o}{.}\PY{n}{zeros}\PY{p}{(}\PY{n}{sweepNum}\PY{p}{)}
            \PY{n}{y} \PY{o}{=} \PY{n}{np}\PY{o}{.}\PY{n}{zeros}\PY{p}{(}\PY{n}{sweepNum}\PY{p}{)}
        
            \PY{k}{for} \PY{n}{i}\PY{p}{,} \PY{n}{c1} \PY{o+ow}{in} \PY{n+nb}{enumerate}\PY{p}{(}\PY{n}{cList}\PY{p}{)}\PY{p}{:}
                \PY{n+nb}{print}\PY{p}{(}\PY{l+s+s2}{\PYZdq{}}\PY{l+s+s2}{Step: }\PY{l+s+s2}{\PYZdq{}} \PY{o}{+} \PY{n+nb}{str}\PY{p}{(}\PY{n}{i}\PY{p}{)}\PY{p}{)}
                \PY{k}{for} \PY{n}{j}\PY{p}{,} \PY{n}{c2} \PY{o+ow}{in} \PY{n+nb}{enumerate}\PY{p}{(}\PY{n}{cList}\PY{p}{)}\PY{p}{:}
                    \PY{n}{x0}\PY{p}{,} \PY{n}{out} \PY{o}{=} \PY{n}{netSim}\PY{p}{(}\PY{p}{[}\PY{n}{c1}\PY{p}{,} \PY{n}{c2}\PY{p}{]}\PY{p}{,}
                        \PY{n}{axonBias}\PY{p}{,} \PY{n}{weightBias}\PY{p}{,} \PY{n}{outBias}\PY{p}{)}
                    \PY{n}{x}\PY{p}{[}\PY{n}{i}\PY{p}{]} \PY{o}{+}\PY{o}{=} \PY{n}{x0}\PY{p}{[}\PY{l+m+mi}{0}\PY{p}{]}
                    \PY{n}{y}\PY{p}{[}\PY{n}{j}\PY{p}{]} \PY{o}{+}\PY{o}{=} \PY{n}{x0}\PY{p}{[}\PY{l+m+mi}{1}\PY{p}{]}
                    \PY{n}{z}\PY{p}{[}\PY{n}{i}\PY{p}{,} \PY{n}{j}\PY{p}{]} \PY{o}{=} \PY{n}{out}
                    
            \PY{c+c1}{\PYZsh{} Average value of x0[0], x0[1] for values of cList}
            \PY{n}{x} \PY{o}{=} \PY{n}{np}\PY{o}{.}\PY{n}{divide}\PY{p}{(}\PY{n}{x}\PY{p}{,} \PY{n}{sweepNum}\PY{p}{)}
            \PY{n}{y} \PY{o}{=} \PY{n}{np}\PY{o}{.}\PY{n}{divide}\PY{p}{(}\PY{n}{y}\PY{p}{,} \PY{n}{sweepNum}\PY{p}{)}
        
            \PY{n}{f} \PY{o}{=} \PY{n}{interpolate}\PY{o}{.}\PY{n}{interp2d}\PY{p}{(}\PY{n}{x}\PY{p}{,}\PY{n}{y}\PY{p}{,}\PY{n}{z}\PY{p}{,}\PY{n}{kind}\PY{o}{=}\PY{l+s+s1}{\PYZsq{}}\PY{l+s+s1}{cubic}\PY{l+s+s1}{\PYZsq{}}\PY{p}{)}
        
            \PY{n}{delta} \PY{o}{=} \PY{l+m+mf}{0.025}
            \PY{n}{maxVal} \PY{o}{=} \PY{l+m+mf}{0.8}
            \PY{n}{X} \PY{o}{=} \PY{n}{Y} \PY{o}{=} \PY{n}{np}\PY{o}{.}\PY{n}{arange}\PY{p}{(}\PY{l+m+mf}{0.0}\PY{p}{,} \PY{n}{maxVal}\PY{p}{,} \PY{n}{delta}\PY{p}{)}
            \PY{n}{Z} \PY{o}{=} \PY{n}{f}\PY{p}{(}\PY{n}{X}\PY{p}{,} \PY{n}{Y}\PY{p}{)}
            \PY{k}{return} \PY{n}{maxVal}\PY{p}{,} \PY{n}{Z}
\end{Verbatim}

    \textbf{Test on some sample weights:}

    \begin{Verbatim}[commandchars=\\\{\}]
{\color{incolor}In [{\color{incolor}7}]:} \PY{c+c1}{\PYZsh{} Global Parameters}
        \PY{n}{Resp} \PY{o}{=} \PY{l+m+mf}{0.9}
        \PY{n}{Rt} \PY{o}{=} \PY{l+m+mi}{7500} \PY{c+c1}{\PYZsh{} (7.5MOhm)}
        \PY{n}{Rt2} \PY{o}{=} \PY{l+m+mi}{7500} \PY{c+c1}{\PYZsh{} (7.5MOhm)}
        \PY{n}{Rs} \PY{o}{=} \PY{l+m+mi}{2} \PY{c+c1}{\PYZsh{}(2kOhm)}
        \PY{n}{bv} \PY{o}{=} \PY{l+m+mi}{5} \PY{c+c1}{\PYZsh{} 5V}
        
        \PY{c+c1}{\PYZsh{} Should be equivalent of 2\PYZhy{}Neuron Perceptron}
        \PY{n}{weight23} \PY{o}{=} \PY{n}{np}\PY{o}{.}\PY{n}{array}\PY{p}{(}\PY{p}{[}\PY{p}{[}\PY{o}{\PYZhy{}}\PY{l+m+mf}{0.8}\PY{p}{,} \PY{l+m+mf}{0.8}\PY{p}{]}\PY{p}{,} \PY{p}{[}\PY{l+m+mf}{0.8}\PY{p}{,} \PY{o}{\PYZhy{}}\PY{l+m+mf}{0.8}\PY{p}{]}\PY{p}{,} \PY{p}{[}\PY{l+m+mf}{0.0}\PY{p}{,} \PY{l+m+mf}{0.0}\PY{p}{]}\PY{p}{]}\PY{p}{)}
        \PY{n}{weight31} \PY{o}{=} \PY{n}{np}\PY{o}{.}\PY{n}{array}\PY{p}{(}\PY{p}{[}\PY{p}{[}\PY{l+m+mf}{0.8}\PY{p}{,} \PY{o}{\PYZhy{}}\PY{l+m+mf}{0.8}\PY{p}{,} \PY{l+m+mf}{0.0}\PY{p}{]}\PY{p}{]}\PY{p}{)}
        \PY{n}{axonBias} \PY{o}{=} \PY{o}{\PYZhy{}}\PY{l+m+mf}{2.8}\PY{o}{*}\PY{n}{np}\PY{o}{.}\PY{n}{array}\PY{p}{(}\PY{p}{[}\PY{l+m+mf}{1.0}\PY{p}{,} \PY{l+m+mf}{1.0}\PY{p}{,} \PY{l+m+mf}{1.0}\PY{p}{]}\PY{p}{)}
        \PY{n}{outBias} \PY{o}{=} \PY{o}{\PYZhy{}}\PY{l+m+mi}{1}
        
        \PY{n}{maxVal}\PY{p}{,} \PY{n}{surface} \PY{o}{=} \PY{n}{sweepNetwork}\PY{p}{(}\PY{n}{weight23}\PY{p}{,}
            \PY{n}{weight31}\PY{p}{,} \PY{n}{axonBias}\PY{p}{,} \PY{n}{outBias}\PY{p}{,} \PY{l+m+mi}{10}\PY{p}{)}
\end{Verbatim}

    \begin{Verbatim}[commandchars=\\\{\}]
{\color{incolor}In [{\color{incolor}8}]:} \PY{n}{im} \PY{o}{=} \PY{n}{plt}\PY{o}{.}\PY{n}{imshow}\PY{p}{(}\PY{n}{surface}\PY{p}{,} \PY{n}{interpolation}\PY{o}{=}\PY{l+s+s1}{\PYZsq{}}\PY{l+s+s1}{bilinear}\PY{l+s+s1}{\PYZsq{}}\PY{p}{,} \PY{n}{cmap}\PY{o}{=}\PY{n}{cm}\PY{o}{.}\PY{n}{RdYlGn}\PY{p}{,}
            \PY{n}{origin}\PY{o}{=}\PY{l+s+s1}{\PYZsq{}}\PY{l+s+s1}{lower}\PY{l+s+s1}{\PYZsq{}}\PY{p}{,} \PY{n}{extent}\PY{o}{=}\PY{p}{[}\PY{l+m+mi}{0}\PY{p}{,} \PY{n}{maxVal}\PY{p}{,} \PY{l+m+mi}{0}\PY{p}{,} \PY{n}{maxVal}\PY{p}{]}\PY{p}{,}
            \PY{n}{vmax}\PY{o}{=}\PY{n+nb}{abs}\PY{p}{(}\PY{n}{surface}\PY{p}{)}\PY{o}{.}\PY{n}{max}\PY{p}{(}\PY{p}{)}\PY{p}{,} \PY{n}{vmin}\PY{o}{=}\PY{o}{\PYZhy{}}\PY{n+nb}{abs}\PY{p}{(}\PY{n}{surface}\PY{p}{)}\PY{o}{.}\PY{n}{max}\PY{p}{(}\PY{p}{)}\PY{p}{)}
        \PY{n}{plt}\PY{o}{.}\PY{n}{colorbar}\PY{p}{(}\PY{p}{)}
        \PY{n}{plt}\PY{o}{.}\PY{n}{show}\PY{p}{(}\PY{p}{)}
        \PY{c+c1}{\PYZsh{} Classification}
        \PY{n+nb+bp}{cls} \PY{o}{=} \PY{n}{np}\PY{o}{.}\PY{n}{sign}\PY{p}{(}\PY{n}{surface}\PY{p}{)}
        \PY{n}{im} \PY{o}{=} \PY{n}{plt}\PY{o}{.}\PY{n}{imshow}\PY{p}{(}\PY{n+nb+bp}{cls}\PY{p}{,} \PY{n}{interpolation}\PY{o}{=}\PY{l+s+s1}{\PYZsq{}}\PY{l+s+s1}{bilinear}\PY{l+s+s1}{\PYZsq{}}\PY{p}{,} \PY{n}{cmap}\PY{o}{=}\PY{n}{cm}\PY{o}{.}\PY{n}{RdYlGn}\PY{p}{,}
            \PY{n}{origin}\PY{o}{=}\PY{l+s+s1}{\PYZsq{}}\PY{l+s+s1}{lower}\PY{l+s+s1}{\PYZsq{}}\PY{p}{,} \PY{n}{extent}\PY{o}{=}\PY{p}{[}\PY{l+m+mi}{0}\PY{p}{,} \PY{n}{maxVal}\PY{p}{,} \PY{l+m+mi}{0}\PY{p}{,} \PY{n}{maxVal}\PY{p}{]}\PY{p}{,}
            \PY{n}{vmax}\PY{o}{=}\PY{l+m+mi}{1}\PY{p}{,} \PY{n}{vmin}\PY{o}{=}\PY{o}{\PYZhy{}}\PY{l+m+mi}{1}\PY{p}{)}
        \PY{n}{plt}\PY{o}{.}\PY{n}{show}\PY{p}{(}\PY{p}{)}
\end{Verbatim}

    
    
    \subsubsection{Step 3: Run Backprop for Virtual Weights and
Biases}\label{step-3-run-backprop-for-virtual-weights-and-biases}

(See 2-3-1\_FFNet\_Backprop)

    \begin{Verbatim}[commandchars=\\\{\}]
{\color{incolor}In [{\color{incolor}11}]:} \PY{c+c1}{\PYZsh{} Pull Network Params}
         \PY{n+nb}{print}\PY{p}{(}\PY{n}{np}\PY{o}{.}\PY{n}{diag}\PY{p}{(}\PY{n}{K}\PY{p}{)}\PY{p}{[}\PY{l+m+mi}{2}\PY{p}{:}\PY{l+m+mi}{5}\PY{p}{]}\PY{p}{)}
         
         \PY{n+nb}{print}\PY{p}{(}\PY{p}{[}\PY{n}{CurrentUnit}\PY{o}{.}\PY{n}{voltTo}\PY{p}{(}\PY{n}{heatBias}\PY{p}{[}\PY{n}{c}\PY{p}{]}\PY{p}{,} \PY{n}{CurrentUnit}\PY{o}{.}\PY{n}{mA}\PY{p}{)}
            \PY{k}{for} \PY{n}{c} \PY{o+ow}{in} \PY{n}{axonChannels}\PY{p}{[}\PY{l+m+mi}{1}\PY{p}{]}\PY{p}{]}\PY{p}{)}
\end{Verbatim}

    \begin{Verbatim}[commandchars=\\\{\}]
[ 20.  20.  20.]
[5.6, 6.4, 7.2]

    \end{Verbatim}

    \begin{Verbatim}[commandchars=\\\{\}]
{\color{incolor}In [{\color{incolor}60}]:} \PY{c+c1}{\PYZsh{}\PYZsh{}\PYZsh{} Output of 231 Backprop:}
         
         \PY{c+c1}{\PYZsh{}\PYZsh{}\PYZsh{} Best Output So Far:}
         \PY{n}{virt23} \PY{o}{=} \PY{n}{np}\PY{o}{.}\PY{n}{array}\PY{p}{(}\PY{p}{[}\PY{p}{[}\PY{l+m+mf}{1.30109513}\PY{p}{,} \PY{l+m+mf}{0.90975827}\PY{p}{]}\PY{p}{,}
            \PY{p}{[}\PY{o}{\PYZhy{}}\PY{l+m+mf}{0.79916418}\PY{p}{,} \PY{o}{\PYZhy{}}\PY{l+m+mf}{0.85981083}\PY{p}{]}\PY{p}{,} \PY{p}{[}\PY{l+m+mf}{0.9742766}\PY{p}{,}  \PY{o}{\PYZhy{}}\PY{l+m+mf}{1.02000749}\PY{p}{]}\PY{p}{]}\PY{p}{)}
         \PY{n}{virt31} \PY{o}{=} \PY{n}{np}\PY{o}{.}\PY{n}{array}\PY{p}{(}\PY{p}{[} \PY{o}{\PYZhy{}}\PY{l+m+mf}{0.416287088}\PY{p}{,} \PY{l+m+mf}{1.124845219}\PY{p}{,} \PY{o}{\PYZhy{}}\PY{l+m+mf}{0.90137167}\PY{p}{]}\PY{p}{)}
         \PY{n}{virtBias} \PY{o}{=} \PY{n}{np}\PY{o}{.}\PY{n}{array}\PY{p}{(}\PY{p}{[} \PY{l+m+mf}{1.0609535}\PY{p}{,} \PY{l+m+mf}{0.65707952}\PY{p}{,}  \PY{o}{\PYZhy{}}\PY{l+m+mf}{0.01618425}\PY{p}{]}\PY{p}{)}
         \PY{n}{outBias} \PY{o}{=} \PY{l+m+mf}{0.4359157047300002}
\end{Verbatim}

    \subsubsection{Step 4: Convert to "Real"
Weights}\label{step-4-convert-to-real-weights}

Basically multiply by physical parameters to get the actual 2t-1 we send
to the network.

    \begin{Verbatim}[commandchars=\\\{\}]
{\color{incolor}In [{\color{incolor}61}]:} \PY{n}{p} \PY{o}{=} \PY{n}{kOSAPwr} \PY{o}{*} \PY{n}{fullAtten}
         \PY{c+c1}{\PYZsh{} Global Parameters}
         \PY{n}{Resp} \PY{o}{=} \PY{l+m+mf}{0.9}
         \PY{n}{Rt} \PY{o}{=} \PY{l+m+mi}{15000} \PY{c+c1}{\PYZsh{} (15MOhm)}
         \PY{n}{Rt2} \PY{o}{=} \PY{l+m+mi}{3000} \PY{c+c1}{\PYZsh{} (3MOhm)}
         \PY{n}{Rs} \PY{o}{=} \PY{l+m+mi}{1} \PY{c+c1}{\PYZsh{}(1kOhm)}
         \PY{n}{bv} \PY{o}{=} \PY{l+m+mi}{4} \PY{c+c1}{\PYZsh{} 4V}
\end{Verbatim}

    \begin{Verbatim}[commandchars=\\\{\}]
{\color{incolor}In [{\color{incolor}62}]:} \PY{n}{weight23} \PY{o}{=} \PY{n}{np}\PY{o}{.}\PY{n}{multiply}\PY{p}{(}\PY{n}{virt23}\PY{p}{,} \PY{n}{Rs}\PY{o}{*}\PY{l+m+mi}{6}\PY{o}{/}\PY{p}{(}\PY{n}{Rt}\PY{o}{*}\PY{n}{Resp}\PY{o}{*}\PY{n}{p}\PY{p}{)}\PY{p}{)}
         \PY{n+nb}{print}\PY{p}{(}\PY{n}{weight23}\PY{p}{)}
\end{Verbatim}

    \begin{Verbatim}[commandchars=\\\{\}]
[[ 0.5782645   0.40433701]
 [-0.35518408 -0.38213815]
 [ 0.43301182 -0.45333666]]

    \end{Verbatim}

    \begin{Verbatim}[commandchars=\\\{\}]
{\color{incolor}In [{\color{incolor}63}]:} \PY{n}{weight31} \PY{o}{=} \PY{n}{np}\PY{o}{.}\PY{n}{multiply}\PY{p}{(}\PY{n}{virt31}\PY{p}{,} \PY{l+m+mi}{2}\PY{o}{/}\PY{p}{(}\PY{n}{Rt2}\PY{o}{*}\PY{n}{Resp}\PY{o}{*}\PY{n}{p}\PY{p}{)}\PY{p}{)}
         \PY{n+nb}{print}\PY{p}{(}\PY{n}{weight31}\PY{p}{)}
\end{Verbatim}

    \begin{Verbatim}[commandchars=\\\{\}]
[-0.30836081  0.83321868 -0.66768272]

    \end{Verbatim}

    \begin{Verbatim}[commandchars=\\\{\}]
{\color{incolor}In [{\color{incolor}64}]:} \PY{n}{axonBias} \PY{o}{=} \PY{n}{np}\PY{o}{.}\PY{n}{zeros}\PY{p}{(}\PY{p}{(}\PY{l+m+mi}{1}\PY{p}{,} \PY{l+m+mi}{3}\PY{p}{)}\PY{p}{)}
         \PY{n}{axonBias} \PY{o}{=} \PY{n}{np}\PY{o}{.}\PY{n}{subtract}\PY{p}{(}\PY{n}{virtBias}\PY{p}{,} \PY{p}{(}\PY{n}{Rt}\PY{o}{*}\PY{n}{Resp}\PY{o}{*}\PY{n}{p}\PY{o}{/}\PY{l+m+mi}{6}\PY{p}{)} \PY{o}{+} \PY{p}{(}\PY{n}{bv}\PY{o}{/}\PY{n}{Rs}\PY{p}{)}\PY{p}{)}
         \PY{n+nb}{print}\PY{p}{(}\PY{n}{axonBias}\PY{p}{)}
\end{Verbatim}

    \begin{Verbatim}[commandchars=\\\{\}]
[-5.1890465  -5.59292048 -6.26618425]

    \end{Verbatim}

    \begin{Verbatim}[commandchars=\\\{\}]
{\color{incolor}In [{\color{incolor}65}]:} \PY{c+c1}{\PYZsh{} Check Virtual Weights}
         \PY{n}{weightBias} \PY{o}{=} \PY{n}{getWeightsBias}\PY{p}{(}\PY{n}{weight23}\PY{p}{,} \PY{n}{weight31}\PY{p}{)}
\end{Verbatim}

    \begin{Verbatim}[commandchars=\\\{\}]
{\color{incolor}In [{\color{incolor}66}]:} \PY{c+c1}{\PYZsh{} Check Virtual Weights}
         \PY{n}{x0}\PY{p}{,} \PY{n}{c0} \PY{o}{=} \PY{n}{netSim}\PY{p}{(}\PY{p}{[}\PY{l+m+mf}{0.1}\PY{p}{,} \PY{l+m+mf}{0.1}\PY{p}{]}\PY{p}{,} \PY{n}{axonBias}\PY{p}{,} \PY{n}{weightBias}\PY{p}{,} \PY{n}{outBias}\PY{p}{,} \PY{k+kc}{True}\PY{p}{)}
         \PY{n+nb}{print}\PY{p}{(}\PY{l+s+s2}{\PYZdq{}}\PY{l+s+s2}{Actual: }\PY{l+s+s2}{\PYZdq{}} \PY{o}{+} \PY{n+nb}{str}\PY{p}{(}\PY{n}{c0}\PY{p}{)}\PY{p}{)}
         \PY{n}{cWant} \PY{o}{=} \PY{n}{np}\PY{o}{.}\PY{n}{dot}\PY{p}{(}\PY{n}{virt23}\PY{p}{,} \PY{n}{x0}\PY{p}{)} \PY{o}{+} \PY{n}{virtBias}
         \PY{n+nb}{print}\PY{p}{(}\PY{l+s+s2}{\PYZdq{}}\PY{l+s+s2}{Target: }\PY{l+s+s2}{\PYZdq{}} \PY{o}{+} \PY{n+nb}{str}\PY{p}{(}\PY{n}{cWant}\PY{p}{)}\PY{p}{)}
\end{Verbatim}

    \begin{Verbatim}[commandchars=\\\{\}]
Actual: [ 2.61785122 -0.48164702 -0.18952191]
Target: [ 2.60217079 -0.52201333 -0.17577987]

    \end{Verbatim}

    \begin{Verbatim}[commandchars=\\\{\}]
{\color{incolor}In [{\color{incolor}69}]:} \PY{n}{maxVal}\PY{p}{,} \PY{n}{surface} \PY{o}{=} 
    \PY{n}{sweepNetwork}\PY{p}{(}\PY{n}{weight23}\PY{p}{,} \PY{n}{weight31}\PY{p}{,} \PY{n}{axonBias}\PY{p}{,} \PY{n}{outBias}\PY{p}{,} \PY{l+m+mi}{20}\PY{p}{)}
\end{Verbatim}

    \begin{Verbatim}[commandchars=\\\{\}]
{\color{incolor}In [{\color{incolor}71}]:} \PY{c+c1}{\PYZsh{} Plot Normalized Output}
         \PY{n}{im} \PY{o}{=} \PY{n}{plt}\PY{o}{.}\PY{n}{imshow}\PY{p}{(}\PY{n}{surface}\PY{p}{,} \PY{n}{interpolation}\PY{o}{=}\PY{l+s+s1}{\PYZsq{}}\PY{l+s+s1}{bilinear}\PY{l+s+s1}{\PYZsq{}}\PY{p}{,} \PY{n}{cmap}\PY{o}{=}\PY{n}{cm}\PY{o}{.}\PY{n}{RdYlGn}\PY{p}{,}
            \PY{n}{origin}\PY{o}{=}\PY{l+s+s1}{\PYZsq{}}\PY{l+s+s1}{lower}\PY{l+s+s1}{\PYZsq{}}\PY{p}{,} \PY{n}{extent}\PY{o}{=}\PY{p}{[}\PY{l+m+mi}{0}\PY{p}{,} \PY{n}{maxVal}\PY{p}{,} \PY{l+m+mi}{0}\PY{p}{,} \PY{n}{maxVal}\PY{p}{]}\PY{p}{,}
            \PY{n}{vmax}\PY{o}{=}\PY{l+m+mi}{1}\PY{p}{,} \PY{n}{vmin}\PY{o}{=}\PY{o}{\PYZhy{}}\PY{l+m+mi}{1}\PY{p}{)}
         \PY{n}{plt}\PY{o}{.}\PY{n}{colorbar}\PY{p}{(}\PY{p}{)}
         \PY{n}{plt}\PY{o}{.}\PY{n}{show}\PY{p}{(}\PY{p}{)}
         \PY{c+c1}{\PYZsh{} Classification}
         \PY{n+nb+bp}{cls} \PY{o}{=} \PY{n}{np}\PY{o}{.}\PY{n}{sign}\PY{p}{(}\PY{n}{surface}\PY{p}{)}
         \PY{n}{im} \PY{o}{=} \PY{n}{plt}\PY{o}{.}\PY{n}{imshow}\PY{p}{(}\PY{n+nb+bp}{cls}\PY{p}{,} \PY{n}{interpolation}\PY{o}{=}\PY{l+s+s1}{\PYZsq{}}\PY{l+s+s1}{bilinear}\PY{l+s+s1}{\PYZsq{}}\PY{p}{,} \PY{n}{cmap}\PY{o}{=}\PY{n}{cm}\PY{o}{.}\PY{n}{RdYlGn}\PY{p}{,}
                \PY{n}{origin}\PY{o}{=}\PY{l+s+s1}{\PYZsq{}}\PY{l+s+s1}{lower}\PY{l+s+s1}{\PYZsq{}}\PY{p}{,} \PY{n}{extent}\PY{o}{=}\PY{p}{[}\PY{l+m+mi}{0}\PY{p}{,} \PY{n}{maxVal}\PY{p}{,} \PY{l+m+mi}{0}\PY{p}{,} \PY{n}{maxVal}\PY{p}{]}\PY{p}{,}
                \PY{n}{vmax}\PY{o}{=}\PY{l+m+mi}{1}\PY{p}{,} \PY{n}{vmin}\PY{o}{=}\PY{o}{\PYZhy{}}\PY{l+m+mi}{1}\PY{p}{)}
         
         \PY{n}{plt}\PY{o}{.}\PY{n}{show}\PY{p}{(}\PY{p}{)}
\end{Verbatim}

    

%% file: main.bbl
\newpage

\begin{thebibliography}{100}

\bibitem{go} Silver, David, et al. "Mastering the game of Go with deep neural networks and tree search." Nature 529.7587 (2016): 484-489.

\bibitem{colah} http://colah.github.io/posts/2014-03-NN-Manifolds-Topology/

\bibitem{truenorth} Akopyan, Filipp, et al. "TrueNorth: Design and tool flow of a 65 mW 1 million neuron programmable neurosynaptic chip." IEEE Transactions on Computer-Aided Design of Integrated Circuits and Systems 34.10 (2015): 1537-1557.

\bibitem{hicann} Friedmann, Simon, et al. "Reward-based learning under hardware constraints-using a RISC processor embedded in a neuromorphic substrate." arXiv preprint arXiv:1303.6708 (2013).

\bibitem{neurogrid} Benjamin, Ben Varkey, et al. "Neurogrid: A mixed-analog-digital multichip system for large-scale neural simulations." Proceedings of the IEEE 102.5 (2014): 699-716.

\bibitem{demo} A. Tait, et al., "Demonstration of a silicon photonic neural network," {\em Photonics Society Summer Topical Meeting Series (SUM)}, IEEE, 2016.

\bibitem{control} A. Tait, et al., "Multi-channel control for microring
weight banks," Opt. Express 24, 8895-8906 (2016).

\bibitem{mdm} L. Luo, et al., "WDM-compatible mode-division multiplexing on silicon chip," Nature Communications, vol. 5, no. 3069, Macmillian Publishers Limited, 2014.

\end{thebibliography}
